\documentclass[11pt]{article}
\usepackage{epsf,amssymb,cite}

\newcommand{\xtra}[1]{{.}}
\renewcommand{\xtra}[1]{{, \tt hep-th/#1.}}
\newcommand{\xxtra}[1]{{, \tt hep-th/#1}}
\newcommand{\xtrac}[1]{{.}}
\renewcommand{\xtrac}[1]{{, \tt cond-mat/#1.}}

\newcommand{\be}{\begin{equation}}
\newcommand{\ee}{\end{equation}}
\newcommand{\eq}{\begin{equation}}
\newcommand{\en}{\end{equation}}
\newcommand{\bea}{\begin{eqnarray}}
\newcommand{\eea}{\end{eqnarray}}

\newcommand\B[3]{{\ensuremath{ {}^{\sscr(#1)}\!B_{#2}^{\sscr #3}}}}
\renewcommand{\bar}{\overline}
\newcommand{\mathematica}[1]{{}}
\newcommand{\bp}[3]{{\ensuremath{ \phi_{#1}^{(#2#3)}}}}

\newcommand\C[3]{{\ensuremath{ C_{#1 #2}{}^{#3} }}}

\newcommand{\cdd}[1]{{{\cdot} 10^{-#1}}}
\newcommand{\cev}[1]{\langle \,#1\,|}
\newcommand{\cg}{{\tilde g}} 
\newcommand{\cG}{{\cal G}}
\newcommand{\cH}{{\cal H}}
\newcommand{\cI}{{\cal I}}

\newcommand{\CK}{{\cal K}}

\newcommand{\ds}{\displaystyle}
\newcommand{\D}{{{\rm d}}}

\newcommand\Eblk{{\cal E}_{\rm bulk}}
\newcommand{\ep}{\varepsilon}

\newcommand{\fract}[2]{{\textstyle\frac{#1}{#2}}}

\newcommand{\Ga}{\Gamma}

\renewcommand\hat{\widehat}
\newcommand{\hyp}{{\ensuremath{{}_2^{\vphantom\phi}{\rm F}_{\!1}^{\vphantom\phi}}}}

\newcommand{\iintd}{\int^{\infty}_{-\infty}\! \D\theta}

\newcommand{\ket}[1]{|#1\rangle}

\def\LY{Lee-Yang}
\def\lr{l}
\def\hc{\hat h_c}
\def\hc{{h_{\rm crit}}}
\def\hhc{{{\hat h}_{\rm crit}}}

\newcommand{\ML}{{\newl}}

\newcommand{\lf}{\left}

\newcommand{\m}{\phantom{-}}
\newcommand{\mm}[2]{{\vphantom{\vbox to 6mm{}}
\renewcommand{\arraystretch}{0.9}
\begin{array}{l}#1\\#2\end{array}}}

\newcommand{\nn}{\nonumber}
\newcommand{\newL}{R}
\newcommand{\newl}{r}

\newcommand{\One}{{\hbox{{\rm 1{\hbox to 1.5pt{\hss\rm1}}}}}}
\renewcommand{\One}{{\mathbb 1}}
\renewcommand{\One}{{\rm 1\!\!1}}

\newcommand{\prtial}{\frac{\partial}{\partial\theta}}

\newcommand{\resection}[1]{\setcounter{equation}{0}\section{#1}}
\newcommand{\ri}{\right}

\newcommand{\sscr}{\scriptscriptstyle}

\newcommand{\te}{\theta}
\renewcommand{\tilde}{\widetilde}

\newcommand{\vac}{\vec 0}
\renewcommand{\vec}[1]{|\,#1\,\rangle}
\newcommand{\vev}[1]{\langle\,#1\,\rangle}
\newcommand{\veev}[2]{\langle\,#1\,|\,#2\,\rangle}

\newcommand{\IJMP}[1]{Int.\ J.\ Mod.\ Phys.\ {\bf #1}}
\newcommand{\JP}[1]{J.\ Phys.\ {\bf #1}}

\newcommand{\NP}[1]{Nucl.\ Phys.\ {\bf #1}}
\newcommand{\PL}[1]{Phys.\ Lett.\ {\bf #1}}
\newcommand{\PR}[1]{Phys.\ Rev.\ {\bf #1}}

\newcommand{\AlBZ}{Al.B.~Zamolodchikov}

\newcommand{\beqcol}{\begin{array}{rcl}}
\newcommand{\eeqcol}{\end{array}}

\begin{document}
%
%
\begin{titlepage}
\vskip 0.5cm
\begin{flushright}
T00/070  \\
KCL-MTH-00-29 \\
ITFA 00-11 \\
DTP/00/51 \\
{\tt hep-th/0007077}\\
Monday, July 10, 2000 \\
\end{flushright}
\vskip .8cm
\begin{center}
{\Large {\bf One-point functions in perturbed}} \\[5pt]
{\Large {\bf boundary conformal field theories } }
\end{center}
\vskip 0.8cm
\centerline{P.~Dorey%
\footnote{e-mail: {\tt P.E.Dorey@durham.ac.uk}},
M.~Pillin\footnote{e-mail: {\tt mathias.pillin@etas.de}},
R.~Tateo\footnote{e-mail: {\tt tateo@wins.uva.nl}}
and G.M.T.~Watts\footnote{e-mail: {\tt gmtw@mth.kcl.ac.uk}}
}
\vskip 0.6cm
\centerline{${}^1$\sl Department of Mathematical Sciences,}
\centerline{\sl  University of Durham, Durham DH1 3LE,
England\,}
\centerline{and}
\centerline{\sl SPhT, CEA-Saclay, F-91191 Gif-sur-Yvette Cedex, France}
\vskip 0.2cm
\centerline{${}^2$\sl 
ETAS, PTS-A, Borsigstr.~10,}
\centerline{\sl
D-70469 Stuttgart, Germany\,}
\vskip 0.2cm
\centerline{${}^3$\sl 
Universiteit van Amsterdam, Inst.~voor Theoretische Fysica\,}
\centerline{\sl
1018 XE Amsterdam, The Netherlands\,}
\vskip 0.2cm
\centerline{${}^{4}$\sl Mathematics Department, }
\centerline{\sl King's College London, Strand, London WC2R 2LS, U.K.}
\vskip 0.9cm
\begin{abstract}
\vskip0.15cm
\noindent
We consider the one-point functions of bulk and boundary fields
in the scaling Lee-Yang model for various combinations of bulk and
boundary perturbations. The one-point functions of the bulk fields are
analysed using the truncated conformal space approach and the
form-factor expansion.
Good agreement is found between the results of the
two methods, though we find that the expression for the general boundary 
state given by Ghoshal and Zamolodchikov has to be corrected slightly. For 
the boundary fields we use thermodynamic Bethe ansatz equations to find
exact expressions for the strip and semi-infinite cylinder geometries.
We also find a novel off-critical identity between the cylinder
partition functions of models with differing boundary conditions, and
use this to investigate the regions of boundary-induced instability
exhibited by the model on a finite strip.

\end{abstract}
\end{titlepage}
\setcounter{footnote}{0}
\def\thefootnote{\fnsymbol{footnote}}

\resection{Introduction}

The purpose of this paper is to make a detailed analysis of the
one-point functions in a particular integrable boundary quantum field
theory, the boundary scaling \LY\ model. Such a study
is of intrinsic interest, allowing one to check the consistency of 
various different approaches to the computation of these quantities.
It also serves to illuminate some issues that have arisen in recent
investigations of the spectra and boundary entropies ($g$--functions)
of integrable models with boundaries.

After a brief review of the pertinent features
of the boundary scaling \LY\ model in section 2, sections 3 and 4 of
the paper are devoted to the one-point
functions of bulk fields in the presence of a boundary.
Since these
generally have a dependence on the distance from the boundary,
their exact determination is a non-trivial problem, of
similar difficulty to the calculation of two-point functions in the
absence of boundaries. For this reason, apart from some simple
cases associated with the Ising model~\cite{KLMu1}, no exact formulae
for these boundary one-point functions are known. Instead, one has to
resort to approximate methods, two of which 
we investigate in some detail.
The first is the truncated conformal space approach, or TCSA.
This has previously been used to study integrable
models both without \cite{YZ,GMagn1}
and with \cite{Us1,Us3} boundaries, but this 
particular application of the technique is new.
The second method is based on a form factor (FF) 
expansion, and makes use of the 
expansion of the boundary states in the basis of infrared multi-particle states
given in~\cite{GZ}. 
Previous work on this topic includes
refs.~\cite{KLMu1,FFextraa,FFextrab,FFextrac}.
The main novelties arising in the discussion of the scaling \LY\ model
presented below are that this model has a non-trivial bulk S-matrix,
and that the boundary one-point functions 
receive contributions from both even- and odd-particle-number components
of the boundary state.
This allows the relative normalisations of the two
sectors to be checked, 
and we find evidence that the prescription given in \cite{GZ}
is out by a factor of $2$. 
Section 3 introduces and discusses the TCSA and FF methods,
culminating in a numerical
comparison between the two which, modulo the small
correction to the boundary state just mentioned, shows an excellent
agreement.

Both the TCSA and the FF methods can be pushed a little further, allowing
expectation values in states other than the ground state to be accessed.
These generalisations are explained in section 4.

The expectation values of boundary fields are also of 
interest, and for these there is more hope to find exact results, at
least for certain geometries. This is the subject of
section \ref{sec:boundary}.
First, in section \ref{sec:boundaryL}, 
we describe the so-called `R-channel' approach,
which allows us to obtain 
the expectation values of boundary fields at the edge
of a strip of finite width, both in the ground state and, in principle,
in any excited state.
By relating the expectation values
to the derivative of the energy with respect to the
boundary field, we find that they can be expressed in terms of the
solutions to TBA equations,
and compare these results with data from the TCSA (obtained by adapting the
method of \cite{GMagn1}), for various different combinations of
boundary conditions and for varying strip width. 
Then in section \ref{sec:boundaryR} we switch to the
`L-channel'. Making use of results from \cite{Us3}, we are
able to relate the expectation value of a
boundary field placed at one end of a semi-infinite cylinder to
the so-called `Y-function' of the \LY\ model.
An interesting relationship between the spectra of certain different
models placed on the same strip
emerges as a by-product of this discussion.

Finally, in section \ref{sec:discussion}, we make use of the results in
the preceding sections to
examine the RG flow from the $\Phi(h)$ boundary to the $\One$ boundary,
first discussed in \cite{Us1}, and to treat some previously obscure
features of the model on a strip of finite width,
with simultaneous perturbations on both boundaries. Section~\ref{concl}
contains
our conclusions, and indicates some directions for future work.

As in previous papers~\cite{Us1,Us2,Us3}, 
we will be using the boundary scaling \LY\
model as our example. We have included a brief
review of the features of this model in the next section, but the
earlier works should be consulted for some more detailed explanations.

\resection{The boundary \LY\ model}

\subsection{The conformal field theory description of the critical
\LY\ model}
\label{sec:UV}

The \LY\ model is the simplest non-unitary conformal field theory,
${\cal M}_{2,5}$, 
and has central charge $-22/5$ and effective central charge 
$2/5$. There are only two representations of the Virasoro algebra of
interest, of weight $0$ and $-1/5$, and consequently only two bulk
primary fields, the identity $\One$ of weight $0$, and $\varphi$ of
weight $x_\varphi\,{=}\,{-}2/5$; equally there are only two
conformally-invariant boundary conditions, which we denote by
$\One$ and $\Phi$.

There are three non-trivial boundary fields $\bp{-1/5}\alpha\beta$
interpolating the various boundary conditions $\alpha$ and $\beta$,
of weight $h_\phi = 1/5$. 
Two of these $(\psi,\psi^\dagger)$ interpolate the two different
conformal boundary conditions%
\footnote{In \cite{Us3} we did not distinguish these fields and denoted
them both by $\psi$}, and one $(\phi)$ lives on the $\Phi$
boundary: %
\be
  \psi \equiv \bp {-1/5}\One\Phi
\;,\;\;\;\;
  \psi^\dagger \equiv \bp {-1/5}\Phi\One   
\;,\;\;\;\;
  \phi
\equiv
  \bp {-1/5}\Phi\Phi
\;.
\ee
The OPEs of interest are the bulk OPE,
\bea
  \varphi(z,\bar z)\;
  \varphi(w,\bar w)
&=&
  \C\varphi\varphi\One \, |z-w|^{4/5}
\;+\; \C\varphi\varphi\varphi \, |z-w|^{2/5}\,\varphi(w,\bar w)
\;+\; \ldots
\;,
\nn\\
\noalign{%
\vskip 1mm%
\noindent the boundary OPEs,%
\vskip 1mm}
  \phi(z)\;\phi(w)
&=&
  \C\phi\phi\One \, |z-w|^{2/5}
\;+\; \C\phi\phi\phi \, |z-w|^{1/5}\,\phi(w)
\;+\; \ldots
\;,
\nn\\
  \psi(z)\;\phi(w)
&=&
  \C\psi\phi\psi
  |z-w|^{1/5}\,
  \psi(w)
\;+\; \ldots
\;,
\nn\\
  \phi(z)\;\psi^\dagger(w)
&=&\!\!
  \C\phi{\psi^\dagger}{\!\!\psi^\dagger}
  |z-w|^{1/5}\,
  \psi^\dagger(w)
\;+\; \ldots
\;,
\nn\\
  \psi(z)\;\psi^\dagger(w)
&=&
  \C\psi{\psi^\dagger}\One
  |z-w|^{2/5}\,
\;+\; \ldots
\;,
\nn\\
  \psi^\dagger(z)\;\psi(w)
&=&
  \C{\psi^\dagger}\psi\One
  |z-w|^{2/5}\,
\;+\; 
  \C{\psi^\dagger}\psi\phi
  |z-w|^{1/5}\,
  \phi(w)
\;+\; \ldots
\;,
\nn\\
\noalign{%
\vskip 1mm%
\noindent and the two bulk-boundary OPEs%
\vskip 2mm}
 \left. \varphi(z)\;\right|_\One
&=&
  \B\One\varphi\One \, |2(z-w)|^{2/5}
\;+\;\ldots
\;,
\nn\\
 \left. \varphi(z)\;\right|_\Phi
&=&
  \B\Phi\varphi\One \, |2(z-w)|^{2/5}
\;+\;
 \B\Phi\varphi\phi \, |2(z-w)|^{1/5}\,\phi(w)
\;+\; \ldots
\;.
\nn
\eea
\vfill~

A suitable choice for the structure constants is
\be
{
\renewcommand{\arraystretch}{1.7}
\begin{array}{rclrcl}
\multicolumn{6}{c}{
\C\varphi\varphi\One
 ~=~ \C\phi\phi\One
 ~=~
  -1
\;,\;\;\;\;
\;\;\;\;\;
\C\psi{\psi^\dagger}\One = 1
\;,\;\;\;\;
\;\;\;\;\;
\C{\psi^\dagger}\psi\One = - \fract{1 + \sqrt 5}2
\;,
}

\\

  \C\varphi\varphi\varphi
&= &
  - \left|
    \fract{2}{1 + \sqrt 5}
    \right|^{1/2}
    \cdot\alpha^2
\;,
&

  \B \One\varphi\One
&= &
  -\left| \fract{2}{1 + \sqrt 5} \right|^{1/2}
\;,
\\

  \C{\psi^\dagger}{\psi}\phi
\;=\;
  \C\phi\phi\phi
&= &
  - \left|
    \fract{1 + \sqrt 5}{2}
    \right|^{1/2}
    \cdot\alpha
\;,
&

  \B \Phi\varphi\One
&= &\m
  \left| \fract{1 + \sqrt 5}{2} \right|^{3/2}
\;,
\\

  \C\phi{\psi^\dagger}{\!\!\psi^\dagger}
\;=\;
  \C\psi\phi\psi
&= &
  - \left|
    \fract{2}{1 + \sqrt 5}
    \right|^{1/2}
    \cdot\alpha
\;,
&

  \B \Phi\varphi\phi
&= &\m
  \left|
    \fract{5 + \sqrt 5}2
  \right|^{1/2}
  \cdot\alpha
\;,
\\

\multicolumn{6}{c}{
  \alpha
 ~=~
  \m
  \left|
  \fract{\Ga(1/5)\,\Ga(6/5)}{\Ga(3/5)\,\Ga(4/5)}
  \right|^{1/2}
\;.
}

\end{array}
}
\label{eq:ylscs}
\ee%
\mathematica{
The coupling constants are:

  \B \One\varphi\One
&=& -0.7861514.. 

  \B \Phi\varphi\One
&=& \m 2.0581710..  

  \B \Phi\varphi\phi\,
  \C\phi\phi\phi
&= & \m 5.8824156..  

  \B \Phi\varphi\phi\,
  \C\psi\phi\psi
&= & -3.6355328..   
}
There are three possible choices for pairs of conformal boundary
conditions on a strip: $(\One,\One)$, $(\Phi,\One)$ and
$(\Phi,\Phi)$. We shall take the strip to be of width $\newL$ with
coordinates $0\,{\leq}\,x\,{\leq}\,\newL$ across the strip and $y$
running along the strip, and normalise all our correlation functions
so that the expectation value of the identity operator
on a strip is always one.

The strip correlation functions between states $\cev\alpha$ and
$\vec\beta$ can be found by mapping the strip to the unit disc and
inserting the appropriate fields $\psi_\alpha$ and $\psi_\beta$.
Since the ground state on the strips with boundary conditions
$(\One,\One)$, $(\One,\Phi)$ and $(\Phi,\Phi)$ correspond to the
fields $\One$, $\psi$ and $\phi$ respectively, one needs to include
the appropriate field insertions to find the ground state expectation
values on these strips. These insertions lead directly to the
particular chiral blocks and structure constants in
(\ref{eq:c1ptfns}) and (\ref{eq:c1ptfnsb}).

The one-point functions of the field $\varphi(x)$ on such a strip are
best expressed in terms of the four strip chiral block 
functions $f_i(\theta)$,
\be
\begin{array}{rcccl}
  f_1(\theta)
&{\!\!=\!\!}&
  f^{\phi\phi\phi\phi}_\One(\theta)
&{\!\!=\!\!}&\ds
  \left( \frac{2 \sin\theta}{\cos^2\!\theta} \right)^{2/5}\!
  \hyp(\fract 4{10}, \fract 8{10}; \fract {11}{10}; - \tan^2\!\theta)
\;,
\\[3mm]
  f_2(\theta)
&{\!\!=\!\!}&
  f^{\phi\phi\phi\phi}_\phi(\theta)
&{\!\!=\!\!}&\ds
  \left( \frac{2 \sin\theta}{\cos^3\!\theta} \right)^{1/5}\!
  \hyp(\fract 3{10}, \fract 8{10}; \fract {9}{10}; - \tan^2\!\theta)
\;,
\\[3mm]
  f_3(\theta)
&{\!\!=\!\!}&
  f^{\phi\phi\One\One}_\One(\theta)
&{\!\!=\!\!}&
  \,\left( {2 \sin\theta} \right)^{2/5}
\;,
\\[2mm]
  f_4(\theta)
&{\!\!=\!\!}&
  f^{\phi\phi\phi\One}_\phi(\theta)
&{\!\!=\!\!}&
  \,\left( {2 \sin\theta} \right)^{1/5}
\;.
\end{array}
\label{eq:cbs}
\ee
In terms of these functions, we have
\be
\begin{array}{rcl}
  \vev{\varphi(x,y)}_{(\One,\One)}
&=&\ds
  \left(\frac \newL\pi \right)^{2/5}\, \B \One\varphi\One\, f_3(\frac{\pi x}\newL)
\;\;=\;\;
  \B \One\varphi\One\,
  \left(
  \frac{2 \newL}{\pi}\, \sin\frac{\pi x}{ \newL }
  \right)^{2/5}
\;,\;\;
\\[4mm]
  \vev{\varphi(x,y)}_{(\Phi,\One)}
&=&\ds
  \left( \frac{\newL}{\pi} \right)^{2/5}
  \,
  \left(\;
  \B \Phi\varphi\One\; f_1( \frac{\pi x}{\newL} )
\;+\;
  \B \Phi\varphi\phi\; \C\psi\phi\psi\,f_2( \frac{\pi x}{\newL} )
  \;\right)
\;,\;\;
\\[4mm]
  \vev{\varphi(x,y)}_{(\Phi,\Phi)}
&=&\ds
  \left( \frac{\newL}{\pi} \right)^{2/5}
  \,
  \left(\;
  \B \Phi\varphi\One\; f_1( \frac{\pi x}{\newL} )
\;+\;
  \B \Phi\varphi\phi\;\C\phi\phi\phi\,f_2( \frac{\pi x}{\newL} )
  \;\right)
\;.
\end{array}
\label{eq:c1ptfns}
\ee
It is only for the latter two pairs of boundary conditions
that
the boundary field $\phi(y)$ exists on the boundary $x\,{=}\,0$. For
these cases the one-point
functions are simply
\be
\begin{array}{rcl}
  \vev{\phi(y)}_{(\Phi,\One)}
=\ds
  \left(\frac \newL\pi \right)^{1/5}\, \C\psi\phi\psi
\;,\;\;\;\;
  \vev{\phi(y)}_{(\Phi,\Phi)}
=\ds
  \left(\frac \newL\pi \right)^{1/5}\, \C\phi\phi\phi
\;.
\end{array}
\label{eq:c1ptfnsb}
\ee

Of the five expectation values (\ref{eq:c1ptfns}),
(\ref{eq:c1ptfnsb}),
only the first has a finite limit as the strip width tends to
$\infty$, the others all diverging. 
None of them depend on $y$, and so when no confusion can arise this variable
will often be omitted, even inside the vacuum expectation values.

\subsection{The scaling \LY\ model}
\label{sec:IR}

The scaling \LY\ (SLY) model can be described as a perturbation of the
critical \LY\ model by the term 
\be
  \lambda \int \varphi(w,\bar w) \; \D^2 w
\;.
\label{eq:bulkpert}
\ee
This leads to a massive scattering theory, comprising
a single particle with two--particle S--matrix \cite{CM}
\eq
  S(\te)
=  -(1)(2)
\;,\;\;
  (x)
=  {\sinh \lf( \fract{\te}{2} +\fract{i \pi x}{6} \ri)
     \over
    \sinh \lf( \fract{\te}{2} -\fract{i \pi x}{6} \ri)}
\;.
\label{SMatrix}
\en
The mass $M$ of the particle is
related to the bulk perturbation parameter $\lambda$ by \cite{Zb,Zg}
\be
  M 
= \kappa \lambda^{5/12} 
\;,
\;\;\;\;
  \kappa
= 2^{19/12}\sqrt{\pi}\,
  \ds\frac
  {  ~\left( \Ga(3/5)\Ga(4/5) \right)^{5/12} }
  {   5^{5/16}\Ga(2/3)\Ga(5/6)  }
= 2.642944\dots
\;.
\mathematica{
  kappa  = 
  2^(19/12) *
  Sqrt[ Pi ] *
 (  Gamma[ 3/5] Gamma[4/5] )^(5/12) /
 (  5^(5/16) Gamma[ 2/3] Gamma[5/6] )
  }
\label{eq:Mlambda}
\ee

  We will also need the form factors of the bulk model. These are the
  matrix elements of the elementary field $\varphi(x)$ in the asymptotic
  $n$--particle states which can be formally written in terms of the
  ZF operators $A(\theta)$ as 
$ \vec{\theta_1,\ldots,\theta_n} = 
  A(\theta_1)\,\ldots\,A(\theta_n)\,\vac.$
  The form-factor $F_n(\theta_i\ldots\theta_n)$ 
  is then given by
\be
\beqcol
  \cev 0 \varphi(x,0) \vec{ \theta_1, \ldots , \theta_n }
= \exp ( - M x \sum_i \cosh\theta_i )\,
   F_n(\theta_1, \ldots , \theta_n )
\;.
\eeqcol
\label{F-def}
\ee
  The form factors of the SLY model were first computed in \cite{SM};
  we, however, adopt the conventions of \cite{ZAM}, modulo the fact
  that for us $\varphi(x)$ is a real field.
The function $F_n$ can be parametrised as \cite{ZAM}:
\be
  F_n(\theta_1, \ldots , \theta_n ) 
= H_n\,
  Q_n(x_1, \ldots , x_n )\,
  \prod_{i<j}^n 
  \frac{ f(\theta_i - \theta_j )}{x_i + x_j}
\;,
\label{F-param}
\ee
where $x_i = \exp ( \theta_i)$, $i=1, \ldots , n$.
The terms in (\ref{F-param}) can be determined through the form factor
bootstrap \cite{SM2}, with the result 
\be
  f( \theta ) 
= \frac{ \cosh \theta -1}{\cosh \theta +1/2}\,
  v(i \pi - \theta )\,
  v(- i \pi + \theta )
\;,
\label{f-def}
\ee
where we take the function $v$ in a form 
suitable for numerical evaluation \cite{ZAM}
\be
\beqcol
    v(\theta ) 
&=& \ds
    \prod_{n=1}^N 
    \left[ 
    \frac{
    (\fract{\theta}{2\pi i} + n + 1/2)  
    (\fract{\theta}{2\pi i} + n - 1/6)  
    (\fract{\theta}{2\pi i} + n - 1/3)}
   {(\fract{\theta}{2\pi i} + n - 1/2)  
    (\fract{\theta}{2\pi i} + n + 1/6)  
    (\fract{\theta}{2\pi i} + n + 1/3)} 
     \right]^n       
\\[6mm]
&\times& \ds
     \exp\left( 2 \int_0^{\infty} 
     \D t\,
     \frac{\sinh (t/2) \sinh (t/3) \sinh (t/6)}{t \sinh^2(t)}\,
     (N+1 - N {\mathrm e}^{-2 t} )\,
     {\mathrm e}^{-2 N t + i\theta t/ \pi} 
     \right)
\;,\\[6mm]
   v(0)
\mathematica{
 N[ Exp[ 2 NIntegrate[ Sinh[t/2] Sinh[ t/3 ] Sinh[ t/6 ]/
             ( t Sinh[t]^2 ) , {t,0,Infinity} ]] , 15 ]
}
&=& 1.111544045... 
\eeqcol
\label{v-def}
\ee
(with $N$ arbitrary) and
\be
  H_n 
=  \psi  
   \left( -\frac{i\,3^{1/4}}{\sqrt{2} v(0)}
   \right)^n
\;.
\label{H-def}
\ee
This overall normalisation of the form factors is taken 
{}from the results of \cite{FLZZ}, where (in our conventions) the
expectation value $\vev\varphi$ in the bulk is 
\be
\beqcol
    \psi = 
    \vev \varphi 
\mathematica{
psi0 = - ( lambda )^(-1/6) *
    (( 5 Gamma[1/3]^2 )/( 2^(19/12) Sqrt[ 3 ] Pi^2)) *
    (( Gamma[7/12]  Gamma[1/12] )/(Gamma[5/12]  Gamma[11/12]) )*
    ( - Pi^2   
     (( Gamma[9/10]^2  Gamma[11/10])/( Gamma[1/10]^2  Gamma[-1/10] ))
    )^(5/12) 
psi1 = - ( lambda )^(-1/6) *
    (( 5 Gamma[1/3]^2 )/( 2^(19/12) Sqrt[ 3 ] Pi^2)) *
    (( Gamma[7/12]  Gamma[1/12] )/(Gamma[5/12]  Gamma[11/12]) )*
    ( (Pi^2  /100)
     (( Gamma[9/10] ) /( Gamma[1/10]   ))
    )^(5/12) 
psi2 = - ( lambda )^(-1/6) *
    (( 5 Gamma[1/3]^2 )/( 2^(19/12) Sqrt[ 3 ] Pi^2)) *
    (( Sqrt[Pi] Gamma[1/6] 2^(1-1/6) )/
     ( Sqrt[Pi] Gamma[5/6] 2^(1-5/6) ) )*
    ( (Pi^2  /100)
     (( Gamma[9/10] ) /( Gamma[1/10]   ))
    )^(5/12) 
psi3 = - ( lambda )^(-1/6) *
    5^(1/6) * Gamma[1/3]^2 * 2^(-7/4) * 3^(-1/2) * Pi^(-7/6) *
    ( Gamma[ 1/6]  /  Gamma[ 5/6] ) * 
    ( Gamma[9/10]  /  Gamma[1/10] )^(5/12) 
 -0.840183746026537
}
&=& \ds
    -| \lambda |^{-1/6} 
    \frac{ 5^{1/6}\, \Gamma(1/3)^2 }{ 2^{7/4}\cdot 3^{1/2}\cdot \pi^{7/6}}
    \frac{ \Gamma(1/6) }{\Gamma(5/6)}
    \left[ 
     \frac{ \Gamma(9/10)}{ \Gamma(1/10) }
     \right]^{5/12} 
\\[4mm]
&=& (- 0.840184 \ldots\, )  
    |  \lambda |^{-1/6} .
\eeqcol 
\label{ff-nn}
\ee
Using the relation between $M$ and $\lambda$ 
in (\ref{eq:Mlambda}), this boils down to
\be
    \psi 
=  \frac{
  - 3^{\frac 9{10}}\,
     \Gamma(\frac 13)^{\frac{36}5} \, M^{-\frac 2 5}
        }
       {
       (2\pi)^{\frac {14}5 }\,
       5^{\frac 14 }\,
      \Gamma(\frac 15)\,
      \Gamma(\frac 25)}
= (-1.239394325...\,) M^{-2/5}
\;.
\label{psi}
\ee
The functions $Q_n(x_1 , \ldots , x_n )$ in 
(\ref{F-param}) are symmetric polynomials 
of degree $n (n-1)/2$ and partial degree $n-1$. 
These polynomials have been determined via the 
form factor bootstrap approach for arbitrary 
particle numbers $n$. They can be nicely written 
in the form of a determinant of a matrix in 
symmetric polynomials \cite{SM,ZAM}, for a related 
formulation see \cite{MP}.
For our purposes it will be sufficient to list 
the first few:
\be
  Q_0
= 1
  \;,\;\; 
  Q_1 
= 1
  \;,\;\; 
  Q_2
= e_1^{(2)}
  \;,\;\;
  Q_3
= e_2^{(3)} e_1^{(3)}
  \;,\;\;
  Q_4
= e^{(4)}_3 e^{(4)}_2 e^{(4)}_1 
\;,
\label{Q-ff}
\ee
where the elementary symmetric polynomials in $n$ variables
$e^{(n)}_r$  are defined by 
\[
  \prod_{i=1}^n ( 1 + t x_i) 
= \sum_{r=0}^n t^r e^{(n)}_r
\;.
\]

The integrable boundary conditions for the 
model were
discussed in detail in \cite{Us1}. 
The allowed boundary conditions are the $\One$
conformal b.c., and the perturbation $\Phi(h)$ of
the conformal $\Phi$ boundary by the integral along the boundary  
\be
  h \int \phi(x) \; \D x
\;.
\label{eq:boundpert}
\ee
The boundary reflection factors corresponding to these two boundary
conditions are
\eq
  R_{\Phi(h)}(\te)
= R_b(\te)
\;,\;\;\;\;
  R_\One(\te)
= R_0(\te)
\;,
\label{eq:ss3}
\en
where 
\eq
  R_b(\te)
= \lf(\fract{1}{2}\ri)\lf(\fract{3}{2}\ri)
  \lf(\fract{4}{2}\ri)^{-1}
  \lf( S(\te+i \pi \fract{b+3}{6})
  S(\te-i \pi \fract{b+3}{6}) \ri)^{-1}\,.
\label{bf}
\label{eq:ss2}
\en
The relation between $b$ and $h$ was conjectured in \cite{Us1} to be
\eq
  h(b)
=
  - \, |\hc|
  \,\sin( \pi (b+1/2)/5) 
\;.
\en
Sometimes we shall find it useful to consider instead the
dimensionless quantity
\[
  \hhc = \hc M^{-6/5}
\;,
\]
a constant which was found in~\cite{Us3}:
\eq
  \hhc
=
   - \pi^{3/5}
  \, 2^{4/5}
  \, 5^{1/4}
  \frac{ \sin \fract{2\pi} 5 }
       { ( \Ga( \fract 35 ) \Ga(\fract 45) )^{1/2} }
  \left( 
  \frac{ \Ga( \fract 23 ) }{\Ga( \fract 16 )}
  \right)^{6/5}
\mathematica{
 - Pi^(3/5) * 
  2^(4/5) *
  5^(1/4) *
  (   Sin[ ( 2 Pi) / 5 ]  / 
      ( Gamma[3/5] Gamma[4/5] )^(1/2) ) * 
  ( Gamma[2/3] / Gamma[1/6] )^(6/5)
}
= -0.68528998399118\ldots
\;.
\label{eq:hexact}
\en
Combining (\ref{eq:Mlambda}) and (\ref{eq:hexact}), we obtain
the more convenient formula
\be
\mathematica{  
 M = kappa * lambda^(5/12);
 h = 
  (
 - Pi^(3/5) * 
  2^(4/5) *
  5^(1/4) *
  (   Sin[ ( 2 Pi) / 5 ]  / 
      ( Gamma[3/5] Gamma[4/5] )^(1/2) ) * 
  ( Gamma[2/3] / Gamma[1/6] )^(6/5) 
  ) * Sin[ Pi (b+1/2)/5 ] * M^(6/5);
h = (-2*2^(1/5)*Sqrt[5 + Sqrt[5]]*Sqrt[lambda]*Pi^(6/5)*Sin[((1/2 + b)*Pi)/5])/
   (5^(1/8)*(Gamma[1/6]*Gamma[5/6])^(6/5)) ;
h = (-2*2^(1/5)*Sqrt[5 + Sqrt[5]]*Sqrt[lambda]*Pi^(6/5)*Sin[((1/2 + b)*Pi)/5])/
   (5^(1/8)*(Pi / Sin[ Pi  / 6 ])^(6/5)) ;
h =  -((Sqrt[5 + Sqrt[5]]*Sqrt[lambda]*Sin[((1/2 + b)*Pi)/5])/5^(1/8)) ;
h =  -((5^(3/8)*Sqrt[lambda]* (Sin[Pi/10 + (b*Pi)/5]/Sin[Pi/5]))/Sqrt[2]);
}
h = 
 - \frac{ 5^{3/8} }{2^{1/2} }\,
  \frac{ \sin( \pi(b+1/2)/5) }{\sin(\pi/5)}\,
  \lambda^{1/2}
\;.
\label{eq:hlambda}
\ee

Finally, we will need the boundary-particle couplings $g^a_{\alpha}$
for the various
boundary conditions.  In \cite{GZ} these
were defined in two different ways, 
either via the residue 
at $\theta=i\pi/2$
of the reflection factor $R^a_{\alpha}(\theta)$ 
for a particle of type
$a$ on the boundary $\alpha$\,:
\eq
R^a_{\alpha}(\theta)\sim \frac{i}{2}\,
\frac{(g^a_{\alpha})^2}{\theta-i\pi/2}
\en
or, in models where bulk fusings occur,
via the residues of certain other poles, divided by the corresponding
bulk couplings. 
(If the boundary scattering is non-diagonal, the formulae
become a little more complicated -- see \cite{GZ}.)
For the \LY\ model,
the consistency of the two definitions was shown to follow
{}from the bootstrap equations and crossing in~\cite{Us2}.
Since for this case
there is only one particle type, the particle index $a$ can be dropped,
and we have

\begin{itemize}

\item{}

For the $\One$ boundary, 
\be
  g_{\One} 
= -i \,2\, \sqrt{\,2\,\sqrt{~3~} - 3~}
\;.
\label{eq:gone}
\ee
\item{}
For the $\Phi(h(b))$ boundary,
\be
  g_\Phi(b) 
= \frac{ \tan((b+2)\pi/12)}{ \tan((b-2)\pi/12)}\, g_{\One}
\;.
\label{eq+gphi}
\ee

\end{itemize}
Notice that while
the reflection factors for the $\One$ and $\Phi(h(0))$
are identical, the corresponding boundary-particle couplings differ
by a sign.
This can be traced to the fact that
when $b=0$,  the residue of $R_b$ 
at $i\pi/6$ 
for the $\Phi(h(0))$ boundary
receives additional
contributions from intermediate states containing boundary
bound states. The net effect is to negate the coupling; indeed, this is the
only option given that the residue at $i\pi/2$ must remain unchanged.
In terms of the boundary states to be discussed in the next section,
this means that the only difference between the (infinite-volume)
boundary states for these two boundary conditions is  in the sign
of the contributions from states of odd particle number.

\newpage
\resection{The one-point functions of the bulk field}
\label{sec:bulk}

\subsection{$\vev{\varphi(x)}$ from form factors}

In this section we consider the one-point function
$\vev{\varphi(x)}$ in the presence of a boundary using the form factor
approach. We will restrict
consideration to a theory with a single scalar particle, as in
the \LY\ model.
Schematically, the 
idea of the form-factor approach is to evaluate the one-point function 
on the upper half plane as
\[
\vev{\varphi(x)}=  \cev 0 \varphi(x) \vec {B_{\alpha}}=
\sum_{n=0}^{\infty}
\;
  \cev 0 \varphi(x) \vec{n}\;
  \vev{n\,|B_{\alpha}}
\;,
\]
where $\vec {B_{\alpha}}$ is a boundary state 
corresponding to the boundary condition $\alpha$
and the sum over asymptotic states has been split into the
contributions from $n=0$, $1$, $2$, $\dots$ particles
(here $\vec n\cev n$ represents the projection onto asymptotic states with
$n$ particles).
We have also taken the boundary state to be
normalised such that $\vev{0\,|B_{\alpha}}=1$.

This boundary state can be expanded in terms of multi-particle states on
the infinite line, using the Zamolodchikov-Faddeev (ZF) operators
$A(\theta)$  which create single particles of rapidity $\theta$.
When the state contains no zero-rapidity particles, it can be written
as~\cite{GZ}
\be
  \vec{ B_{\alpha} } 
= \exp \left[ \int^{\infty}_{-\infty} 
    \frac{\D\theta}{4 \pi}
    K_{\alpha}(\theta )\,
    A(-\theta)\,
    A(\theta)\, \right]  
    \vac
\;,
\label{Bexp1}
\ee
where $K_{\alpha}(\theta)$ is related to the reflection amplitude 
$R_{\alpha}(\theta)$ for
the $\alpha$ boundary condition by
\eq
  K_{\alpha}(\theta) = R_{\alpha}(\frac{i \pi}2 - \theta)
\;.
\label{K-B}
\en
In general there may also be contributions to the boundary state
involving  zero-momentum particles, which can be associated with 
couplings of single bulk particles to the boundary.
Up to the three-particle contribution,
the appropriate boundary state was given in~\cite{GZ} as
\be
\beqcol
     \vec{B_\alpha}
&=&  \ds
      \Big[
     1 
   \; \;+\;\;  
     \cg_{\alpha}\, A(0) 
   \; \;+\;\; 
     \int^{\infty}_{-\infty} \frac{\D\theta}{4\pi}
     K_{\alpha}(\theta)\,A(-\theta )\,A(\theta )\,
\\[4mm]
&+&  \ds
     \cg_{\alpha}\, A(0) \int^{\infty}_{-\infty} \frac{\D\theta}{4\pi}
     K_{\alpha}(\theta)\,A(-\theta )\,A(\theta )  
     + \ldots  
   \Big]
   \vac
\;,
\eeqcol
\label{Bexp2}
\ee
and it is natural to suppose that the full expression 
is
\be
  \vec{ B_{\alpha} } 
= \exp \left[ 
    \;\;
    \tilde g_\alpha\, A(0) 
\;+\;
    \int^{\infty}_{-\infty} 
    \frac{\D\theta}{4 \pi}
    K_{\alpha}(\theta )\,
    A(-\theta)\,
    A(\theta)\, \right]  
    \vac
\;.
\label{Bexp1b}
\ee
In 
\cite{GZ}, the 
factor $\cg_{\alpha}$ was identified with the boundary-particle
coupling $g_{\alpha}$, as defined at the end of the last section.
However, our numerical results (and also an examination
of the reflection factor and boundary state for the 
Ising model with free boundary conditions given 
in~\cite{GZ}) cast 
doubt on this
suggestion.
As will be explained in section~\ref{ssec:TCSAFF} below, we found that
we had rather to set
\be
  \cg_{\alpha} = g_{\alpha}/2
\label{cset}
\ee
in order to obtain a successful match with TCSA data.

The form factors for the \LY\ model were given in the previous section,
and substituting these into (\ref{Bexp2}), we find the leading
large-$x$ behaviour of $\vev{\varphi(x)}$ is
\be
\beqcol
  \vev{\varphi(x)}
&{\!\!=\!\!}& 
  \cev 0 \varphi(x) \vec{B_\alpha}
\nn\\[2mm]
&{\!\!=\!\!}&\ds
  \psi\,
  \Big(
  1 
\; -i \;
  \cg_{\alpha}\,\frac{3^{1/4}}{\sqrt2\, v(0)}\, {\mathrm e}^{-M\,x}
\; - \;
  \frac{3^{1/2}}{2\, v(0)^2}\,
  \int_{-\infty}^{\infty}\!
  \frac{\D\theta}{4\pi}\,
  K_{\alpha}(\theta)\,
  f(-2\theta)\,
  {\mathrm e}^{-2M\,x\,\cosh(\theta)}\,
\\[5mm]
&&\ds
\!\!\!\!\!\!\!\!\!\!\!\!\!
\!\!\!\!\!\!\!\!\!\!\!\!\!
 + \;
 i \, 
   \cg_{\alpha} \,
   \frac{ 3^{3/4} }{2\sqrt 2\, v(0)^3}\,
  \int_{-\infty}^{\infty}\!
  \frac{\D\theta}{4\pi}\,
  K_{\alpha}(\theta)\,
  \frac{ f(\theta)f(-\theta)f(-2\theta)\, 
         (1 {+} 2 \cosh\theta)^2}
       {8\cosh\theta\,\cosh(\theta/2)^2}\,
  {\mathrm e}^{-Mx(1 {+} 2 \cosh(\theta))}
\;+\ldots
  \Big)
\eeqcol
\label{eq:FF}
\ee
This will be compared with TCSA data in section \ref{ssec:TCSAFF}
below.

\subsection{Estimating $\vev{\varphi(x)}$ using the TCSA}

The first step in the 
calculation of $\vev{\varphi(x)}$ using the TCSA is 
the numerical evaluation of the ground state
$\vec{\hat 0}$ of the perturbed Hamiltonian
with dimensionless strip width $\ML\,{\equiv}\,M\newL$.
\[
\begin{array}{rl}
&
  \hat H(\newL,\lambda,h_l,h_r)
\\
= 
& \ds
\frac{\pi}{\newL} 
    \Big(
  L_0 - \fract{c}{24}
\;\;+\;\;
    \lambda 
    \left| \fract{\newL}{\pi} \right|^{12/5}
    \!\!\int\limits_{\theta = 0}^\pi \!\!
    \hat\varphi(\exp(i\theta))
    \,\D\theta
\;\;+\;\;
     h_l
    \left| \fract{\newL}{\pi} \right|^{6/5}
    \hat\phi_l(-1)
\;\;+\;\;
    h_r
    \left| \fract{\newL}{\pi} \right|^{6/5}
    \hat\phi_r(1)
  \Big)
\;,
\end{array}
\]
where $\hat{\cal O}$ represents the operator ${\cal O}$ on the
upper half plane restricted to the
conformal space truncated to level $N$. 
The parameter $\lambda$ determines the bulk mass and we have allowed
the possibility of boundary fields on the left and right edges of the
strip, with strengths $h_l$ and $h_r$.
The second step is then to estimate the expectation value of the
(dimensionless) operator $M^{2/5}\,\varphi(x)$ in terms of the matrix
elements of the operator $\hat\varphi$ as
\be
 \vev{M^{2/5}\varphi(x)}
\;\sim\;
  \left(\frac{\ML}{\pi}  \right)^{2/5}
\frac{ \cev{\hat 0} \; \hat\varphi(\exp(i\pi x/\newL))\; \vec{\hat 0} }
       { \veev{\hat 0}{\hat 0}}
\;.
\label{eq:tcsavev}
\ee
The state $\vec{\hat 0}$ can be expanded in Virasoro primary and
descendent states;
by repeatedly commuting Virasoro algebra elements through
$\hat\varphi$, the general matrix elements of $\hat\varphi$ can be
expressed in terms of the matrix elements between Virasoro primary
states, and their derivatives.
In the rest of this section we discuss the TCSA method in more
detail; a comparison of the TCSA and FF results is given in section 
\ref{ssec:TCSAFF}. 
All the TCSA results in this paper were calculated on a workstation
in Mathematica with truncation levels up to 18 and on spaces with up
to 161 states. 

\newpage
\subsubsection{The strip with $(\One,\One)$ boundary conditions}

The simplest case is the strip with boundary conditions $\One$ on both
sides. The unperturbed conformal field theory expectation value is
given in (\ref{eq:c1ptfns}).
If we denote the scaled position of the field by $\xi$ and the
normalised strip width by $\ML$ where
\[
  \xi = x\, M
\;,\;\;\;\;
  \ML = M\newL
\;,\;\;\;\;
\hbox{ so that }
 \;\;
  0 \leq \xi \leq \ML
\;,
\] 
then the TCSA estimate of the expectation value in the bulk  perturbed
model truncated to level $N$ takes the form 
\be
  \cG^{(N,\ML)}(\xi)
\equiv
 \left.
  M^{2/5}\,
  \vev{\varphi(x)}_{(\One,\One)}
 \right|_{N,\ML}
= 
  \left(
  \frac{2\ML}{\pi}\,\sin\frac{\pi\xi}\ML
  \right)^{2/5}\,
  \sum_{n=0}^N
  f_n^{(N,\ML)}\,\cos\left(\frac{2n\pi\xi}\ML \right)
\;.
\label{eq:cG}
\ee
The coefficients $f^{(N,\ML)}_n$ are determined by the expansion of
the ground state $\vec{\hat 0}$ and have to be calculated numerically.
Since the state $\vec{\hat 0}$ lies in the $h\,{=}\,0$ representation,
the matrix elements of $\hat\varphi$ are given in terms of the chiral
block $f_3$; furthermore, since this representation has a null state
at level 1,  one can eliminate all states containing $L_{-1}$, 
and so one does not need any terms with derivatives of
this chiral block in (\ref{eq:cG}). 
To compare with the form-factor calculation we will need to take the
simultaneous limits $N\to\infty$, $\ML\to\infty$ while keeping 
$\xi$ fixed.
We shall often drop the labels $(N,\ML)$ and write simply $\cG(\xi)$
for the TCSA estimates.

\begin{figure}[ht]
\refstepcounter{figure}
\label{fig:Graph2}
\label{fig:Graph7}
\label{fig:Graph8}
\[
\begin{array}{cc}
\epsfxsize=.85\linewidth
\parbox[t]{.85\linewidth}{
\epsfxsize=\linewidth
\epsfbox[72 272 540 520]{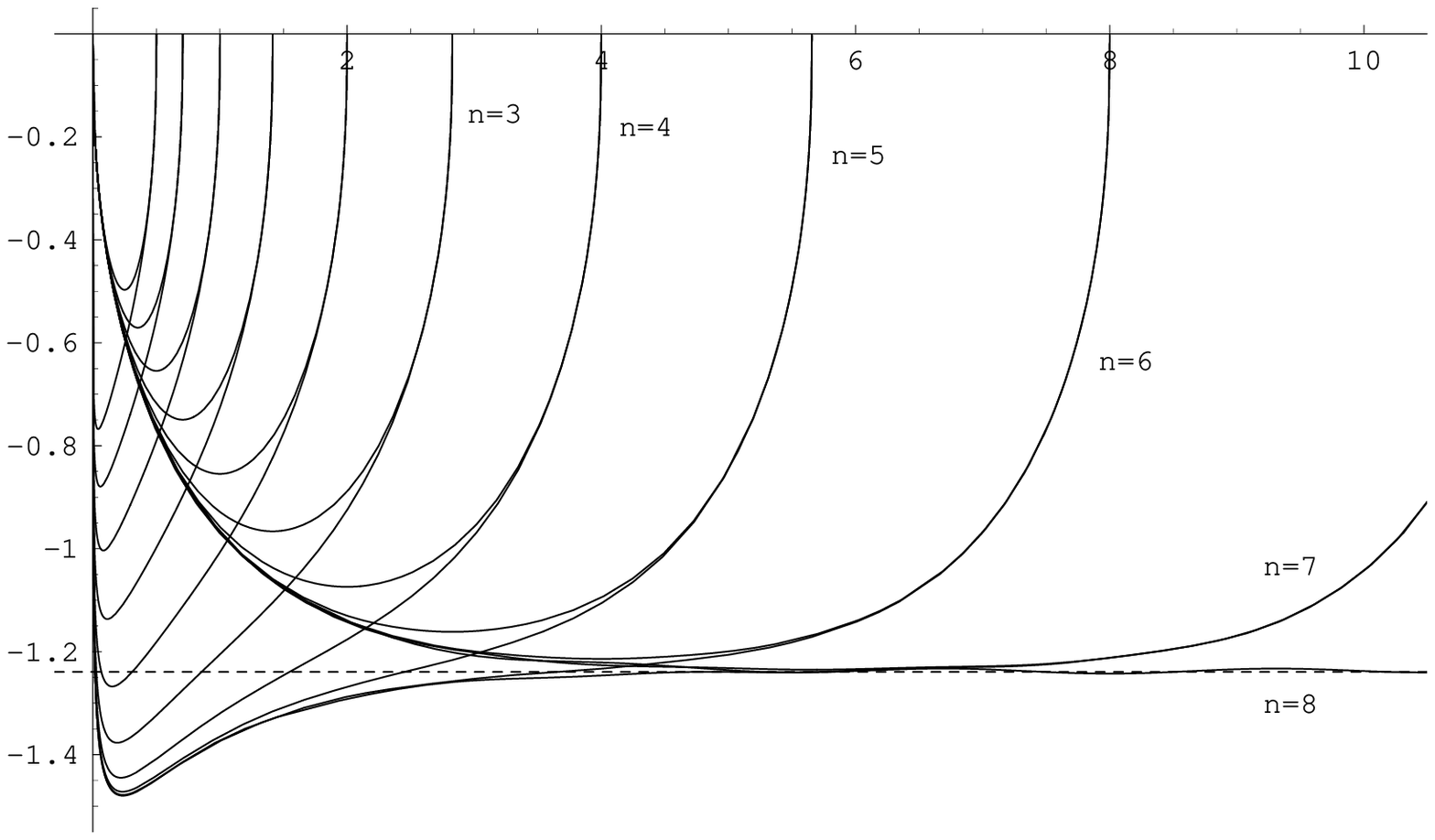}
}
\\
\\
\parbox[t]{.95\linewidth}{\raggedright%
Figure \ref{fig:Graph8}\\\small
Plots of $\cG(\xi)$ and $\cI(\xi)$
against $\xi$ from TCSA truncated to level 12.
The upper set of lines are the (symmetric) functions $\cG(\xi)$, and
the lower set of lines are the (asymmetric) functions $\cI(\xi)$
for $b\,{=}\,{-}1/2$, $h\,{=}\,0$.
These are plotted for $\ML = 2^{n/2}$ with 
${-}2\,{\leq}\,n\,{\leq}\,8$.
As $\ML$ increases, the functions $\cG(\xi)$ and $\cI(\xi)$ both
approach universal functions which we take to be the expectation
values of  $M^{2/5}\varphi(x)$ on a half-line with boundary conditions
$\One$ and $\Phi(0)$ respectively.
For large $\ML$, truncation effects start to intervene, as can be seen
in the slight `ripple' discernible for $n=8$.
}
\end{array}
\]
\end{figure}

To show the typical behaviour of these quantities, 
we give the values of $f_n^{(N,\ML)}$
for $\ML=8$ and $4\leq N\leq 12$ in table  \ref{tab:Table2a},
and 
for $N=12$ and $2\leq \ML\leq 12$ in table \ref{tab:Table2b}.
In figure \ref{fig:Graph2} we plot $\cG^{(12,\ML)}(\xi)$ for fixed
truncation level $N=12$ and for varying values of $\ML$.
We see that on increasing $\ML$, $\cG^{(12,\ML)}(\xi)$ approaches a
universal form until truncation effects take over and the TCSA
approximation breaks down. 

There are two sources of error in the TCSA estimate
$\cG^{(N,\ML)}(\xi)$.  
Firstly, TCSA gives the function
$ \cG^{(N,\ML)}(\xi)\times\big( (2l/\pi)\sin(\pi\xi/l)\big)^{-2/5}$
in equation (\ref{eq:cG})
as a Fourier series truncated at the 
$2N^{\rm th}$ term, 
which leads to the
usual errors associated with truncation of Fourier series.
Secondly, the coefficients $f^{(N,\ML)}_n$ appearing in the truncated
Fourier series are only approximately calculated.

\refstepcounter{table}
\label{tab:Table2a}
{\small
\[
\mathematica{
%
%
Table2a = {{{{4, 0}, -1.598158346234007}, {{4, 1}, -0.17134813008764}, 
    {{4, 2}, -0.03547005696379001}, {{4, 3}, -0.001023825574489247}, 
    {{4, 4}, -0.000101229957281888}, {{4, 5}, 0}, {{4, 6}, 0}, {{4, 7}, 0}, 
    {{4, 8}, 0}, {{4, 9}, 0}, {{4, 10}, 0}, {{4, 11}, 0}, {{4, 12}, 0}}, 
   {{{6, 0}, -1.597162801840318}, {{6, 1}, -0.1672980925514791}, 
    {{6, 2}, -0.03075692245246309}, {{6, 3}, -0.01048076379827189}, 
    {{6, 4}, -0.0003486020480341202}, {{6, 5}, -0.0000461519493070493}, 
    {{6, 6}, -8.254177336039582*^-6}, {{6, 7}, 0}, {{6, 8}, 0}, {{6, 9}, 0}, 
    {{6, 10}, 0}, {{6, 11}, 0}, {{6, 12}, 0}}, 
   {{{8, 0}, -1.596789124818479}, {{8, 1}, -0.1663240273726197}, 
    {{8, 2}, -0.02959643176036319}, {{8, 3}, -0.009033590479785637}, 
    {{8, 4}, -0.004182641117578388}, {{8, 5}, -0.0001464954430720569}, 
    {{8, 6}, -0.00002271638306669595}, {{8, 7}, -5.274754334247155*^-6}, 
    {{8, 8}, -1.286687910716844*^-6}, {{8, 9}, 0}, {{8, 10}, 0}, 
    {{8, 11}, 0}, {{8, 12}, 0}}, 
   {{{10, 0}, -1.596633614179739}, {{10, 1}, -0.1657840018846929}, 
    {{10, 2}, -0.02917158543743268}, {{10, 3}, -0.008795233983032426}, 
    {{10, 4}, -0.003611425939663798}, {{10, 5}, -0.002016722134046272}, 
    {{10, 6}, -0.00007226348779028356}, {{10, 7}, -0.00001214578718413219}, 
    {{10, 8}, -3.284087348554003*^-6}, {{10, 9}, -1.015532924139923*^-6}, 
    {{10, 10}, -2.963633557582563*^-7}, {{10, 11}, 0}, {{10, 12}, 0}}, 
   {{{12, 0}, -1.596642541299763}, {{12, 1}, -0.165535525215146}, 
    {{12, 2}, -0.02898764326374237}, {{12, 3}, -0.008533993562707021}, 
    {{12, 4}, -0.003501800712571309}, {{12, 5}, -0.001747004129943395}, 
    {{12, 6}, -0.001103066067117593}, {{12, 7}, -0.00003983260190607214}, 
    {{12, 8}, -7.034582419495929*^-6}, {{12, 9}, -2.055099430179556*^-6}, 
    {{12, 10}, -7.342359547772959*^-7}, {{12, 11}, -2.697638838322589*^-7}, 
    {{12, 12}, -8.828262531897326*^-8}}}
Table2a2 = {{{{4, 0}, -0.6956388242752089}, {{4, 1}, -0.07458360558376631}, 
    {{4, 2}, -0.0154392390349865}, {{4, 3}, -0.0004456459652942962}, 
    {{4, 4}, -0.00004406289816709512}, {{4, 5}, 0}, {{4, 6}, 0}, 
    {{4, 7}, 0}, {{4, 8}, 0}, {{4, 9}, 0}, {{4, 10}, 0}, {{4, 11}, 0}, 
    {{4, 12}, 0}}, {{{6, 0}, -0.6952054884088526}, 
    {{6, 1}, -0.07282072435452864}, {{6, 2}, -0.01338772808312348}, 
    {{6, 3}, -0.004562017414179614}, {{6, 4}, -0.0001517378546411429}, 
    {{6, 5}, -0.00002008880273323297}, {{6, 6}, -3.59283936471768*^-6}, 
    {{6, 7}, 0}, {{6, 8}, 0}, {{6, 9}, 0}, {{6, 10}, 0}, {{6, 11}, 0}, 
    {{6, 12}, 0}}, {{{8, 0}, -0.6950428360379262}, 
    {{8, 1}, -0.07239673785945704}, {{8, 2}, -0.01288259517026973}, 
    {{8, 3}, -0.003932098640381946}, {{8, 4}, -0.001820600290486698}, 
    {{8, 5}, -0.00006376584524334725}, {{8, 6}, -9.887880037381352*^-6}, 
    {{8, 7}, -2.295970178463767*^-6}, {{8, 8}, -5.600634427304309*^-7}, 
    {{8, 9}, 0}, {{8, 10}, 0}, {{8, 11}, 0}, {{8, 12}, 0}}, 
   {{{10, 0}, -0.6949751461008492}, {{10, 1}, -0.07216167811310256}, 
    {{10, 2}, -0.01269767006739901}, {{10, 3}, -0.003828347949125046}, 
    {{10, 4}, -0.001571964443038277}, {{10, 5}, -0.000877829294902872}, 
    {{10, 6}, -0.00003145451000078697}, {{10, 7}, -5.286760937410564*^-6}, 
    {{10, 8}, -1.429482045598681*^-6}, {{10, 9}, -4.420363795778853*^-7}, 
   {{10, 10}, -1.289996431478398*^-7}, {{10, 11}, 0}, {{10, 12}, 0}}, 
   {{{12, 0}, -0.694979031855532}, {{12, 1}, -0.07205352236078255}, 
    {{12, 2}, -0.01261760458593901}, {{12, 3}, -0.003714636451590047}, 
    {{12, 4}, -0.001524247291439861}, {{12, 5}, -0.0007604277147014392}, 
    {{12, 6}, -0.0004801373930410307}, {{12, 7}, -0.00001733814701344069}, 
    {{12, 8}, -3.061979843922597*^-6}, {{12, 9}, -8.945339832861781*^-7}, 
    {{12, 10}, -3.195947620118212*^-7}, {{12, 11}, -1.174215505135616*^-7}, 
    {{12, 12}, -3.842724460034641*^-8}}}
}
\begin{array}{r|lllll}
     & \multicolumn{1}{c}{N}   \\
   n & \multicolumn{1}{c}{4}   
     & \multicolumn{1}{c}{6}   
     & \multicolumn{1}{c}{8}   
     & \multicolumn{1}{c}{10}   
     & \multicolumn{1}{c}{12} \\
\hline
   0 & -0.695639 & -0.695205 & -0.695043  
     & -0.694975 & -0.694979 \\
   1 & -0.074584 & -0.072820 & -0.072397 
     & -0.072162 & -0.072053 \\
   2 & -0.015439 & -0.013388 & -0.003932 
     & -0.012698 & -0.012618 \\
   3 & -0.0004456 & -0.004562 & -0.009036 
     & -0.003828 & -0.003715 \\
   4 & -4.406 \cdd 5& -0.0001517 & -0.001821 
     & -0.001572 & -0.001524 \\
   5 &           & -2.009 \cdd 5 &-6.377 \cdd 5
     & -0.002017 & -0.0007604 \\
   6 &           & -3.59 \cdd 6 & -9.89 \cdd 6
     & -3.15 \cdd 5 & -0.000480  \\
   7 &           &             & -2.30 \cdd 6 
     & -5.29 \cdd 6 & -1.74 \cdd 5 \\
   8 &           &             & -5.60 \cdd 7 
     & -1.43 \cdd 6 & -3.06 \cdd 6 \\
   9 &           &             &  
     & -4.42 \cdd 7 & -8.95 \cdd 7  \\
   10&           &             &  
     & -1.29 \cdd 7 & -3.20 \cdd 7 \\
   11&           &             &  
     &              & -1.17  \cdd 7\\
   12&           &             &  
     &              & -3.84 \cdd 8 
\\
\multicolumn{1}{c}{} &
\multicolumn{5}{l}{\parbox{.6\linewidth}{
 ~ \\
Table \ref{tab:Table2a}:\\
The TCSA coefficients $f^{(N,\ML)}_n$ for $\ML=8$ and $4\leq N\leq 12$.
}}
\end{array}
\]
}

\refstepcounter{table}
\label{tab:Table2b}
{\small
\mathematica{
Table2 = {{{{2, 0}, -1.029790218815312}, {{2, 1}, -0.006525175476855644}, 
    {{2, 2}, -0.0006999559327896099}, {{2, 3}, -0.0001853925649235116}, 
    {{2, 4}, -0.0000736101610893111}, {{2, 5}, -0.0000368961639713377}, 
    {{2, 6}, -0.00002166427638943371}, {{2, 7}, -4.087116545194628*^-8}, 
    {{2, 8}, -5.455012336482715*^-9}, {{2, 9}, -1.522059182213267*^-9}, 
    {{2, 10}, -5.293722793222491*^-10}, {{2, 11}, -1.902363816978438*^-10}, 
    {{2, 12}, -5.675280850665517*^-11}}, 
   {{{4, 0}, -1.322335631507487}, {{4, 1}, -0.03945533773799424}, 
    {{4, 2}, -0.004772332457783045}, {{4, 3}, -0.001284746807550296}, 
    {{4, 4}, -0.0005122599982365599}, {{4, 5}, -0.00025574578307775}, 
    {{4, 6}, -0.0001513182265431643}, {{4, 7}, -1.394136969527658*^-6}, 
    {{4, 8}, -1.984658876680367*^-7}, {{4, 9}, -5.577994322816985*^-8}, 
    {{4, 10}, -1.945752610678077*^-8}, {{4, 11}, -7.0023855756757*^-9}, 
    {{4, 12}, -2.124113465259632*^-9}}, 
   {{{6, 0}, -1.491371239546741}, {{6, 1}, -0.0974965093783418}, 
    {{6, 2}, -0.01411079989920577}, {{6, 3}, -0.003936644448148431}, 
    {{6, 4}, -0.001586039202307934}, {{6, 5}, -0.0007893822358945965}, 
    {{6, 6}, -0.0004774421373675481}, {{6, 7}, -0.00001020475526756612}, 
    {{6, 8}, -1.607497633736968*^-6}, {{6, 9}, -4.586776367383765*^-7}, 
    {{6, 10}, -1.61306022887582*^-7}, {{6, 11}, -5.841707374968121*^-8}, 
    {{6, 12}, -1.827928980943349*^-8}}, 
   {{{8, 0}, -1.596642541299763}, {{8, 1}, -0.165535525215146}, 
    {{8, 2}, -0.02898764326374237}, {{8, 3}, -0.008533993562707021}, 
    {{8, 4}, -0.003501800712571309}, {{8, 5}, -0.001747004129943395}, 
    {{8, 6}, -0.001103066067117593}, {{8, 7}, -0.00003983260190607214}, 
    {{8, 8}, -7.034582419495929*^-6}, {{8, 9}, -2.055099430179556*^-6}, 
    {{8, 10}, -7.342359547772959*^-7}, {{8, 11}, -2.697638838322589*^-7}, 
    {{8, 12}, -8.828262531897326*^-8}}, 
   {{{10, 0}, -1.667338153628752}, {{10, 1}, -0.2320020717534375}, 
    {{10, 2}, -0.0483000618861303}, {{10, 3}, -0.01516566239344002}, 
    {{10, 4}, -0.006382225506259051}, {{10, 5}, -0.003218671975323074}, 
    {{10, 6}, -0.002172300541964051}, {{10, 7}, -0.0001115396654492206}, 
    {{10, 8}, -0.00002200424820700038}, {{10, 9}, -6.62984831802727*^-6}, 
    {{10, 10}, -2.427335073679809*^-6}, {{10, 11}, -9.133241341088814*^-7}, 
    {{10, 12}, -3.167949465985897*^-7}}, 
   {{{12, 0}, -1.718271571373209}, {{12, 1}, -0.2920622546250419}, 
    {{12, 2}, -0.07039112588231146}, {{12, 3}, -0.02364201613087307}, 
    {{12, 4}, -0.01023937113034857}, {{12, 5}, -0.005273698275721853}, 
    {{12, 6}, -0.003898653870621236}, {{12, 7}, -0.0002561108936624071}, 
    {{12, 8}, -0.00005582860838111455}, {{12, 9}, -0.00001743865884441132}, 
    {{12, 10}, -6.598188695867891*^-6}, {{12, 11}, -2.566582156565421*^-6}, 
    {{12, 12}, -9.569365361724988*^-7}}}
Table2b2 = {{{{2, 0}, -0.7804350473443685}, {{2, 1}, -0.006525175476855644}, 
    {{2, 2}, -0.0006999559327896099}, {{2, 3}, -0.0001853925649235116}, 
    {{2, 4}, -0.0000736101610893111}, {{2, 5}, -0.0000368961639713377}, 
    {{2, 6}, -0.00002166427638943371}, {{2, 7}, -4.087116545194628*^-8}, 
    {{2, 8}, -5.455012336482715*^-9}, {{2, 9}, -1.522059182213267*^-9}, 
    {{2, 10}, -5.293722793222491*^-10}, {{2, 11}, -1.902363816978438*^-10}, 
    {{2, 12}, -5.675280850665517*^-11}}, 
   {{{4, 0}, -0.7594823823333082}, {{4, 1}, -0.03945533773799424}, 
    {{4, 2}, -0.004772332457783045}, {{4, 3}, -0.001284746807550296}, 
    {{4, 4}, -0.0005122599982365599}, {{4, 5}, -0.00025574578307775}, 
    {{4, 6}, -0.0001513182265431643}, {{4, 7}, -1.394136969527658*^-6}, 
    {{4, 8}, -1.984658876680367*^-7}, {{4, 9}, -5.577994322816985*^-8}, 
    {{4, 10}, -1.945752610678077*^-8}, {{4, 11}, -7.0023855756757*^-9}, 
    {{4, 12}, -2.124113465259632*^-9}}, 
   {{{6, 0}, -0.72832507711925}, {{6, 1}, -0.0974965093783418}, 
    {{6, 2}, -0.01411079989920577}, {{6, 3}, -0.003936644448148431}, 
    {{6, 4}, -0.001586039202307934}, {{6, 5}, -0.0007893822358945965}, 
    {{6, 6}, -0.0004774421373675481}, {{6, 7}, -0.00001020475526756612}, 
    {{6, 8}, -1.607497633736968*^-6}, {{6, 9}, -4.586776367383765*^-7}, 
    {{6, 10}, -1.61306022887582*^-7}, {{6, 11}, -5.841707374968121*^-8}, 
    {{6, 12}, -1.827928980943349*^-8}}, 
   {{{8, 0}, -0.694979031855532}, {{8, 1}, -0.165535525215146}, 
    {{8, 2}, -0.02898764326374237}, {{8, 3}, -0.008533993562707021}, 
    {{8, 4}, -0.003501800712571309}, {{8, 5}, -0.001747004129943395}, 
    {{8, 6}, -0.001103066067117593}, {{8, 7}, -0.00003983260190607214}, 
    {{8, 8}, -7.034582419495929*^-6}, {{8, 9}, -2.055099430179556*^-6}, 
    {{8, 10}, -7.342359547772959*^-7}, {{8, 11}, -2.697638838322589*^-7}, 
    {{8, 12}, -8.828262531897326*^-8}}, 
   {{{10, 0}, -0.6637792746970348}, {{10, 1}, -0.2320020717534375}, 
    {{10, 2}, -0.0483000618861303}, {{10, 3}, -0.01516566239344002}, 
    {{10, 4}, -0.006382225506259051}, {{10, 5}, -0.003218671975323074}, 
    {{10, 6}, -0.002172300541964051}, {{10, 7}, -0.0001115396654492206}, 
    {{10, 8}, -0.00002200424820700038}, {{10, 9}, -6.62984831802727*^-6}, 
    {{10, 10}, -2.427335073679809*^-6}, {{10, 11}, -9.133241341088814*^-7}, 
    {{10, 12}, -3.167949465985897*^-7}}, 
   {{{12, 0}, -0.6359446328459963}, {{12, 1}, -0.2920622546250419}, 
    {{12, 2}, -0.07039112588231146}, {{12, 3}, -0.02364201613087307}, 
    {{12, 4}, -0.01023937113034857}, {{12, 5}, -0.005273698275721853}, 
    {{12, 6}, -0.003898653870621236}, {{12, 7}, -0.0002561108936624071}, 
    {{12, 8}, -0.00005582860838111455}, {{12, 9}, -0.00001743865884441132}, 
    {{12, 10}, -6.598188695867891*^-6}, {{12, 11}, -2.566582156565421*^-6}, 
    {{12, 12}, -9.569365361724988*^-7}}}
}
\[
\begin{array}{r|llllll}
     & \multicolumn{1}{c}{\ML}   \\
   n & \multicolumn{1}{c}{2}
     & \multicolumn{1}{c}{4}   
     & \multicolumn{1}{c}{6}   
     & \multicolumn{1}{c}{8}   
     & \multicolumn{1}{c}{10}   
     & \multicolumn{1}{c}{12} \\
\hline
   0 &-0.780435       &-0.759482     &-0.728325     &-0.694979     
     &-0.663779       &-0.635945  \\
   1 &-0.0049452      &-0.0226611    &-0.0476133    &-0.0720535     
     &-0.0923617      &-0.108094  \\
   2 &-0.0005305      &-0.0027410    &-0.0068911    &-0.0126176     
     &-0.0192286      &-0.026052  \\
   3 &-0.0001405      &-0.0007379    &-0.0019225    &-0.0037146     
     &-0.0060376      &-0.0087501  \\
   4 &-5.579 \cdd 5   &-0.0002942    &-0.0007746    &-0.0015242     
     &-0.0025408      &-0.0037897  \\
   5 &-2.796 \cdd 5   &-0.0001469    &-0.0003855    &-0.0007604     
     &-0.0012814      &-0.001952  \\
   6 &-1.642 \cdd 5   &-8.691 \cdd 5 &-0.0002332    &-0.0004801     
     &-0.0008648      &-0.001443  \\
   7 &-3.10 \cdd 8    &-8.01 \cdd 7  &-4.98 \cdd 7  &-1.73 \cdd 5
     &-4.44 \cdd 5    &-9.48 \cdd 5  \\
   8 &-4.13 \cdd 9    &-1.14 \cdd 7  &-7.85 \cdd 7  &-3.06 \cdd 6 
     &-8.76 \cdd 6    &-2.07 \cdd 5 \\
   9 &-1.15 \cdd 9    &-3.20 \cdd 8  &-2.24 \cdd 7  &-8.95 \cdd 7 
     &-2.64 \cdd 6    &-6.45 \cdd 6 \\
  10 &-4.01 \cdd{10}  &-1.12 \cdd 8  &-7.88 \cdd 8  &-3.20 \cdd 7 
     &-9.66 \cdd 7    &-2.44 \cdd 6  \\
  11 &-1.44 \cdd{10}  &-4.02 \cdd 9  &-2.85 \cdd 8  &-1.17 \cdd 7 
     &-3.64 \cdd 7    &-9.50 \cdd 7  \\
  12 &-4.30 \cdd{11}  &-1.22 \cdd 9  &-8.93 \cdd 9  &-3.84 \cdd 8
     &-1.26 \cdd 7    &-3.54 \cdd 7  \\
\hline
\hbox{sum} 
    &-0.786151        &-0.786151 &-0.786151 &-0.786151 &-0.786151 & -0.786151
\\
\multicolumn{1}{c}{} &
\multicolumn{6}{l}{\parbox{.6\linewidth}{
 ~ \\
Table \ref{tab:Table2b}:\\
The TCSA coefficients $f^{(N,\ML)}_n$ for $N=12$ and $2\leq \ML\leq 12$.
}}
\end{array}
\]
}

\subsubsection{The strip with $(\Phi(h),\One)$ boundary conditions}

This case is only slightly more complicated.
The unperturbed expectation value is again given in (\ref{eq:c1ptfns}).
The TCSA estimation of the expectation value in the massive model with
boundary perturbations then takes the general form
\bea
  \cI^{(N,\ML)}(\xi)
&{\!\!\!\!\equiv\!\!\!\!}&
 \left.
 M^{2/5} \vev{ \varphi(x) }_{(\Phi(h),\One)}
 \right|_{N}
\label{eq:cI}
\\
&{\!\!\!\!=\!\!\!\!}&
  \left( \frac{2\,\ML}\pi \right)^{\!\!2/5}
  \sum_{n=0}^{2N}
  \sum_{j=1}^2
  \left(
  g_{nj}^{(N,\ML)}(h)\,
  f_j\!\!\left( \frac{\pi\xi}{\ML} \right)
  \cos\!\left(\frac{n\pi\xi}\ML\right) 
\,+\,
  h_{nj}^{(N,\ML)}(h)\,
  f_j'\!\!\left( \frac{\pi\xi}{\ML} \right)
  \sin\!\left(\frac{n\pi\xi}\ML\right) 
  \right)
\nn
\eea
where $h$ is related to the reflection factor parameter $b$
by $h\,{=}\,h(b)$, and where the functions 
$g_{nj}^{(N,\ML)}(h)$ and $h_{nj}^{(N,\ML)}(h)$ have to be
evaluated numerically.
Since the ground state $\vec{\hat 0}$ of the TCSA Hamiltonian lies in
the $h\,{=}\,{-}1/5$ representation, the matrix elements of
$\hat\varphi$ are given in terms of the two chiral blocks $f_1,f_2$;
furthermore, since this representation has a null state at level 2, 
one can eliminate all states with more than one mode $L_{-1}$, which
leads to the fact that one does not need to use higher than the first
derivative of the chiral blocks in (\ref{eq:cI}).
Since there are rather many coefficients $g_{n}^{(N,\ML)}$,
$h_n^{(N,\ML)}$, we shall not give any explicit examples.

In figure \ref{fig:Graph7} we 
also plot $\cI(\xi)$ against $\xi$ for various values of $l$ between
$0.5$ and $16$, for the fixed value $h\,{=}\,0$, $b\,{=}\,{-}1/2$.
The excellent agreement between $\cI(\xi)$ and $\cG(\xi)$
that can be seen on the half of the strip $\ML/2<\xi<\ML$ for the larger
values of $\ML$ is a good sign that the TCSA estimates
of the functions are converging to their correct values.
Table~\ref{tab:Table10} below 
includes results on the convergence in $N$ of
$\cI^{(N)}(\xi)$ for values of $\xi$ between $0.01$ and $1.0\,$.

\subsubsection{The strip with $(\Phi(h_l),\Phi(h_r))$ boundary conditions}

The calculation of the expectation value of $\varphi$ on the strip
with two perturbed boundary conditions $(\Phi(h_l),\Phi(h_r))$ is in
principle the same, except that the functional form is rather more
involved.
The unperturbed expectation value is again given in
(\ref{eq:c1ptfns}),
\[
  \vev{\varphi(x)}_{(\Phi(0),\Phi(0))}
= \left( \frac{\newL}{\pi} \right)^{2/5}
  \,
  \left(\;
  \B \Phi\varphi\One\; f_1( \frac{\pi x}{\newL} )
\;+\;
  \B \Phi\varphi\phi\;\C\phi\phi\phi\,f_2( \frac{\pi x}{\newL} )
  \;\right)
\;
\]
where the functions $f_i(\theta)$ are the strip chiral blocks
(\ref{eq:cbs}).
However, the massive perturbation introduces terms proportional to the
other two chiral blocks and their derivatives, so that 
the TCSA estimation of the expectation value in the massive model with
boundary perturbations takes the general form
\bea
  \cH^{(N,\ML)}(\xi)
&{\!\!\!\!=\!\!\!\!}&
 \left.
 M^{2/5} \vev{ \varphi(x) }_{(\Phi(h_l),\Phi(h_r))}
 \right|_{N}
\label{eq:cH}
\\
&{\!\!\!\!=\!\!\!\!}&\!\!
  \left( \frac{\ML}\pi \right)^{\!\!2/5}
  \sum_{n=0}^{2N}
  \sum_{k=1}^4
  \left(
  j_{kn}^{(N,\ML)}(h_l,h_r)\,
  f_k\!\!\left( \fract{\pi\xi}{\ML} \right)
  \cos\!\left(\fract{n\pi\xi}\ML\right) 
\,+\,
  k_{kn}^{(N,\ML)}(h_l,h_r)\,
  f_k'\!\!\left( \fract{\pi\xi}{\ML} \right)
  \sin\!\left(\fract{n\pi\xi}\ML\right) 
  \right)
\nn
\eea
where $h_l\,{=}\,h(b_l)$, $h_r\,{=}\,h(b_r)$ and the functions 
$j_{in}^{(N,\ML)}(h_l,h_r)$ and $k_{in}^{(N,\ML)}(h_l,h_r)$ are
evaluated numerically.
Again, since there are rather many coefficients, we shall not give
any explicit examples.
For large values of $\ML$, the two boundaries are essentially
non-interacting -- this was already seen in figure \ref{fig:Graph7}.
We therefore see no new phenomena over those seen already;
the TCSA estimates of $\vev{M^{2/5}\varphi(x)}$
near the left boundary from the system  with boundary conditions
$(\Phi(h_l),\Phi(h_r))$  are barely distinguishable from those from the
system with boundary conditions $(\Phi(h),\One)$.
Further confirmation is contained in table \ref{tab:Table10}, which
presents data from the two situations.

However, for small $\ML$, the presence of two perturbed
boundaries can destabilise the vacuum even for values of the parameters 
$h_l$ and $h_r$ which are less negative than 
  $-|\hc|$,
the value for which a single boundary destabilises the 
bulk vacuum on a half line \cite{Us1}. 
For $b_l+b_r=0$, $|b_l|<2$,
the ground state and first excited state have an exact crossing at a 
finite value of $\ML$, while for $b_l+b_r>0$ there is a finite range of
$\ML$ for which they become complex.
We discuss this in section \ref{rgfs},
where we give examples of the spectra, and plots of
the boundary field expectation values
$\vev{M^{1/5}\phi}$, for systems with two perturbed boundaries.

\vspace{2mm}
\begin{table}[ht]
\refstepcounter{table}
\label{tab:Table10}
{\small
\[
\mathematica{
%
%
%
  {{0.01,-1.61719015605},
   {0.03,-1.6097092974},
   {0.1, -1.58467371675},
   {0.3, -1.52208518161},
   {1.0, -1.3797744504}}
  {{0.01,-1.21289371013},
   {0.03,-1.34553783152},
   {0.1, -1.4543815195},
   {0.3, -1.47810017212},
   {1.0, -1.37548697616}}
  {{0.01 ,-1.14850765784},
   {0.03 ,-1.31311309464},
   {0.1  ,-1.44450624328},
   {0.3  ,-1.47657109985},
   {1.0  ,-1.37545817048}}
}
\begin{array}{c}
\begin{array}{l|llll|lll}
     & \multicolumn{4}{l|}{\hbox{\ \ \ \ TCSA truncation level $N$}}   
     & \multicolumn{3}{l}{\hbox{\ \ \ \ FF   truncation level $n$}}   
\\
       \multicolumn{1}{c|}{\xi} 
     & \multicolumn{1}{c}{4}   
     & \multicolumn{1}{c}{8}   
     & \multicolumn{1}{c}{12}   
     & \multicolumn{1}{c|}{16} 
     & \multicolumn{1}{c}{1} 
     & \multicolumn{1}{c}{2} 
     & \multicolumn{1}{c}{3}
\\
\hline
0.01& \mm{-1.19134}{-1.16788}
    & \mm{-1.16743}{-1.16261}
    & \mm{-1.16362}{-1.16234}
    & \mm{-1.16306}{-1.16239} 
    & -1.61719 & -1.21289 & -1.14851 \\
0.03& \mm{-1.35236}{-1.32317}
    & \mm{-1.32259}{-1.31662}
    & \mm{-1.31786}{-1.31627}
    & \mm{-1.31716}{-1.31633} 
    & -1.60971 & -1.34554 & -1.31311 \\
0.1 & \mm{-1.48965}{-1.45280}
    & \mm{-1.45199}{-1.44456}
    & \mm{-1.44604}{-1.44414}
    & \mm{-1.44517}{-1.44421} 
    & -1.58467 & -1.45438 & -1.44451 \\
0.3 & \mm{-1.52882}{-1.48563}
    & \mm{-1.48386}{-1.47625}
    & \mm{-1.47722}{-1.47586}
    & \mm{-1.47633}{-1.47592} 
    & -1.52209 & -1.47810 & -1.47657 \\
1.0 & \mm{-1.40584}{-1.37786}
    & \mm{-1.37064}{-1.37380}
    & \mm{-1.37158}{-1.37424}
    & \mm{-1.3738}{-1.37447} 
    & -1.37977 & -1.37549 & -1.37546 
\end{array}
\\
\parbox{.95\linewidth}{
 ~ \\
Table \ref{tab:Table10}:\\
The TCSA and FF estimates of $\vev{\varphi(\xi)}_{\Phi(0)}$,
for varying distance $\xi$ from the boundary 
{}from the Form-Factor approach truncated at particle number $n$ 
and from TCSA truncated to level $N$ with $\ML\,{=}\,12$ on strips
with b.c.'s $(\Phi(0),\One)$ (upper line) and
$(\Phi(0),\Phi(0))$ (lower line).
}
\end{array}
\]
}
\end{table}
\vspace{1mm}

\newpage
\subsection{The comparison of the TCSA and Form-Factor results}
\label{ssec:TCSAFF}

The Form-Factor method gives the expectation value of $\varphi(x)$ 
on the half-plane. While we can think of the half-plane the
infinite-width limit of a finite strip, this is not accessible
directly using the TCSA method, which 
is limited to strips of finite width $\ML$ and
is expected to perform best near to $\ML=0$.
To enable a comparison of the two methods, we shall simply take the
TCSA results on a strip of width $\ML\,{=}\,12$, rather than extrapolating
finite width TCSA results to infinite width.
(The error for small values of $\xi$ from taking TCSA results at
$\ML\,{=}\,12$ should be much less than the one-particle FF
contribution in the middle of the strip, which is $\sim 0.2\%$).

In figure \ref{fig:PlotAC} we show results for two
cases: the boundary conditions $(\One)$ and $(\Phi(0))$.
We give the Form-Factor expansion (with $\tilde g\,{=}\,g/2$) 
truncated to one--, two-- and three-- particle states, 
and the TCSA data from 
truncation to level 16 (with $\ML \,{=}\, 12$), and they are clearly
in excellent agreement.
We also show the Form-Factor expansion assuming $\tilde g\,{=}\,g$,
and it is clear that this is wrong.
In table \ref{tab:Table10} we give some numbers illustrating the 
convergence in TCSA truncation level and in form-factor truncation
level for various values of $.01\,{\leq}\,\xi\,{\leq}\,1$.
\[
\begin{array}{c}
\refstepcounter{figure}
\label{fig:PlotAC}
\epsfxsize=.6\linewidth
\epsfbox[72 162 540 640]{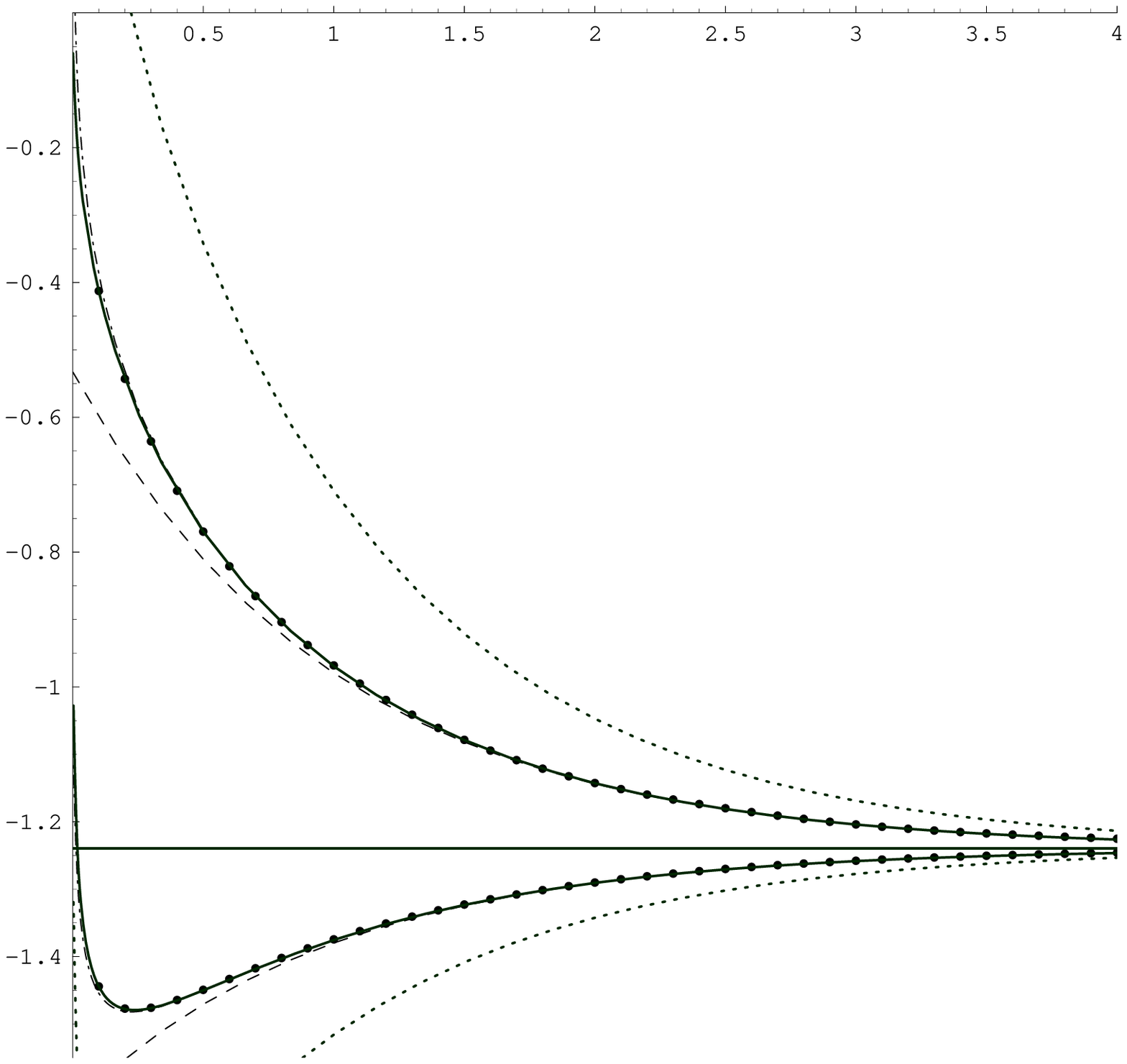}
\\
\parbox{.91\linewidth}{\raggedright
Figure \ref{fig:PlotAC}\\ \small
Comparisons of $\vev{M^{2/5}\varphi(x)}$ on a half-plane with
boundary condition $(\One)$ (upper lines) and $(\Phi(0))$ (lower lines).
The points are the TCSA data, 
the dashed lines the FF result up to 1 particle with $\cg= g/2$,
the dot-dashed line the FF result up to 2 particles,
the solid line the FF result up to 3 particles.
The dotted lines are the FF results up to 3 particles with $\cg= g$.
}
\end{array}
\]

The boundary condition $\Phi(0)$ ($b=-1/2$) was chosen because it is
for this value of $b$ that the accuracy of the TCSA is highest;
comparisons between FF and TCSA results for the further
values $b=-1$ and $b=0$ can be found in figure
\ref{fig:Graph9}, in section~\ref{bndryfl} below. The curve there for
$b=0$ is particularly interesting, as this is the case, mentioned at
the end of  section~2, for which the only difference  between the
$\Phi(h)$ and the $\One$ boundary states is the sign of the
odd-particle-number contributions.

The two-- and three-- particle form factor
expressions 
in figure \ref{fig:PlotAC}
are barely distinguishable from the TCSA data, and so in
figures
\ref{fig:PlotA1} and \ref{fig:PlotC1}
we plot $\log(\vev{M^{2/5}\varphi(x)})$ 
against $\log(\xi)$ for the $\One$ and $\Phi(0)$
boundary conditions.
We also show the leading behaviour 
\bea
  \vev{M^{2/5}\varphi(x)}_{\Phi(0)}
&=& 
  (2\xi)^{1/5}\,\B\Phi\varphi\phi\,\vev{\phi}_{\Phi(0)}
\;+\;
  (2\xi)^{2/5}\,\B\Phi\varphi\One\,\vev{\One}_{\Phi(0)}
\;+\;
  O(\xi^{12/5})
\;,
\label{eq:exp1}
\\
  \vev{M^{2/5}\varphi(x)}_{\One}
&=& 
  (2\xi)^{2/5}\,\B\One\varphi\One\,\vev{\One}_{\One}
\;+\;
  O(\xi^{12/5})
\;,
\label{eq:exp2}
\eea
of small--$\xi$ expansions obtained from a perturbative treatment of the
structure functions. 
(We intend to report on this approach elsewhere~\cite{Usnext}.)
We see that the three-particle FF approximation already agrees very
well with these expansions
for $-3 \lesssim \log\xi \lesssim -2 $.

\vspace{4mm}

\[
\begin{array}{cc}
\refstepcounter{figure}
\label{fig:PlotA1}
\epsfxsize= .45\linewidth
\epsfbox[72 162 540 650]{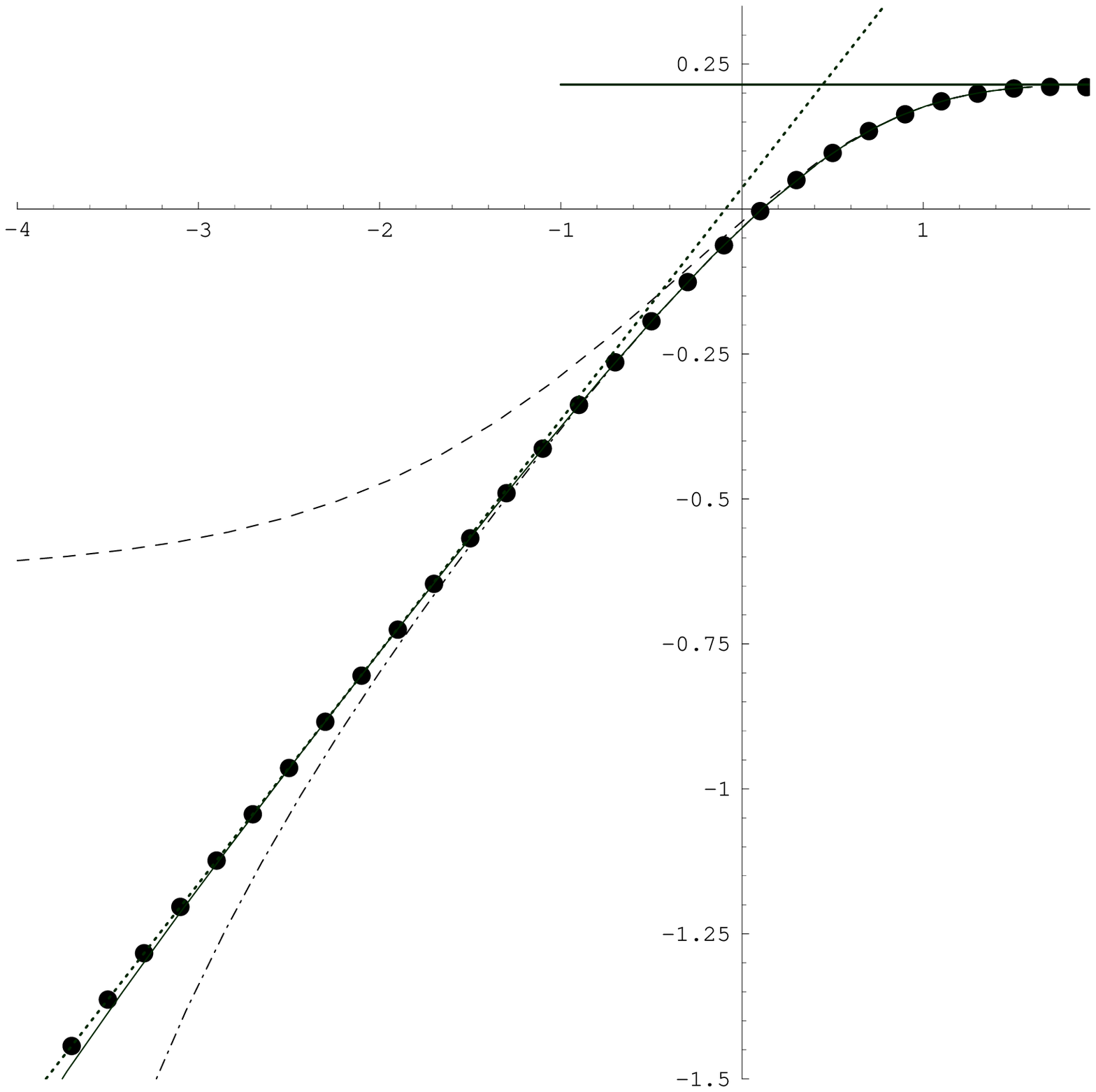} 
{~~~}&{}
\refstepcounter{figure}
\label{fig:PlotC1}
\epsfxsize= .45\linewidth
\epsfbox[72 162 540 650]{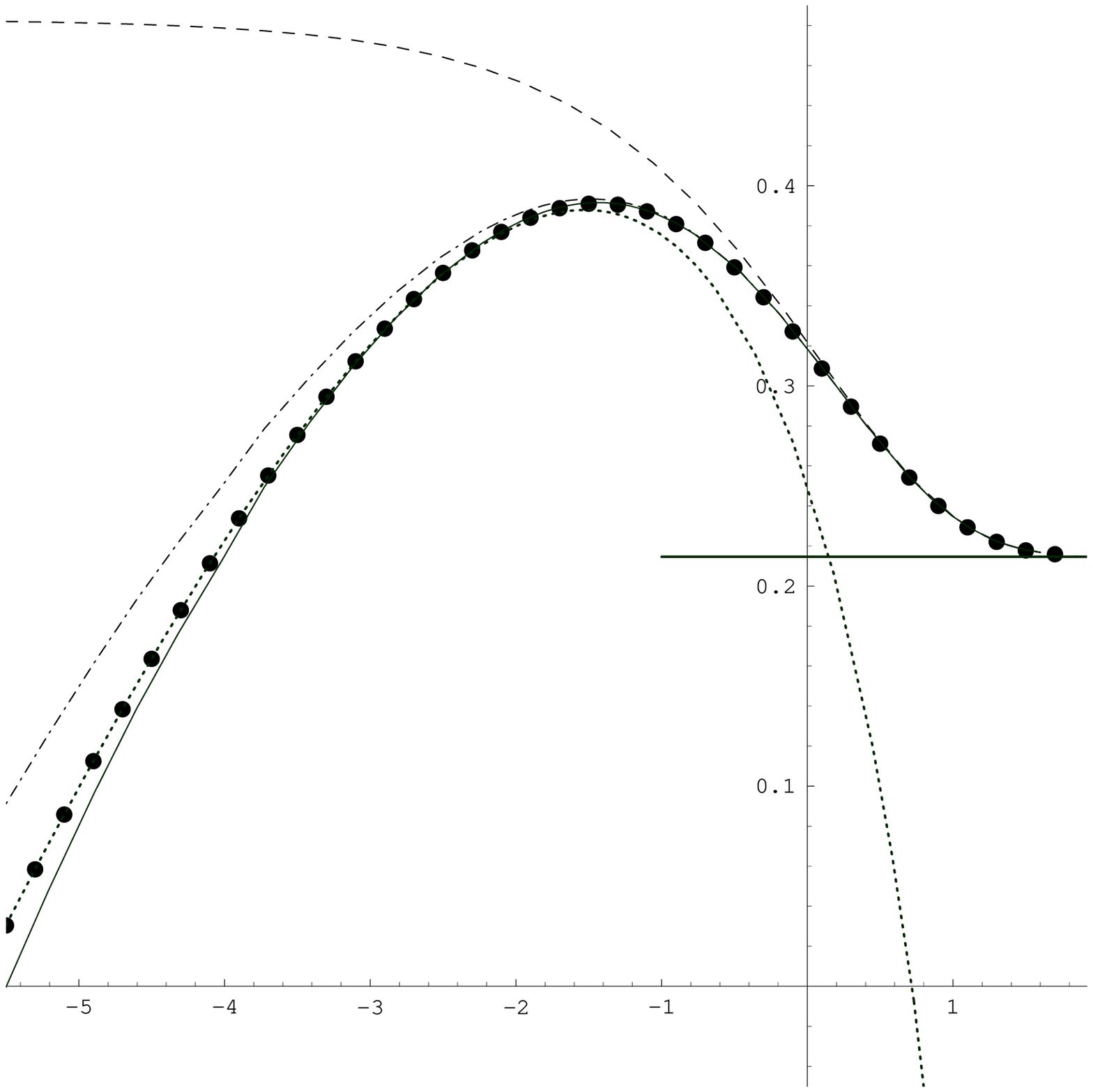} 
\\[2mm]
\parbox[t]{.45\linewidth}{\raggedright%
Figure \ref{fig:PlotA1}:\\ \small
Plots of $\log|\vev{M^{2/5}\varphi(x)}_{\One}|$ vs.\ $\log(\xi)$.
The points are TCSA data, the dashed, dot-dashed and solid lines are
{}from the FF expansion truncated to one, two and three particles resp.,
and the dotted line is (\ref{eq:exp2}).
}
{~~~}&{~}
\parbox[t]{.45\linewidth}{\raggedright%
Figure \ref{fig:PlotC1}:\\ \small
Plots of $\log|\vev{M^{2/5}\varphi(x)}_{\Phi(0)}|$ vs.\ $\log(\xi)$.
The points are TCSA data, the dashed, dot-dashed and solid lines are
{}from the FF expansion truncated to one, two and three particles resp.,
and the dotted line is (\ref{eq:exp1}).
}
\\
\end{array}
\]

\newpage
\resection{Excited states and energy density}
\label{sec:excited}

The TCSA and FF methods are not
restricted to the ground state expectation values of the field
$\varphi(x,y)$. In this section we give a couple of examples of their
wider applicability.

\subsection{The expectation value of $\varphi$ in the first excited state}
\label{tevovitfes}

It is just as easy in the TCSA to find the first excited
eigenstate $\vec{\hat 1}$ as to find the ground state and to find the
corresponding expectation value of $\varphi$. The result is
somewhat less accurate than in the ground state, and this accuracy
decreases as higher and higher excited levels are considered.
In figure \ref{fig:excited1} we show (as points) the TCSA result near
the $\Phi(0)$ boundary of the strip with boundary conditions 
$(\Phi(0),\One)$ with $\ML=14$ and truncation level 14.
This state corresponds to the boundary bound state of energy 
$e_1 = M \cos((b+1)\pi/6) = \sqrt 3 M / 2$,
and so we expect that near the $\Phi(0)$ boundary the expectation
value is approximately given by the expectation value 
in the first excited state for the semi-infinite geometry.
It turns out that this expectation value can also be obtained using FF
techniques, with results that are shown in the various curves on the
figure. 
The calculation relies on an idea of analytic continuation between
states; similar methods were used to find TBA equations for excited
states in~\cite{DTa}.

\[
\begin{array}{c}
\refstepcounter{figure}
\label{fig:excited1}
\epsfxsize=.8\linewidth
\epsfbox[72 242 540 560]{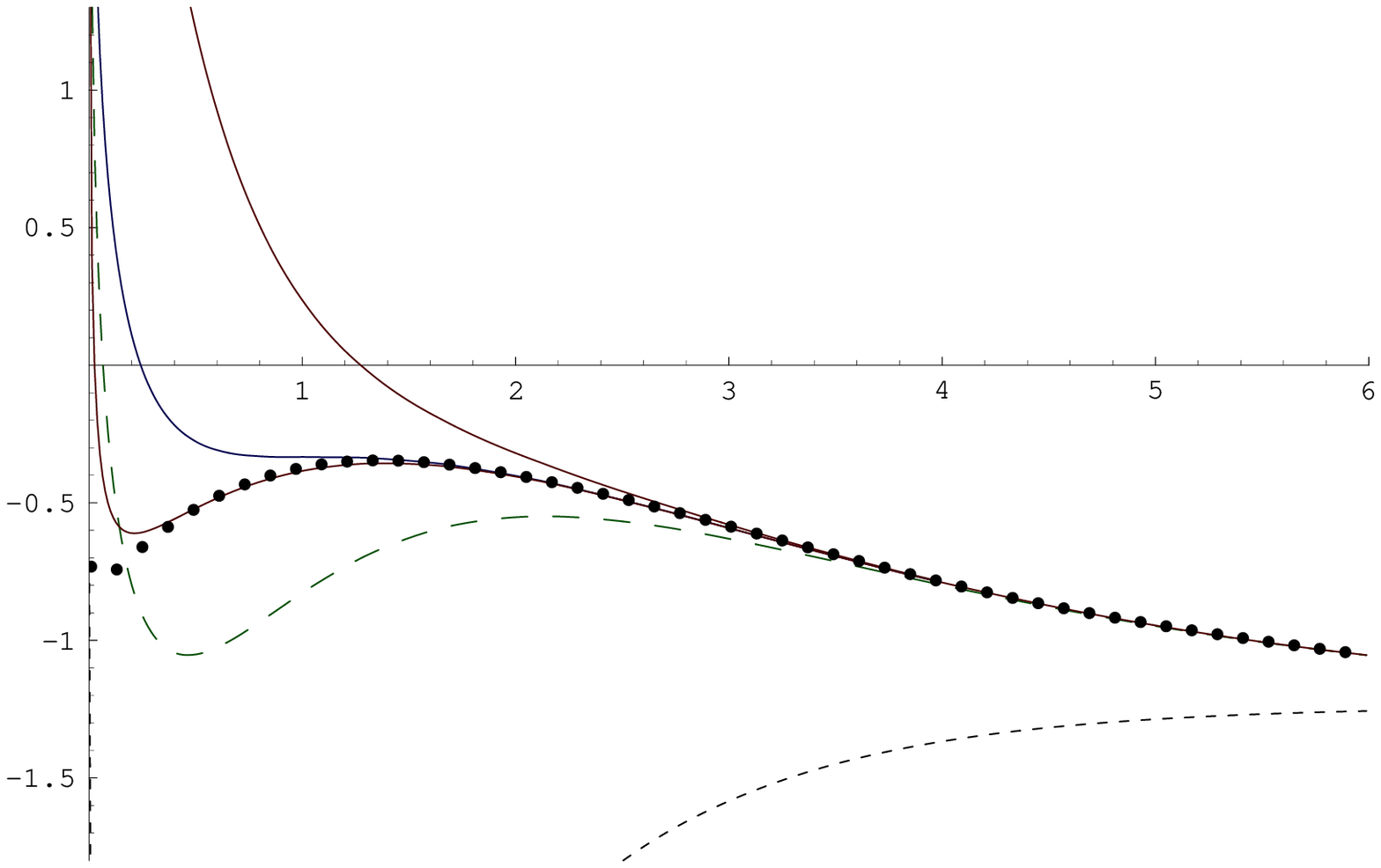}
\\
\parbox{.9\linewidth}{\raggedright
Figure \ref{fig:excited1}:\\ \small
Comparisons of $\vev{M^{2/5}\varphi(x)}_{\Phi(0)}$ in the first
excited state. 
The points are the TCSA data from truncation level 14 and $\ML=14$.
The dotted line is the FF expression (\ref{eq:FF}) with $b=9/2$;
the dashed line is the correction by the terms in (\ref{eq:Corr1})
{}from the poles at $\theta={\pm}(2{-}b)\pi/6$,
and the solid lines are the results of including all corrections by
the addition of the terms up to 
(\ref{eq:Corr1}), up to
(\ref{eq:Corr2}) and up to (\ref{eq:Corr4}) respectively.
}
\end{array}
\]

The fact that the reflection factors $R_b(\theta)$ obey
a curious continuation property
was already observed in \cite{Us2}.
While the physical parameter $h(b)$ is invariant under under 
$b \to 4{-}b$,
the reflection factor $R_b$ is not; instead, it is interchanged with
the reflection factor of the first boundary bound state.
This suggests that the expectation value of the
field $\varphi$ in the first excited state can be obtained
by continuing the FF expression (\ref{eq:FF}) from the domain
${-}3<b<2$ to the region $2<b<5$.

In figure \ref{fig:excited1} the result of the substitution of $b=9/2$
in (\ref{eq:FF}) is shown as a dotted line. Clearly, it is a long way
{}from the corresponding TCSA results for $\vev\varphi^{(1)}$.
The explanation is simple: there are poles in $K_b(\theta)$ whose
positions depend on $b$, in particular at $\theta = \pm i (2{-}b)\pi/6$
and $\theta = \pm i(4{-}b)\pi/6$.
These two pairs of poles cross the integration contour (the real axis) at
$b=2$ and $b=4$ respectively, 
and contributions from both pairs must be added in explicitly 
to recover the correct analytic
continuation in $b$ of $\vev\varphi$ to  $b=9/2$, corresponding to $h=0$.

We can regard the contributions from these poles as directly affecting
the exponential in (\ref{Bexp1b}). We 
denote the positions of the `active' poles (those which have crossed the
integration contour during the continuation)
by $\theta_i(b)$, and the
contribution to the contour integral from the pole in 
$K_\alpha(\theta)$ at $\theta_i(b)$ by $k_i(b)$ -- this will be $\pm 1$
times the relevant residue, depending on whether the contour was crossed
{}from above or below when the pole became active.
Then we can associate the following state in the full-line Hilbert
space with the first excited state on the half-line:
\be
  \vec{ B_{\alpha}' } 
= \exp \left[ 
    \;\;
    \tilde g_\alpha\, A(0) 
\;+\;
\frac i2 \sum_i k_i(b) A({-}\theta_i(b))A(\theta_i(b))
\;+\;
    \int^{\infty}_{-\infty} 
    \frac{\D\theta}{4 \pi}
    K_{\alpha}(\theta )\,
    A({-}\theta)\,
    A(\theta)\, \right]  
    \vac
\;.
\label{Bexp1c}
\ee
Expanding this out and inserting the appropriate form factors,
the first corrections to the expectation value coming from the 
residue terms are
\bea
  \frac i2 
  \sum_i k_i(b)
\hspace{-6.2mm}
&&
  \cev 0 \varphi(0) 
  \vec{ {-}\theta_i(b),\theta_i(b) }
\,e^{ - 2 M x \cosh\theta_i(b) }
\;,
\label{eq:Corr1}
\\
  \frac {i\tilde g_\alpha}2 
  \sum_i k_i(b)
\hspace{-6.2mm}
&&
  \cev 0 \varphi(0) 
  \vec{ 0,{-}\theta_i(b),\theta_i(b) }
\,e^{ - M x (1 + 2 \cosh\theta_i(b))}
\;,
\label{eq:Corr2}
\\
  -\frac 14
  \sum_{i,j} k_i(b) k_j(b)
\hspace{-6.2mm}
&&
  \cev 0 \varphi(0) 
  \vec{ {-}\theta_i(b),\theta_i(b),{-}\theta_j(b),\theta_j(b) }
\,e^{ - 2 M x (\cosh\theta_i(b) +  \cosh\theta_j(b)) }
\;,
\label{eq:Corr3}
\\
  \frac i2
  \sum_{i} k_i(b) 
\hspace{1.0mm}
&&
\hspace{-7.5mm}
  \int_{-\infty}^{\infty}
  \frac{\D\theta}{4\pi}
  K_\alpha(\theta)
  \cev 0 \varphi(0) 
  \vec{ {-}\theta_i(b),\theta_i(b),{-}\theta,\theta} 
\,e^{ - 2 M x (\cosh\theta_i(b) +  \cosh\theta)}
\;,
\label{eq:Corr4}
\eea
where the last two terms can be seen as coming from poles in the
four-particle contribution to the boundary state which hitherto we
have neglected. 
Also, it turns out that the term (\ref{eq:Corr3}) always gives zero, due
to the particular relative positions of the poles at $\theta_1$ and
$\theta_2$.

We have shown the result of correcting (\ref{Bexp1b}) by the
dominant correction
(the term in (\ref{eq:Corr1}) coming from the poles at
$\theta={\pm}(2-b)\pi/6$) as a dashed line on figure
\ref{fig:excited1}, and the result of adding all terms up to
(\ref{eq:Corr1}), up to (\ref{eq:Corr2}) and up to (\ref{eq:Corr4}) as
solid lines. It is clear that these are converging rapidly to the
TCSA value.

\subsection{The expectation value of the energy density $\ep$}
\label{sec:exx}

As a second example, we use the TCSA to find the expectation
values of the energy density $\ep(x)$ on the strip for which
\[
  H = \int_0^\newL \! \ep(x)\,\D x
\;,
\;\;\;\;
  \ep(x)
= 
  - \frac{1}{2\pi}\left[ T(x) + \bar T(x) \right]
\;+\;
  \lambda\,\varphi(x)
\;.
\]
One has to be rather careful
about the specification of the operators in this expression.
Here, we mean by $T(x)$ and $\bar T(x)$ the `bare TCSA'
quantities -- in other words, their expectation values are computed in any
given state using the matrix elements of the CFT
operators $T$ and $\bar T$ between the eigenstates $\ket{\hat n}$ of the
perturbed Hamiltonian, themselves
expanded in the basis of CFT states. This is the
same procedure as was used for computations of $\vev{\varphi(x)}$ earlier, 
but some new issues arise in this case, 
to which we hope to return in~\cite{Usnext}.

\begin{figure}[ht]
\refstepcounter{figure}
\label{fig:Graph10}
\refstepcounter{figure}
\label{fig:bounce}
\[
\begin{array}{cr}
\epsfxsize=.47\linewidth
\epsfbox[72 252 540 540]{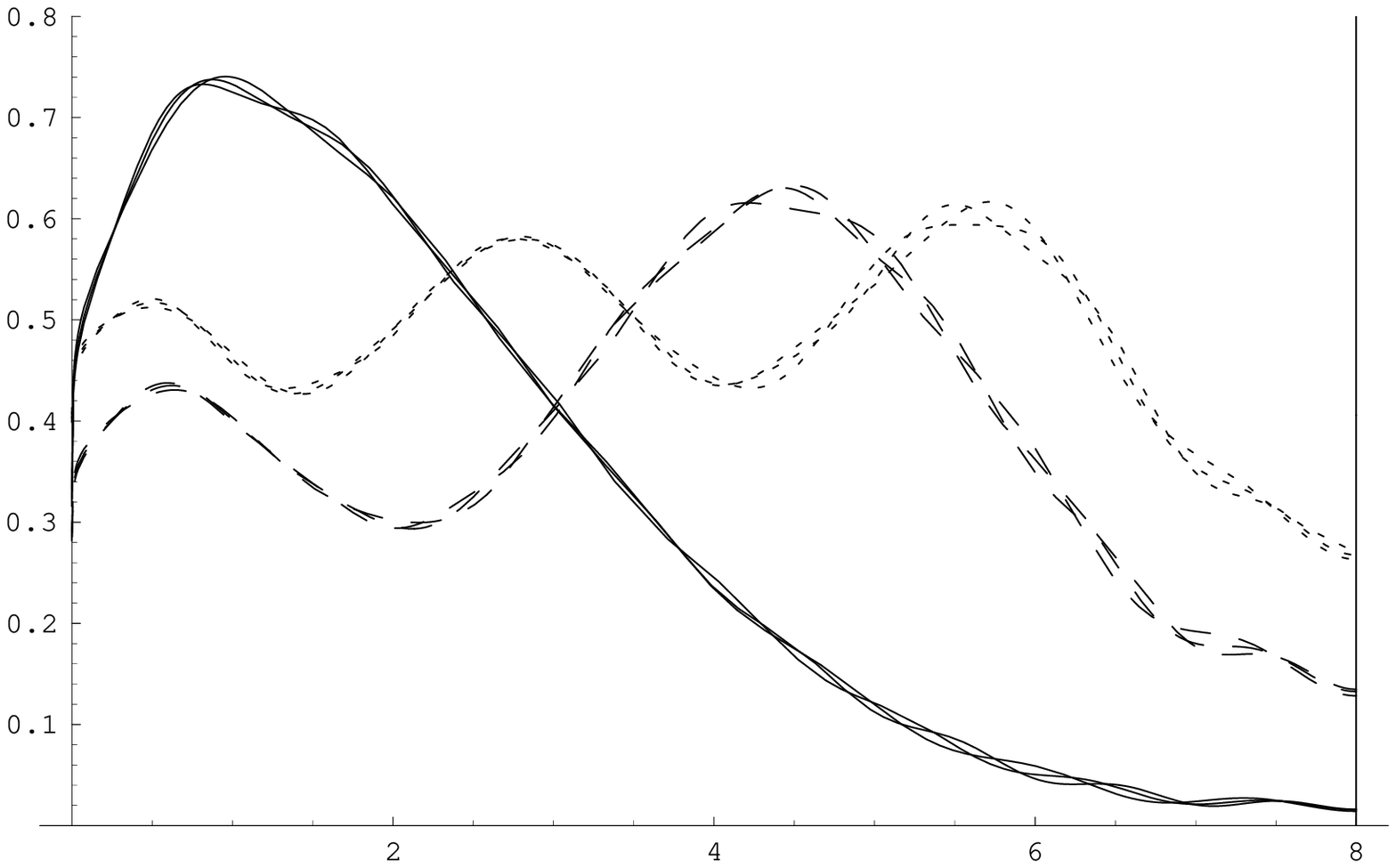}
&
\epsfxsize=.27\linewidth
\epsfbox[0 -30 417 400]{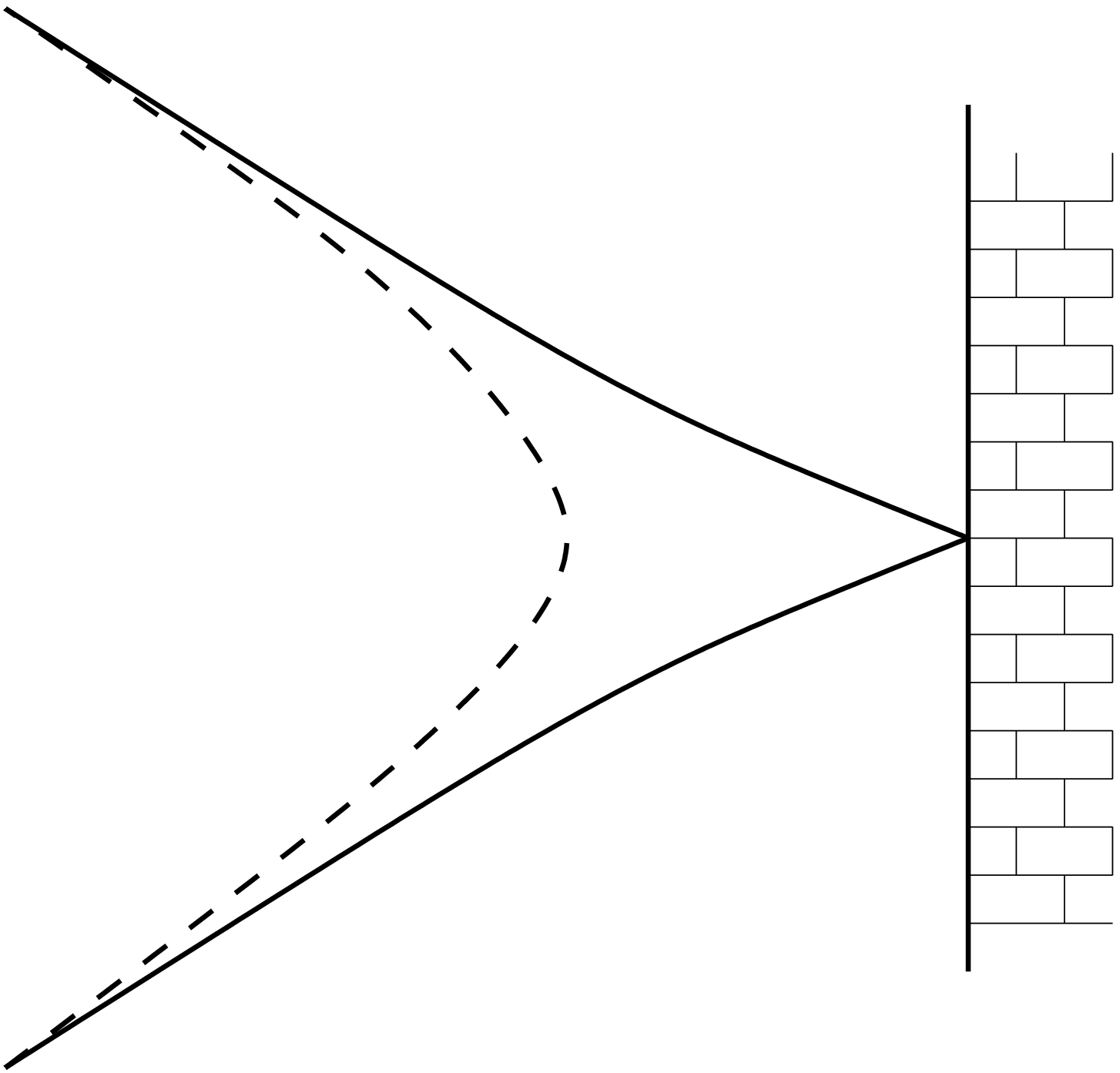}
\\
\parbox[t]{.6\linewidth}{\small\raggedright%
Figure \ref{fig:Graph10}:
Plots of 
$M^{-2}(\vev{\ep(\xi)}_n - \vev{\ep(\xi)}_0)$
vs.\ $\xi$ for the first few excited states on the system with
boundary conditions $(\Phi(0),\One)$ and $\ML\,{=}\,8$.
The solid line is the 1st excited state, the dash-dotted line the
2nd, and the dotted line the 3rd.
For each line,
TCSA results truncated to levels 12, 14 and 16 are superimposed to
give an idea of the errors.
}~~
&
 ~~\parbox[t]{.3\linewidth}{
Figure \ref{fig:bounce}:
\small\raggedright%
An attractive and a repulsive scattering process with the same
classical time delay.
}
\end{array}
\]
\end{figure}

Leaving these questions to one side, the operator $\varepsilon$ 
allows us to see directly that `boundary bound states'
are indeed localised at the boundary. 
In figure \ref{fig:Graph10} we plot the difference in the energy
density between the first few  excited states and the ground state 
for the system with boundary conditions $(\Phi(0),\One)$ for
$\ML\,{=}\,8$, calculated using TCSA truncated to 
levels 12, 14 and 16. 
The result of truncating the Fourier-like expansions
is very evident here, the TCSA estimates 
having distinct high-frequency ripples (varying with truncation level)
superposed on the overall function.

{}From the analysis in \cite{Us1,Us2}, for large $\ML$ the first excited
state is a boundary bound state, and the next several excited states
are single-particle scattering states. 
This is  borne out by figure \ref{fig:Graph10}, where we see very
clearly that the first excited state corresponds to a particle trapped
on the left ($\Phi(0)$) boundary and decaying exponentially across the
strip, while the  higher excited modes are well spread across the
strip, attracted to the $\Phi(0)$ boundary and repelled by the $\One$
boundary.

\noindent
It is impossible to decide whether a boundary $(B_\alpha)$ is
repulsive or attractive purely given the reflection factors
$R_\alpha(\theta)$.
The $\Phi(h(0))$ and $\One$ boundaries have the same reflection factor
but from figure \ref{fig:Graph10}, the $\Phi(h(0))$ boundary is 
attractive and the $\One$ boundary repulsive. This is the quantum
analogue of the inability in a classical theory
to determine whether a boundary is attractive or repulsive given only
the time-delays -- the two processes illustrated in figure
\ref{fig:bounce} have the same time delay but clearly one describes
attraction to the boundary and the other repulsion.


\resection{The one-point functions of the boundary field}
\label{sec:boundary}

We now turn to the one-point functions 
involving the boundary perturbing field $\phi$. 
In preparation for the main calculations, 
consider first the partition function $Z_{\alpha\beta}$ of
the model on cylinder of length $R$ and circumference $L$, with
boundary conditions $\alpha$ and $\beta$ imposed at the two ends. 
Formally, we can write
\eq
Z_{\alpha\beta}=
\int [{\cal D}\Psi]\, e^{-{\cal A}_{BLY}}\,,
\en
where
$\int[{\cal D}\Psi]$ implies that
a functional integral over all bulk and boundary degrees of freedom is
taken, and
${\cal A}_{BLY}$ denotes the
combined bulk and boundary action.
For the pair of boundary conditions
$(\alpha,\beta)=(\Phi(h_l),\Phi(h_r))$, this
can be written as
\eq
{\cal A}_{BLY}
= {\cal A}_{BCFT}+\lambda\int_0^R \D x\int_0^L \D y\,\varphi(x,y)
+h_l\int_0^L \D y\,\phi(0,y)
+h_r\int_0^L \D y\,\phi(R,y)
\en
where ${\cal A}_{BCFT}$ is an action for the ${\cal M}_{2,5}$ 
conformal field theory on the cylinder with conformal
boundary condition $\Phi$ at the two ends. (For the other two
pairs of boundary conditions the expression is similar,
but lacks one or both perturbing boundary fields.)
 
The behaviour of $Z_{\alpha\beta}$ as a function of $R$ and $L$ is
complicated, but if
both are much larger than all bulk and boundary
scales, then, up to exponentially-small corrections, 
\eq
\log Z_{\alpha\beta}\sim 
 -RL\Eblk
 -Lf_{\alpha}
 -Lf_{\beta}
\label{exten}
\en
where $f_{\alpha}$ and $f_{\beta}$ are the extensive parts of the
boundary free energies, and $\Eblk$ the extensive part of the bulk free
energy. 
For the scaling Lee-Yang model, $\Eblk=-M^2/(4\sqrt{3})$~\cite{Zb}.

Given  $Z_{\alpha\beta}\,$, the (normalised)
one point functions of the field $\phi$ 
can be simply  obtained by differentiation:
\eq
 \vev{\phi_{l/r}}_{\rm cyl}=
 -{1 \over L} {\partial \over \partial h_{l/r}}\log Z_{\alpha\beta} \,.
\en

However, while it was  shown in \cite{Us1} that 
the partition function  was 
numerically accessible via the TCSA, 
at the current
state of technology  $Z_{\alpha\beta}$  
(and hence   $\vev{\phi}_{\rm cyl}$)
is not directly computable by means of the TBA.
Contact with this `exact' method can instead be made in certain limits,
and these are best discussed using a Hamiltonian formulation.

In fact, there are two alternative  Hamiltonian descriptions of 
the partition
function. 
In the  so-called L-channel representation the r{\^o}le of time 
is taken by $L$:
\bea
  Z_{\alpha\beta}
&=&{\rm Tr}_{\cH_{(\alpha,\beta)}} e^{-LH_{\alpha\beta}(M,R)}
\label{lchantr}
\\[5pt]
&=&\!\!\!\!
 \sum_{E_n\in\,{\rm spec}(H_{\alpha\beta})}\!\!\!\!\!\!
 \exp( - L E_n^{\rm strip}(M,R))
\;,
\label{rrchan}
\eea
while  in  the R-channel representation  the r{\^o}le of time is taken 
by $R$:
\bea
  Z_{\alpha\beta}
&=&   \cev {\alpha}
  \,\exp(-RH_{\rm circ}(M,L))\,
  \vec {\beta}\nn\\[5pt]
&=&\!\!\!\!
 \sum_{E_n\in\,{\rm spec}(H_{\rm circ})}
{  \vev{\alpha|\psi_n}\vev{\psi_n|\beta} \over \vev{\psi_n|\psi_n} }
 \exp( - R E_n^{\rm circ}(M,L) )
\;.
\label{llchan}
\eea
In (\ref{llchan}) we have used the boundary states $\ket{\alpha}$ and
$\ket{\beta}$, and the  eigenbasis $\{ \ket{\psi_n} \}$
of $H_{\rm circ}\,,$
the Hamiltonian propagating states living on a circle of circumference
$L$. 
The two decompositions are illustrated in figures \ref{fig:lchan} and
\ref{fig:rchan}.

\medskip

\[
\begin{array}{cc}
\refstepcounter{figure}
\label{fig:lchan}
\epsfxsize=.44\linewidth
\epsfbox{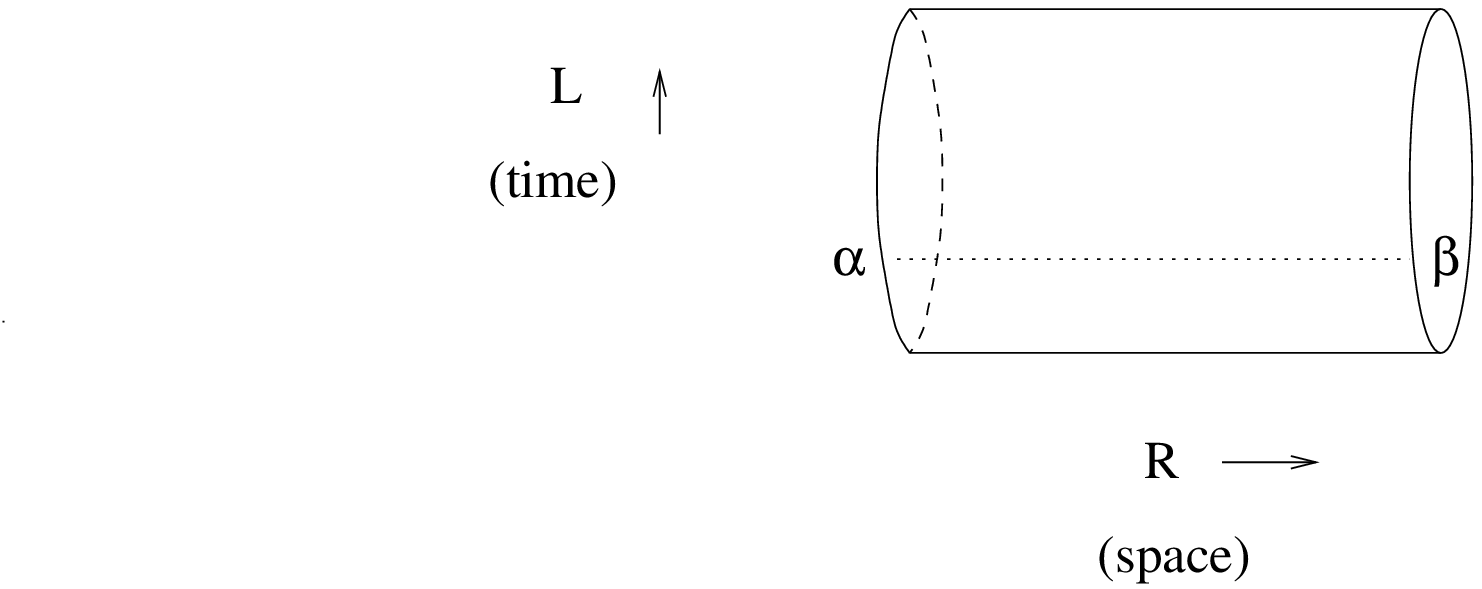}{~~~~} 
  &
\refstepcounter{figure}
\label{fig:rchan}
\epsfxsize=.44\linewidth
{}\epsfbox{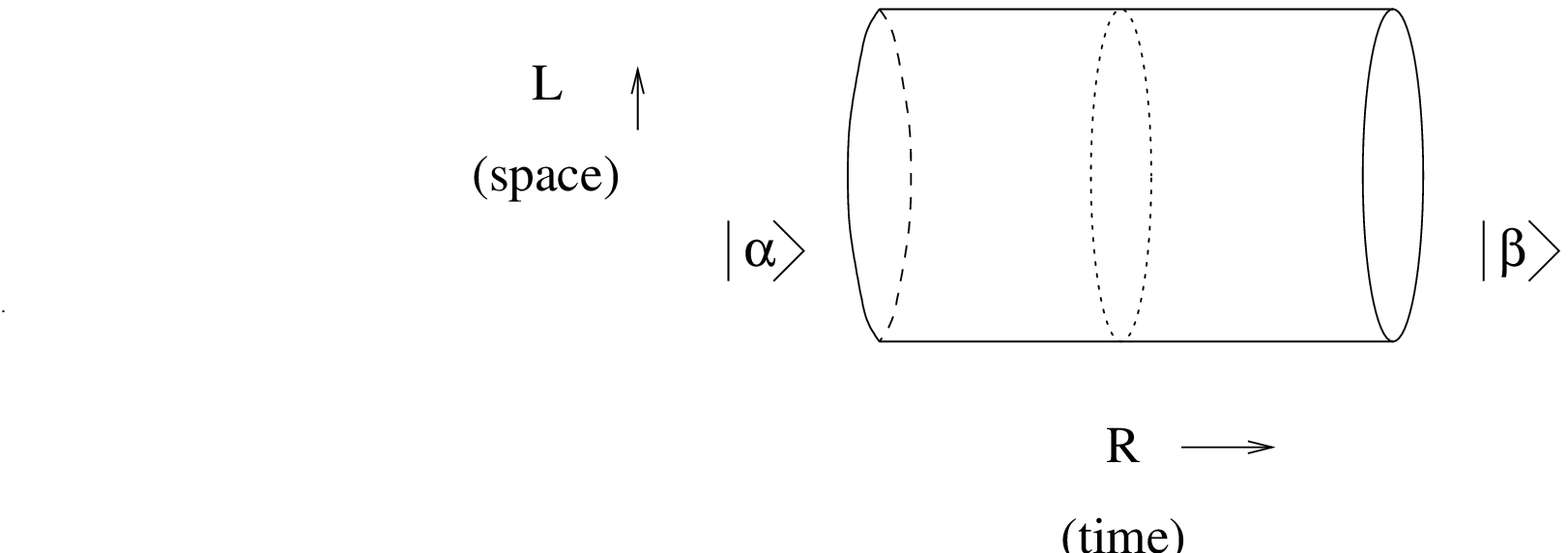}{~~~}
\\
\parbox[t]{.47\linewidth}{\small\raggedright%
Figure \ref{fig:lchan}:\\ 
The L-channel decomposition: states $\ket{\chi_n}$ live on the dotted
line segment across the cylinder.
}~
  &
\parbox[t]{.47\linewidth}{\small\raggedright%
Figure \ref{fig:rchan}:\\
The R-channel decomposition: states $\ket{\psi_n}$ live on the dotted
circle around the cylinder.
}
\end{array}
\]

\medskip

For the rest of this section we will focus on results for the boundary
field $\phi_l$ on the left-hand end of the cylinder, but with a trivial
relabelling it is clear that analogous results for $\phi_r$ can be
found.

\subsection{L-channel  decomposition}
\label{sec:boundaryL}
First we consider  the L-channel representation, depicted in 
figure~\ref{fig:lchan}.
This will enable us to find exact formulae for the expectation value of
$\phi$ on strips of finite width.
Introducing the eigenbasis $\{ \ket{\chi_n} \}$
of $H_{(\alpha,\beta)}$ we differentiate inside the trace (\ref{lchantr})
to find
\eq
\vev{\phi_{\lr}}_{\rm cyl }
=  
{ 1 \over   Z_{\alpha\beta}}
\sum_n {\vev{\chi_n|\phi_{\lr}|\chi_n}  \over \vev{\chi_n|\chi_n}}
\exp({-L \,{E^{\rm strip}_{n}(M,R)}}) \,.
\en
Comparing with (\ref{rrchan}),
\eq
{\vev{\chi_n|\phi_{\lr}|\chi_n}
 \over  \vev{\chi_n|\chi_n}}
=\frac{\partial}{\partial h_{\lr}}
 E^{\rm strip}_{n}(M,R) \,,
\label{vevfromE}
\en
where   
$\vev{\chi_n|\phi_{\lr}|\chi_n}/\vev{\chi_n|\chi_n}$ 
is the expectation value of the  field  
$\phi_{l}$ on the left boundary,
taken in the $n^{\rm th}$ excited state.
 
It was shown explicitly in \cite{Us1} that, at least for small $n$,
$E^{\rm strip}_{n}$
can be computed using generalisations of the boundary TBA
equations of~\cite{LMSS}. For the scaling
\LY\ model this involves a
non-linear integral equation for a single function $\ep(\theta)$:
\eq
  \ep(\theta)
= 2r\cosh\theta-\log\lambda_{\alpha \beta }(\theta)
+ \sum_p \log {S(\theta - \theta_p) \over {S(\theta - \bar{\theta}_p)}}
 - \CK {*}L(\theta) \,,
\label{kermit}
\en
and an associated set of equations for the (possibly empty)
set $\{\theta_p,\bar{\theta}_p\}$ of so-called `active'
singularities~(cf. \cite{DTa,Us1}):
\eq
e^{\ep(\theta_p)}=e^{\ep(\bar{\theta}_p)}=-1~~~~~(\forall p) \,.
\label{kermitaux}
\en
Here, $r=M\!R$ as in earlier sections,
$L(\theta)=\log\bigl(1{+}e^{-\ep(\theta)}\bigr)$,
$f{*}g(\theta)=\frac{1}{2\pi}\iintd'f(\theta{-}\theta')g(\theta')$,
and, for the $(\Phi(h(b_l)),\Phi(h(b_r)))$ boundary conditions,
\eq
\lambda_{\alpha\beta}(\theta)=K_{b_l}(\theta)K_{b_r}(-\theta)~
,\qquad
\CK(\theta)=-i\prtial\log S(\theta)\,,
\en
with 
$K_{b}(\theta)$ defined in (\ref{K-B}), and
$S(\theta)$ the bulk S-matrix~(\ref{SMatrix}).
The number of active singularities depends on the particular 
energy level under consideration; for some pairs of boundary conditions
on the strip it is nonzero even for the ground state~\cite{Us1}, in contrast
to the
situation for the more familiar TBA equations for periodic 
boundary conditions.

The solution to~(\ref{kermit}) for
a given value of $r$ 
determines the  function $c_n(r)$:
\eq
c_n(r)=\frac{6}{\pi^2}\iintd\, r\cosh\theta L(\theta) + i {12 r \over \pi}
\sum_p (\sinh \theta_p -\sinh\bar{\theta}_p) \, , 
\label{piggy}
\en
in terms of which $E_n^{\rm strip}(M,R)$ is
\eq
   E_n^{\rm strip}(M,R)
= \Eblk R
+  f_{b_l}+f_{b_r}
- \frac{\pi}{24 R}c_n(r)
\;,
\label{bath}
\en
and, rewriting (\ref{vevfromE}) in terms of $b_{\lr}$,
\eq
{ \vev{\chi_n|\phi_{\lr}|\chi_n} \over \vev{\chi_n|\chi_n}}  = 
  \left(\frac{\D h}{\D b}\right)^{-1}\!
\frac{\partial}{\partial b_{\lr}}
 E^{\rm strip}_{n}
=\frac{-5~}{\pi |h_{\rm crit}|\cos(\pi(b_{\lr}{+}1/2)/5)}\,
\frac{\partial}{\partial b_{\lr}}
 E^{\rm strip}_{n} 
 \,.
\label{vevfromEb} 
\en

In the expression (\ref{bath}),
$f_{b_l}$ and $f_{b_r}$ are $R$-independent contributions to 
$E_n^{\rm strip}(M,R)$ 
{}from the two boundaries, and $\Eblk$ is the bulk energy per unit length.
If the equations for the particular state under consideration contain
active singularities $\{\theta_p,\bar\theta_p\}$ whose positions do not
tend to zero as $r\to\infty$, then there will be further $R$-independent
contributions to
$E_n^{\rm strip}(M,R)$ coming from the second term on the RHS of
(\ref{piggy}) -- these will be described shortly.
However for ground states such contributions are always
absent, and so (as anticipated by the notation) $f_{b_l}$
and $f_{b_r}$ can be identified with the
extensive parts of the boundary free energies as defined in (\ref{exten}).
The exact values of these
quantities were extracted from the (ground state) TBA equations 
in~\cite{Us1}, both for the 
$\Phi(h(b))$ and the $\One$ boundaries, and are
\be
  {f_b}
= \left( \fract{\sqrt 3 - 1}4 + \sin\fract{\pi b}6 \right)\, M
\;\;\;,\;\;\;  {f_\One}= f_{b=0} \;. 
\label{eq:ftba}
\ee

States $\ket{\chi_n}$ lying above the ground state $\ket{\chi_0}$
will generally be separated from $\ket{\chi_0}$ by a finite energy
gap, even at large $R$. 
At the level of the TBA, this gap is seen in the presence of the
active singularities, mentioned in the last paragraph,
whose positions do not tend
to zero as $r\to\infty$. These give $c_n(r)$ a linear growth in $r=M\!R$,
which via (\ref{bath})
yields an extra constant term in the large-$R$ asymptotic of
$E^{\rm strip}_n(M,R)$.
Physically, the gap arises from two sources: the state 
$\ket{\chi_n}$ may contain a number, $k(n)$ say, of bulk
particles bouncing between
the two edges of the strip, and in addition there may be a boundary bound
state sitting at one or both of the boundaries.
Suppose that the possible boundary bound states for the left-hand boundary
are indexed by $k=0,1,\dots k_{\rm max}(b_l)$ with $k=0$
the boundary ground state,
and likewise for the right boundary, and that for
the state $\ket{\chi_n}$ the left and right boundaries are in states $k_l(n)$,
$k_r(n)$ respectively.
Then the general behaviour for $R\to\infty$ is as follows:
\eq
      E_n^{\rm strip}(M,R)
\sim \Eblk R +M k(n)
   +  f_{b_l}+f_{b_r}
   +  e_{k_l(n)}(b_l)
   +  e_{k_r(n)}(b_r)
\;,
\label{largeR}
\en
where $e_k(b)$ is the energy of the $k^{\rm th}$ boundary bound state of the
$\Phi(h(b))$ boundary condition (with $e_0(b)=0$, since $k=0$ is just the
boundary ground state).

Taking the limit $R\to\infty$ of the ground state expectation value
allows the boundary
expectation values on the semi-infinite plane to be recovered. 
For the state $\chi_0\,$, $k(0)=k_l(0)=k_r(0)=0$, and
substituting 
(\ref{largeR}) into 
(\ref{vevfromEb}) using (\ref{eq:ftba}),
we obtain the  exact value of the 
dimensionless expectation value 
$\vev{M^{1/5} \phi}$ in a semi-infinite geometry:
\mathematica{
 M = kappa lambda^(5/12);
 vp =  
 (Pi/6) *
 Cos[ Pi b / 6 ] * M / D[h,b];
}    
\eq
 \vev{M^{1/5} \phi}
= - \frac{5}{6\,|\hhc|}\,
    \frac{ \cos(\pi b/6)}
         { \cos( \pi(b+1/2)/5)}~.
\label{eq:vevphi}
\en
For now we take $b$ to be restricted to the `fundamental region'
$-3<b<2$, in which case the above formula is indeed correct for the
ground state. Its interpretation as $b$ moves outside this region
will be described in section~\ref{bndryfl} below.

Returning to finite values of $R$, 
the asymptotic (\ref{largeR}) no longer
suffices to obtain the expectation value of the boundary field, and one has
rather to differentiate the exact formula (\ref{bath}), so that
(\ref{eq:vevphi}) becomes
\eq
 \vev{M^{1/5} \phi_l}(r)
= - \frac{5}{6\,|\hhc|}\,
    \frac{ \cos(\pi b_l/6)-\frac{1}{4r}\,\frac{\partial}{\partial
    b_l}c(r)}
         { \cos( \pi(b_l+1/2)/5)}~,
\label{eq:vevphib}
\en
where $\frac{\partial}{\partial b_l}c(r)\equiv 
\frac{\partial}{\partial b_l}c_0(r)$ can be obtained by
differentiating the full TBA equation (\ref{kermit}) 
(a similar idea was applied to the case of
periodic boundary conditions in \cite{Zb}). 
Restricting, for simplicity, to the ground state energy 
in the region where no active singularities are present, the
terms involving the $\theta_p$ in (\ref{kermit}) are absent, and
\eq
\frac{\partial}{\partial b_l} c(r)= -
\frac{6}{\pi^2}\iintd\, r\cosh\theta {\eta(\theta) \over
1+\exp(\ep(\theta))}~,
\label{c1}
\en  
where $\ep(\theta)$
solves~(\ref{kermit}),  
while $\eta(\theta)=\frac{\partial}{\partial b_l} \ep(\theta)$
can be obtained from the linear integral 
equation
\eq
\eta(\theta) 
= 
 -\frac{\partial}{\partial b_l} \log\lambda_{\alpha \beta }(\theta)
+\CK {*}{\eta \over 1+\exp(\ep)}(\theta)~.
\label{l1}
\en 
In this way the final estimates for $ \eta(\theta)$ 
and $\ep(\theta)$ have roughly the same accuracy.
 
We also estimated $\vev\phi$ using the TCSA.
For this, we used a strip geometry with specific
boundary conditions $(\alpha,\beta)$ on the two edges. 
We then calculated the dimensionless expectation value
$\vev{\,M^{1/5}\phi_{\lr}\,}_{\alpha,\beta}(\ML) \;,$
as a function of the strip width $\ML$ for finite truncation
level $N$.
While $\vev{M^{1/5}\phi_l}_{\alpha,\beta}$ can depend strongly on
$\beta$ for small $\ML$, as $\ML$ increases, this dependence
decreases, and as $\ML\to\infty$ it approaches the half-plane value
$\vev{M^{1/5}\phi}_\alpha$.
In table \ref{tab:Table1_1} we compare the TCSA estimates
for  $\vev{M^{1/5}\phi}_{\alpha,\beta}$ for  
$\alpha = \Phi(0)$ ,  $\beta = \One$ and  $\ML$ small 
with  the 
numerical solution of the 
ground state  
TBA equations~(\ref{vevfromE},\ref{kermit},\ref{kermitaux}) and 
(\ref{c1},\ref{l1}). A similar agreement was found 
for other pairs of boundary conditions.

In table \ref{tab:Table1} we also give 
$\vev{M^{1/5}\phi}$ for the same boundary conditions for larger values
of $\ML$ to show the convergence to the IR value, and 
include several plots of $\vev{M^{1/5}\phi}$ for various boundary
conditions in figure \ref{fig:Graph3b3}.

\begin{table}[ht]
\renewcommand{\arraystretch}{1.2}
\refstepcounter{table}
\label{tab:Table1_1}
\[
\begin{array}{r|lllll}
\multicolumn{1}{c}{}
     & \multicolumn{5}{c}{\log(\ML)}   \\
     & \multicolumn{1}{c}{-8}   
     & \multicolumn{1}{c}{-6}   
     & \multicolumn{1}{c}{-4}   
     & \multicolumn{1}{c}{-2}   
     & \multicolumn{1}{c}{0}   \\  
\hline

\hbox{TCSA} & {-0.19684202180}    
        & {-0.29365378867}
        & {-0.4380797960}
        & {-0.65350591}
        & {-0.969216} \\ 

\hbox{TBA}  & {-0.19684202181}
        & {-0.29365378869}
        & {-0.4380797961}
        & {-0.65350592}
        & {-0.969217}                      \\
\hline
\multicolumn{1}{c}{}&
\multicolumn{5}{c}{\parbox{.85\linewidth}{\raggedright
 ~ \\
Table \ref{tab:Table1_1}:\\
The TCSA estimates, with truncation level $N=14$, of 
 $\vev{M^{1/5}\phi}_{\alpha,\beta}$ for  
$\alpha = \Phi(0)$ ,  $\beta = \One$ in the small $\ML$ region
compared with the  numerical prediction from the TBA  
}}
\end{array}
\]
\end{table}

\begin{table}[ht]
\refstepcounter{table}
\label{tab:Table1}
\[
\begin{array}{rr|lllll}
 &   & \multicolumn{5}{c}{\log(\ML)}   \\
 &   & \multicolumn{1}{c}{0}   
     & \multicolumn{1}{c}{1}   
     & \multicolumn{1}{c}{3/2}   
     & \multicolumn{1}{c}{2}   
     & \multicolumn{1}{c}{5/2}   \\
\hline
 & 6 & \mm{-0.969212}{-1.321108} & \mm{-1.129791}{-1.202314}
     & \mm{-1.166399}{-1.179004} & \mm{-1.175226}{-1.174604} & \mm{-1.182737}{-1.175250} \\ 
 & 8 & \mm{-0.969214}{-1.321116} & \mm{-1.129788}{-1.202350} 
     & \mm{-1.166291}{-1.179061} & \mm{-1.174404}{-1.174612} & \mm{-1.178406}{-1.174563} \\
N&10 & \mm{-0.969215}{-1.321120} & \mm{-1.129795}{-1.202370}  
     & \mm{-1.166281}{-1.179099} & \mm{-1.174229}{-1.174643} & \mm{-1.176402}{-1.174357} \\
 &12 & \mm{-0.969216}{-1.321122} & \mm{-1.129801}{-1.202382} 
     & \mm{-1.166281}{-1.179124} & \mm{-1.174121}{-1.174679} & \mm{-1.175346}{-1.174337} \\
 &14 & \mm{-0.969216}{-1.321124} & \mm{-1.129806}{-1.175194} 
     & \mm{-1.166288}{-1.179142} & \mm{-1.174104}{-1.174709} & \mm{-1.175196}{-1.174342} \\
\multicolumn{7}{c}{\parbox{.7\linewidth}{\raggedright
 ~ \\
Table \ref{tab:Table1}:\\
The TCSA estimates of
$\vev{M^{1/5}\phi}_{\Phi(0),\One}(\ML)$ (upper data)
and 
$\vev{M^{1/5}\phi}_{\Phi(0),\Phi(0)}(\ML)$ (lower data).
These can be compared with the `exact' IR value of 
$-1.17459499975...$
}}
\end{array}
\]
\end{table}

\newpage

\subsection{R-channel  decomposition}
\label{sec:boundaryR}

We now turn to the R-channel representation, depicted in
figure~\ref{fig:rchan}.  This will ultimately lead to expressions for
expectation values when the boundary field is placed at one end of a
semi-infinite cylinder. We shall use the   notation of \cite{Us3} and
denote the boundary states $\vec{\alpha}$ and  $\vec{\beta}$ 
as $\vec{\Phi(h_l)}$ and $\vec{\Phi(h_r)}$ respectively.

If 
$R$ is taken to infinity 
in (\ref{llchan}) 
with all other variables held 
fixed, then the contribution of the ground state 
$\ket{\psi_0}\equiv\ket{\Omega}$
will dominate the spectral
sum. Thus
\eq
  Z_{\alpha\beta}
\sim
  A_{\alpha\beta}(M,L)\,  \exp( - R E_0^{\rm circ}(M,L) )
\;,
\label{lrasympt}
\en
where $E^{\rm circ}_0(M,L)$ is the ground state energy of 
$H_{\rm circ}$
and
\eq
  A_{\alpha\beta}(M,L)
=   \frac
  {  \veev{\alpha}{\Omega}
  \, \veev{\Omega}{\beta}  }
  {\veev \Omega\Omega}
\;.
\label{eq:aml}
\en
If we now let $L$ grow as well and compare with 
(\ref{exten}), we see that
the inner products appearing in (\ref{eq:aml})
will, in general, contain a  term corresponding to a boundary
free energy per
unit length:
\eq
  \log(\, \frac{ \veev\Omega{\alpha} }{\veev\Omega\Omega^{1/2}}\, )
= -Lf_\alpha
+ \log(g_\alpha(M,L))
\;.
\label{logg}
\en
On the other hand, this linear term can be extracted from the small-$L$
behaviour of the functions $\log(g_\alpha(M,L))$, for which `L-channel' TBA
equations\footnote{the terminology is
unfortunately, but unavoidably,
a little confusing -- these equations are called `L-channel'
because their derivation proceeds via the L-channel representation of the
partition function.}
were proposed in \cite{LMSS}. 
This is explained in \cite{Us3}, where a precise match with the 
earlier result (\ref{eq:ftba}) was found.
The consistency between these two determinations of $f_b$ is in some
respects a mystery, since, from other results reported in~\cite{Us3},
there  are good reasons  to doubt the ability of the L-channel TBA
equations of \cite{LMSS} to describe the full variation of
$\log(g_\alpha(M,L))$ as a function of $ML$.

Returning to the R-channel decomposition of the full partition function,
we have
\eq
\vev{\phi_{l}}_{\rm cyl }
=  
 -{ 1 \over  L  Z_{\alpha\beta}}
\sum_n  
{{\rm d}\over {\rm d} h_{l}} 
\vev{\Phi(h_l)|\psi_n} 
{ \vev{\psi_n|\Phi(h_r)} 
\over   \vev{\psi_n|\psi_n}}  
\exp({-R \,{E^{\rm circ}_{n}(M,L)}}) \,
\en
The following identification was made in the massless \cite{BLZ1} and
massive \cite{Us3} cases:
\eq
\vev{\Phi(h(b))|  \psi_n}=
Y_n(i \pi \fract{b+3}{6}) \vev{\One|\psi_n} \,, 
\label{g=y}
\en  
where 
$ Y_n(\theta) = \exp(\ep_n(\theta))$, and
$\ep_n(\theta)$ is the solution of the excited-state 
TBA equation (with periodic boundary conditions) for the state
$\ket{\psi_n}$ (see \cite{BLZ2,DTa}).
Taking the limit $R \rightarrow \infty$ one deduces that 
\eq
    \vev{\phi_l}
=   \frac {\vev{\Phi(h)\,|\,\phi_l\,| \Omega}}
          {\vev{\Phi(h)\,     |      \,\Omega}}
= -{1 \over L} 
\left(\frac{\D h}{\D b}\right)^{-1}\partial_b \log Y(i \pi \fract{b+3}{6}) \,
\label{fermat}
\en
which gives  the 1-point  expectation value of $\phi$ acting  on the 
end of an 
infinite cylinder of circumference $L$ in terms of the 
function $Y$. 
One check on this formula is easily made: from the large-$L$ limit of
the TBA equation for $Y$, we have
\eq
 \log Y(i \pi \fract{b+3}{6}) \sim  -ML \sin \fract{\pi b}{6} = -L 
(f_b-f_{\One}) \qquad(L\to\infty)
\en
using (\ref{eq:ftba}) in the second equality; differentiating,
the expectation value  on the upper half-plane quoted 
in~(\ref{eq:vevphi}) is recovered.

There is evidence (see~\cite{DTa}) that the
full set of ``excited'' $Y_n$'s can be 
obtained from  $Y$  via a process of
analytic continuation in the 
bulk perturbing parameter $\lambda$. 
It then seems reasonable to suppose  that 
(\ref{fermat}) can also be continued, leading to the following general 
relation:
\eq
   \vev{\phi_l}^{(n)}
=  \frac {\vev{\Phi(h)  \,|\,\phi_l\,|\,\psi_n}}
         {\vev{\Phi(h) \,     |    \,\psi_n}}
= 
 - \frac 1L \left(\frac{\D h}{\D b}\right)^{-1}
   \partial_b \log Y_n(i \pi \fract{b+3}{6})
 \;.
\en       
(Alternatively, one can obtain this simply by differentiating 
(\ref{g=y}).)

It turns out that there is a further consequence of the R-channel
decomposition. Note first
that  (\ref{g=y}) can be used to write the 
partition function as
\eq
Z_{\Phi(h(b_r)),\Phi(h(b_l))} =
\sum_n  
{\vev{\One|\psi_n}\vev{\psi_n|\One} 
\over
\vev{\psi_n|\psi_n}}
Y_n(i \pi \fract{b_r+3}{6})
Y_n(i \pi \fract{b_l+3}{6})
\exp({-R \,{E^{\rm circ}_{n}(M,L)}}) \,.
\en
Now, the  Y's satisfy the  functional 
relation~\cite{AlZY}
\eq
Y_n(\theta+ i {\pi \over 3} )Y_n(\theta- i {\pi \over 3} )  
=1 + Y_n(\theta) \,,
\en
and this
suggests the following identity 
\eq
Z_{\Phi(h(b+2)),\Phi(h(b-2))}=Z_{\One,\One}+
Z_{\One,\Phi(h(b))}\,.
\label{zid}
\en
This should hold for all $M$, $R$ and $L$; from the L-channel
decomposition (\ref{rrchan}), it is equivalent to the following 
relation between the spectra of models on strips of equal widths
but different boundary conditions:
\eq
\{  E^{\rm strip}_n(M,R)   \}_{\Phi(h(b+2)),\Phi(h(b-2))} = 
\{  E^{\rm strip}_n(M,R)   \}_{\One,\One}
\cup
\{  E^{\rm strip}_n(M,R)   \}_{\One,\Phi(h(b))} \, .
\label{eid}
\en
Preliminary numerical work confirms this rather surprising identity, but as
yet we do not have a good physical understanding of its origin. However, in
section \ref{rgfs} below
it will be used to formulate an exact conjecture concerning
the regions of the $(b,b')$ plane for which the model with
$(\Phi(h(b)),\Phi(h(b')))$ boundary conditions develops a boundary-induced
vacuum instability.

\newpage
\resection{Applications}
\label{sec:discussion}

In this section we apply some of the results obtained above
to elaborate a few further aspects of the boundary scaling
\LY\ model. We start with the boundary flows of the semi-infinite
system, and then turn to the way that the boundary-induced
vacuum instability of the model 
is affected when the system is confined to a finite strip.

We first recall the way the parametrisations of the
TCSA, TBA and FF calculations are related.
The physical parameter describing the $\Phi(h)$ boundary is, of
course, $h$, which is related to $b$ by
\eq
  h
=
  - \, |\hc|
  \,\sin( \pi (b+1/2)/5) 
\;.
\en
This physical parameter $h$ is periodic in $b$ with  period 10, and we
choose the fundamental domain to be $-3\leq b\leq 2$. Thus for real
$b$, $h$ is restricted to  the range 
  $-|\hc| \leq h \leq |\hc|$. 
To reach real values of 
  $h>|\hc|$ 
we can continue in $b$ to values
$b=-3+i\hat b$ with $\hat b$ real; to reach real values of 
  $h<-|\hc|$ 
we
can continue in $b$ to values $b=2+i\hat b$ with $\hat b$ real. 

As we pointed out in section \ref{tevovitfes}, if we formally continue
in $b$ outside the fundamental region we have to be careful, as
quantities may not continue naively -- in particular
the continuation of the `fundamental' reflection factor to $2<b<5$ in
fact describes the reflection properties of the boundary with the
addition of (the lowest) boundary bound state.

\subsection{Boundary flows on the semi-infinite system}
\label{bndryfl}

It is quite straightforward to see the boundary flow, at the level of
the one-point functions, using the TCSA.
In \cite{Us1} the spectra for the strip with boundary
conditions $(\Phi(h),\One)$ were calculated
for several values of $h$, and they are
consistent with the idea that this spectrum is real for all real
   $h>-|\hc|$.
Thus we can calculate the expectation values of $\varphi(x)$ and
$\phi$ on the boundary $\Phi(h)$ by looking at the large $\ML$ limit
of calculations on the strip with bcs $(\Phi(h),\One)$.
The only restriction is that the TCSA errors increase sharply
with $|h|$, so that TCSA results are restricted to a small range of
$b$ values centred on $b=-1/2$.

The FF and `exact' (TBA) results have more interesting properties.
Recall that the FF calculations are formally functions of 
\be
    g_\Phi(b) 
= \frac{ \tan((b+2)\pi/12)}{ \tan((b-2)\pi/12)}\, g_{\One}
\;,\;\;\;\;
    K_b(\theta)
= R_b(i\fract\pi 2 - \theta)
\;.
\label{eq:FFdata}
\ee
The continuation to large positive value of $h$ through 
$b=-3 + i\hat b$ works well, as one can easily verify that
none of the $b$-dependent poles in $K_b(\te)$
cross the integration contours in the FF integrals (\ref{eq:FF}), 
and so none of the subtleties described in section~\ref{tevovitfes}
above arise. 
The 
formulae
\[
\ds \lim_{\hat b \to \infty}\;
   g_\Phi(-3 + i \hat b) 
=
  g_\One
\;,\;\;
\ds  \lim_{\hat b \to \infty} \;
  K_{(-3 + i \hat b)}(\te) 
=
   K_0(\te) = K_\One(\te)
\;,
\]
can therefore
be substituted directly into the form factor expansion, 
establishing the result
\eq
\lim_{h\to+\infty}\vev{M^{2/5}\varphi(x)}_{\Phi(h)}=
\vev{M^{2/5}\varphi(x)}_{\One}
\label{vevcon}
\en
as an exact identity.
One can similarly discuss the continuation of the expectation value of
the boundary field, finding, as expected,
\[
\ds  \lim_{\hat b \to \infty} \; 
  \vev{M^{1/5}\phi}_{\Phi(h(-3 + i\hat b))}
= 0 =
  \vev{M^{1/5}\phi}_{\One}
\;.
\]

To illustrate the result (\ref{vevcon}),
in figure \ref{fig:Graph9} we plot $\vev{M^{2/5}\varphi(x)}$ close to
the $\Phi(h)$ boundary for various values of $h$.
The FF results smoothly interpolate between the $\One$ and $\Phi(h)$
boundaries, and give good agreement with the TCSA results for small
value of $|h|$.
The $h\to\infty$ limit is not directly accessible in TCSA, but we
have extrapolated our results in $h$ and we can see that (modulo an
amplification of the Fourier-type truncation errors) it shows every
sign of converging on the expectation value in the $\One$ bc.

As was discussed in \cite{Us1}, for $h>h(-1)$, in the massive
system on the half-plane the $\Phi(h)$ boundary
has no boundary bound states, and flows
to the $\One$ boundary condition as $h\to+\infty$.
For $h(1)<h<h(-1)$ this boundary has one bound state, and for
   $-|\hc|=h(2)<h<h(1)$
  there are two boundary bound states; at 
   $h=-|\hc|$
  the
ground state and first excited states become degenerate and for 
$h$ less than this critical value 
   $-|\hc|$,
the system does not have a
real vacuum. 
The presence of the bound states can be understood as particles being
trapped near the boundary, as we have seen in section \ref{sec:exx}.

\[
\begin{array}{c}
\refstepcounter{figure}
\label{fig:Graph9}
\epsfxsize=.85\linewidth
\epsfbox[72 262 540 530]{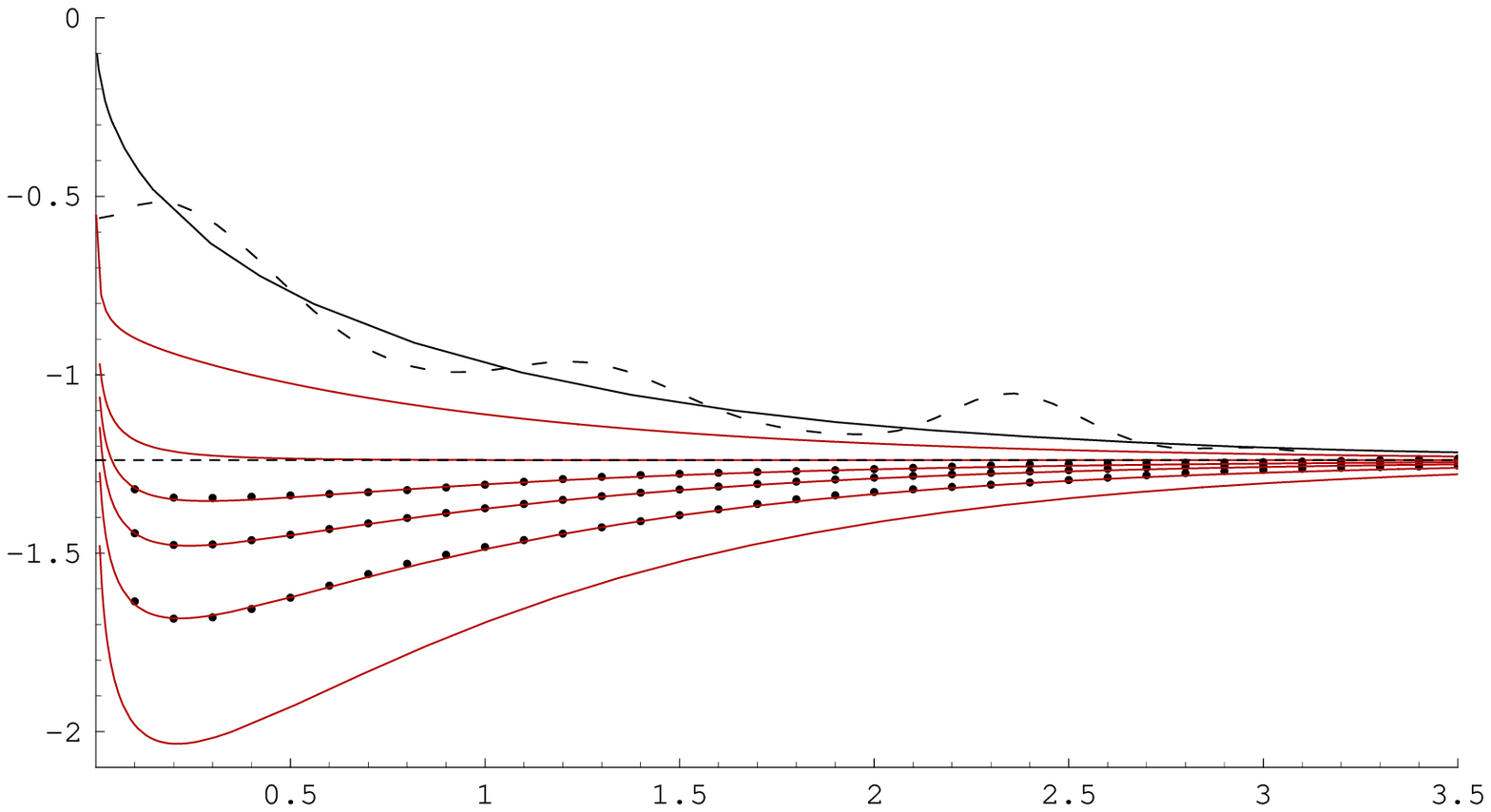}
\\
\parbox[t]{.8\linewidth}{\raggedright%
Figure \ref{fig:Graph9}\\\small
Plots of 
$\vev{\;M^{2/5}\varphi(x)\;}_{\Phi(h(b))}$
with boundary condition $\Phi(h(b))$.
The solid lines are the Form-Factor results 
truncated at three particles with $b$ taking values (from second line
{}from top to bottom) $-3 + 3 i, -2,-1,-1/2,0,1/2$.
The points are the TCSA results $\cI(\xi)$ with truncation level 14
and $\ML\,{=}\,8$.
We have also included a plot of 
$\vev{M^{2/5}\varphi(x)}_{\One}$ (top line), 
the extrapolation of the TCSA data to $h=\infty$ (long dashed line) 
and the bulk value (short dashed line) for comparison.
}
\end{array}
\]

The continuation of the FF results beyond $b=2$ was already discussed
in section \ref{tevovitfes}. The boundary expectation values can
also be continued, and here we find a repetition
of the reflection factor results -- the
ground state and first excited state expectation values swap under 
$b \to 4-b$, and the second excited state is invariant.
To recall, the boundary free energy and the excitation energies of the
two lowest lying states (for $1\,{\leq}\,b\,{\leq}\,2$) are
\[
  f_b = \left( \fract{\sqrt 3 - 1}4 + \sin(\pi b/6)\right) M
\;,\;\;\;
  e_1(b) = M\,\cos(\pi(b+1)/6)
\;,\;\;\;
  e_2(b) = M\,\cos(\pi(b-1)/6)
\;.
\]
This gives~(cf.~eqs.~(\ref{vevfromEb}) and (\ref{largeR})) 
the expectation value 
$\vev{M^{1/5}\phi}^{(n)}$ in these three lowest lying states as
\[
\renewcommand{\arraystretch}{0.9}
\begin{array}{rclcl}
  \vev{M^{1/5}\phi}^{(0)}
&\!\!=\!\!& M^{1/5}
 \frac{\D}{\D h} f_b
&\!\!=\!\!& \ds
   ~~
   \left(\frac{5}{6|\hhc|}\right) 
   \frac{ \cos(\pi b/6 ) }{ \sin(\pi(b-2)/5 ) }
\;,
\\[4mm]
  \vev{M^{1/5}\phi}^{(1)}
&\!\!=\!\!&  M^{1/5}
 \frac{\D}{\D h} ( f_b + e_1 )
&\!\!=\!\!& \ds
   -
   \left(\frac{5}{6|\hhc|}\right) 
   \frac{ \cos(\pi (4-b)/6 ) }{ \sin(\pi(b-2)/5 ) }
\;,
\\[4mm]
  \vev{M^{1/5}\phi}^{(2)}
&\!\!=\!\!&   M^{1/5}
 \frac{\D}{\D h} ( f_b + e_2 )
&\!\!=\!\!& \ds
   -
   \left(\frac{5 \sqrt{3}}{6|\hhc|}\right) 
   \frac{ \sin(\pi (b-2)/6 ) }{ \sin(\pi(b-2)/5 ) }
\;.
\end{array}
\]
Note that $\vev\phi^{(0)} = \vev\phi^{(1)}$ at the threshold for the first
excited state $b\,{=}\,-1$, and  
$\vev\phi^{(0)} = \vev\phi^{(2)} $ at the threshold for the second
excited state $b\,{=}\,1$.

Unfortunately the TCSA does not give very good results for the
expectation values in the excited states for large $\ML$, and while
extrapolations in $\ML$ and truncation level indicate that these
results are indeed correct, there seems little point in showing any
plots. 

\subsection{RG flows on the finite size strip}
\label{rgfs}

In \cite{Us1} the spectrum of the model with $(\One,\Phi)$ boundary
conditions was described in some detail, but results for
the $(\Phi,\Phi)$ system were not presented. The analysis of these
results provides a nice application of the spectral identity found in
section~\ref{sec:boundaryR} above, and in this section we describe how this
goes.  We begin with some results from the TCSA.

In figures  
\ref{fig:Graph4c}--\ref{fig:Graph4} 
the finite size spectrum  is plotted for the system on a strip with
boundary conditions
$(\Phi(h(b_l)),\Phi(h(b_r)))$ with 
$b_l=b_r=b$ taking values $-1/2,0,1/2$. 
Observe that the ground and first excited states cross for $b=0$,
and for $b=1/2$ an interval appears in which the ground state has left
the real spectrum. 
However, so long as $h>-|\hc|$ the large-$\ML$ spectrum stays real.

In figure \ref{fig:Graph3b3} we plot
the log of the expectation value of the boundary field
$\vev{M^{1/5}\phi}_{\alpha,\beta}$ against $\log(\ML)$,
for the two pairs of boundary conditions
$(\alpha,\beta) = (\Phi(h(b)),\One)$ 
and 
$(\alpha,\beta) = (\Phi(h(b)),\Phi(h(b)))$ 
for $b$ taking the values ${-}1/2, 0, 1/2$.
For large $\ML$, the expectation values in the two boundary
conditions converge to the same value, as we expect; for large $\ML$
the influence of the right boundary on the left boundary decreases and
the expectation value tends to the half-plane value,
$\vev{\phi_l}_{\alpha,\beta} \to_{\ML\to\infty} \vev{\phi}_\alpha$,
independent of $\beta$ (provided that the system is not destabilised
by the boundary condition $\beta$).
Conversely, for small $\ML$, $\vev{\phi}_{\alpha,\beta}$ tend to the
conformal limits (\ref{eq:c1ptfnsb}) which are governed by the UV fixed
point of the boundary flow.
On figure \ref{fig:Graph3b3} the conformal limits are shown as dotted
straight lines -- the expectation value for $(\alpha,\beta) =
(\Phi(h(b)),\One)$  all converge to the lower straight line in the UV,
and those for $(\alpha,\beta) = (\Phi(h(b)),\Phi(h(b)))$  all
converge to the upper line.

The most interesting behaviour is that shown by
$\vev{M^{1/5}\phi}_{\Phi(h(b)),\Phi(h(b))}$
for intermediate values of $\ML$.
For $b\,{=}\,0$ we saw in figure \ref{fig:Graph4b} that the ground
state and first excited states cross at one point; at the same point 
$\vev\phi$ diverges with a characteristic `$\lambda$' behaviour.
For $b\,{=}\,1/2$ the ground state drops out of the real spectrum for
a finite range of $\ML$; $\vev\phi$ diverges at the edge points of this
range.

\[
\begin{array}{cc}
\refstepcounter{figure}
\label{fig:Graph4c}
\epsfxsize=.45\linewidth
\epsfbox[72 142 540 650]{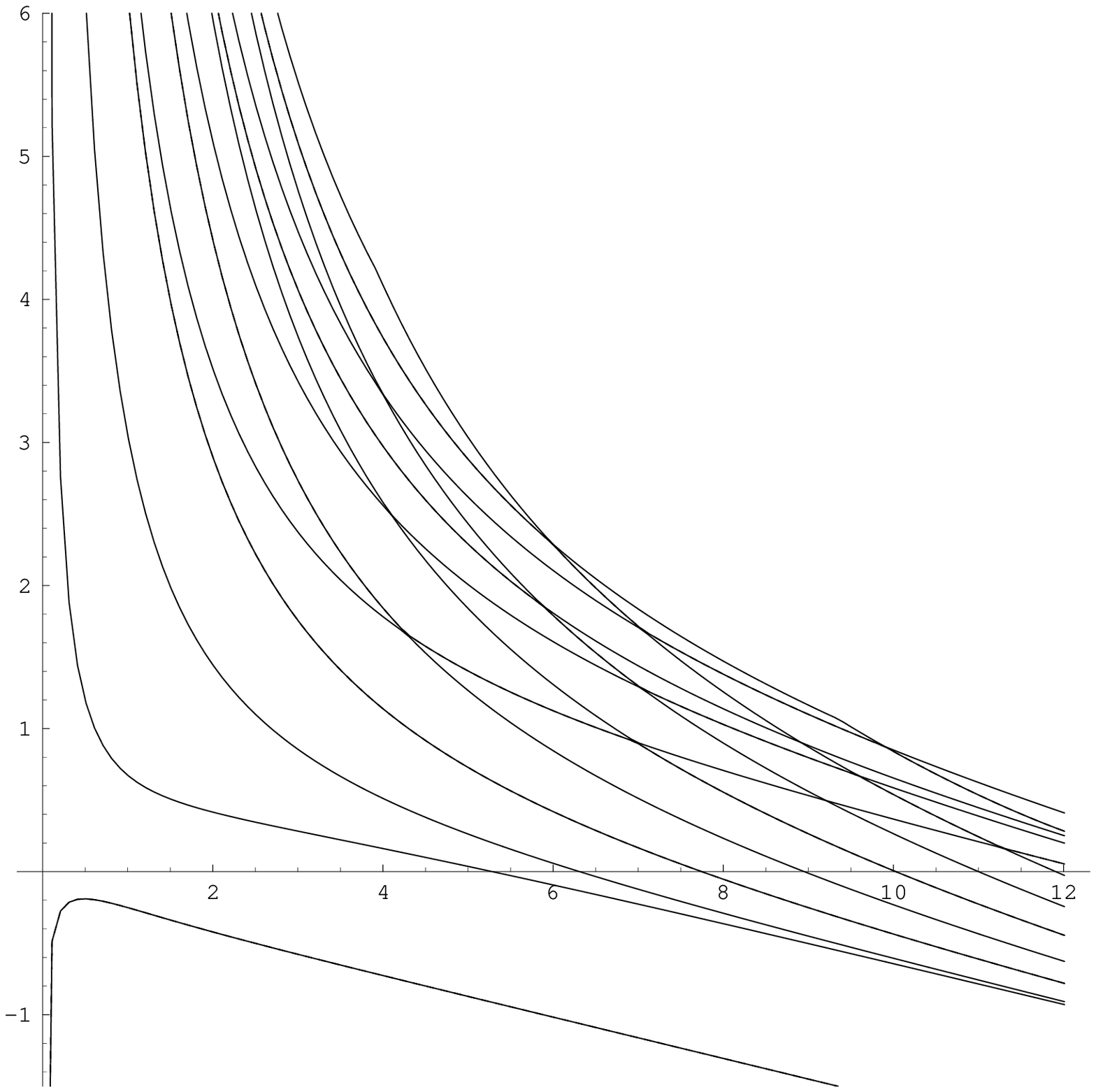} 
  &
\refstepcounter{figure}
\label{fig:Graph4b}
\epsfxsize=.45\linewidth
\epsfbox[72 142 540 650]{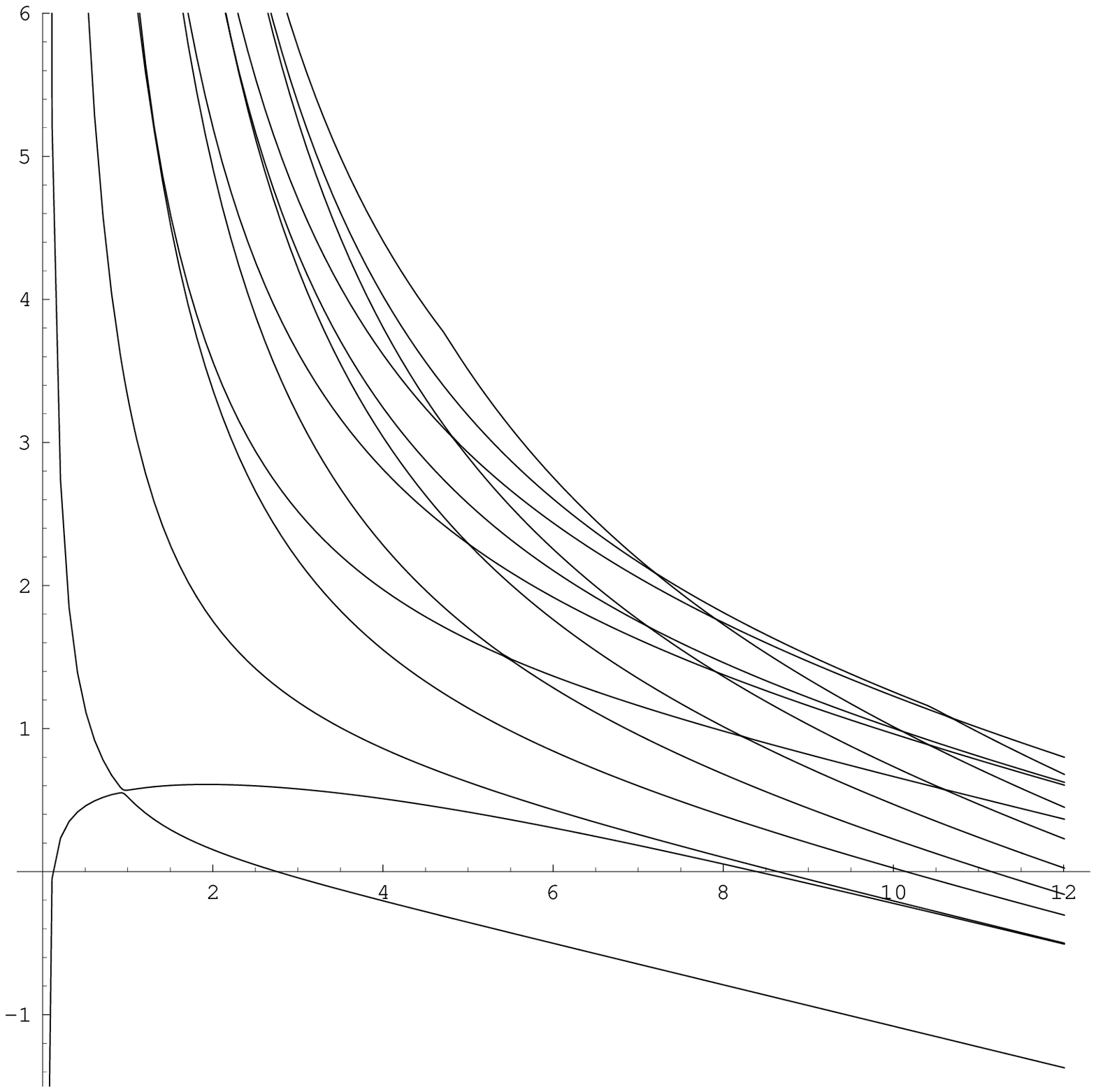}
\\
\parbox[t]{.4\linewidth}{\small\raggedright%
Figure \ref{fig:Graph4c}:
The spectrum of the model on a strip with boundary conditions
$b_l\,{=}\,b_r\,{=}\,{-}1/2$, plotted against $\ML$.
}
  &
\parbox[t]{.4\linewidth}{\small\raggedright%
Figure \ref{fig:Graph4b}:
The spectrum of the model on a strip with boundary conditions
$b_l\,{=}\,b_r\,{=}\,0$, plotted against $\ML$.
}
\\
\refstepcounter{figure}
\label{fig:Graph4}
\epsfxsize=.45\linewidth
\epsfbox[72 142 540 650]{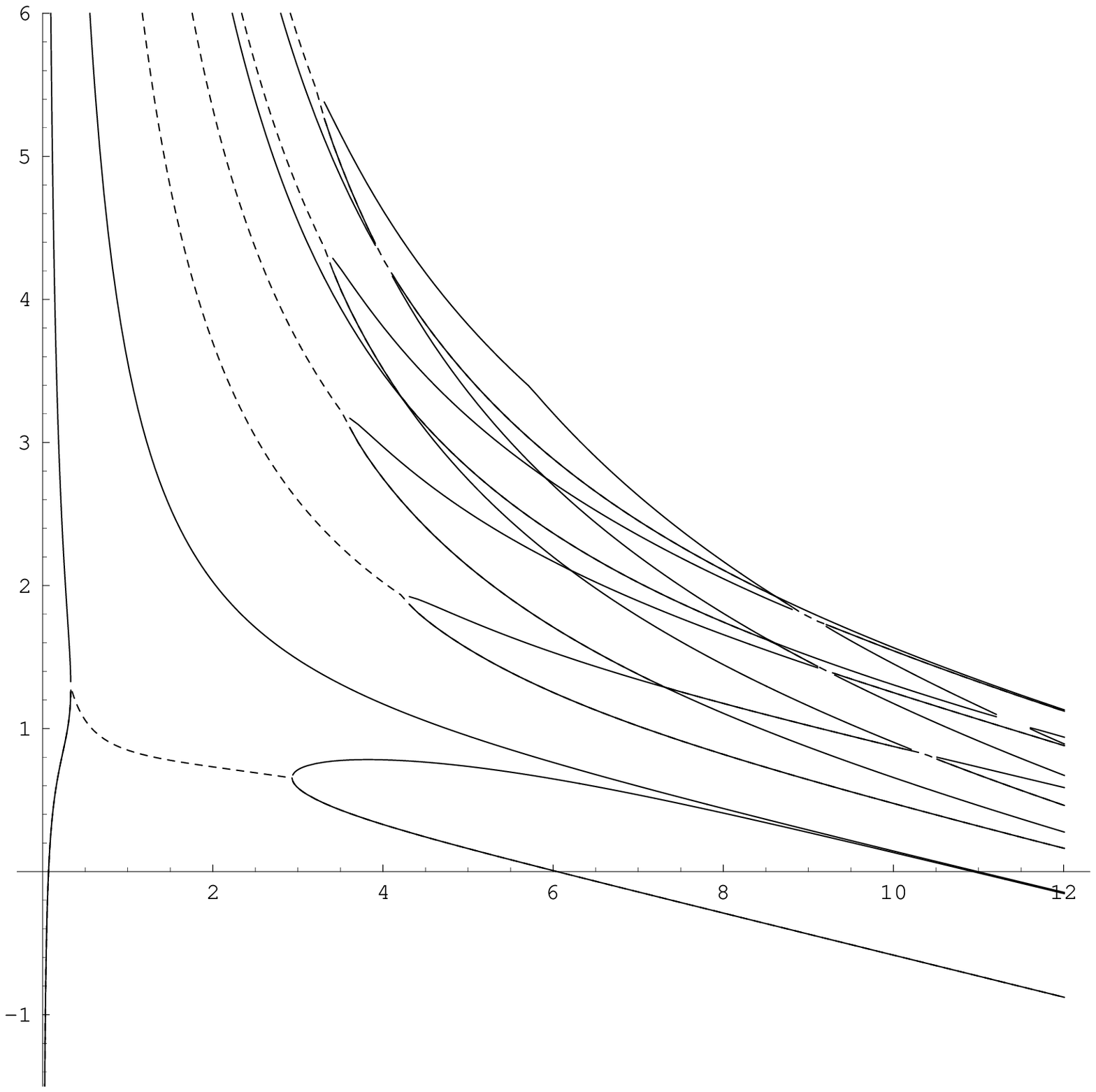}
  &
\refstepcounter{figure}
\label{fig:Graph3b3}
\epsfxsize=.45\linewidth
\epsfbox[72 142 540 650]{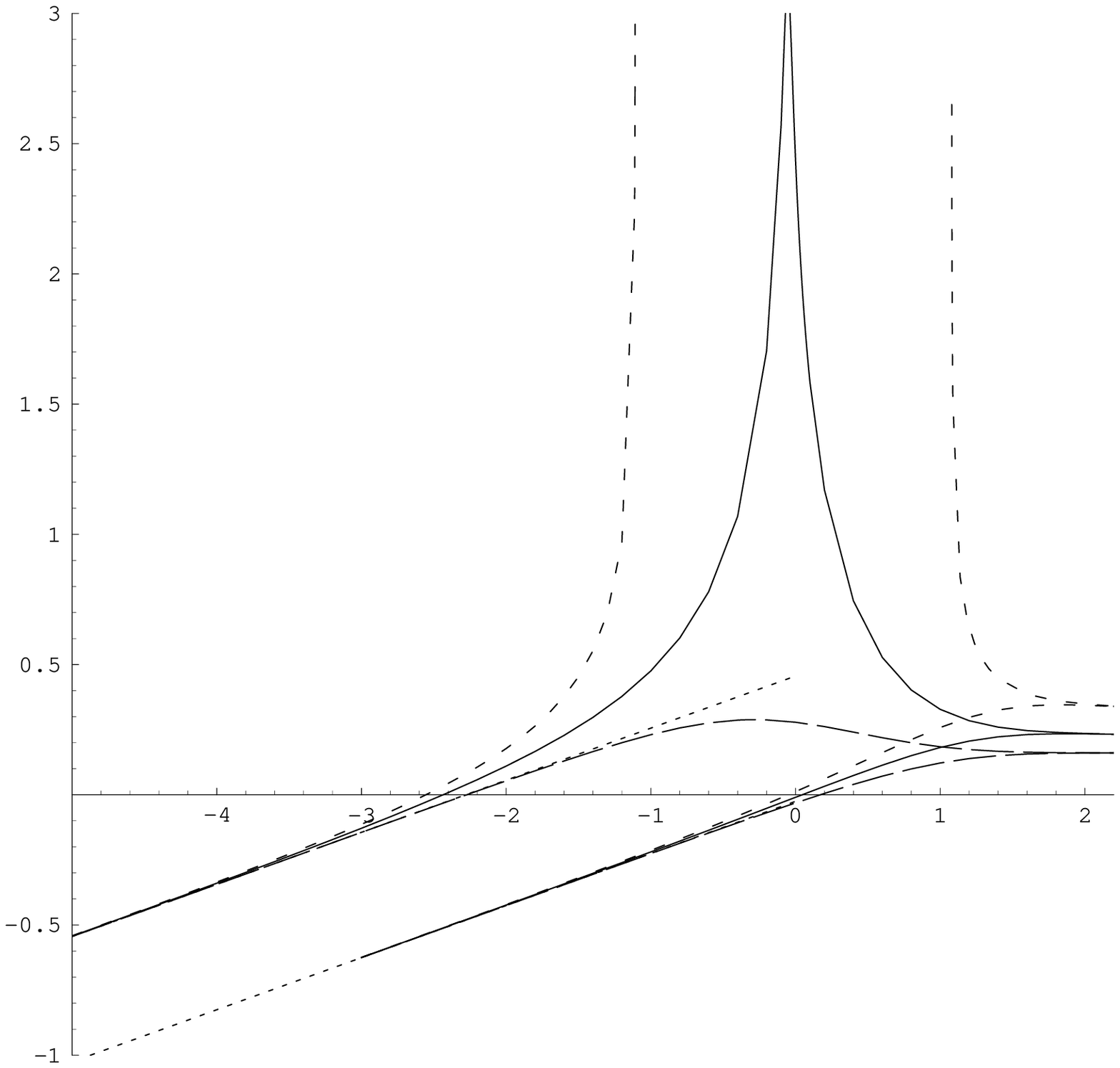}
\\
\parbox[t]{.4\linewidth}{\small\raggedright%
Figure \ref{fig:Graph4}:
The spectrum of the model on a strip with boundary conditions
$b_l\,{=}\,b_r\,{=}\,1/2$, plotted against $\ML$.\\
The dashed lines indicate the real part of a pair of complex conjugate
eigenvalues.
}
  &
\parbox[t]{.45\linewidth}{\small\raggedright%
Figure \ref{fig:Graph3b3}:
Plots of
$\log(-\vev{M^{1/5}\phi}_{\alpha,\beta})$
against $\log\ML$ for pairs of boundary conditions
$(\Phi(h(b)),\Phi(h(b)))$ and $(\Phi(h(b)),\One)$
for
$b\,{=}\,{-}1/2$ (long dashed lines),
$b\,{=}\,0$ (solid lines) and
$b\,{=}\,1/2$ (short dashed lines).
Also shown (dotted lines) are the conformal expressions
(\ref{eq:c1ptfnsb}).
See text for more details.
}
\end{array}
\]

\newpage

The appearance of level crossing can be understood by making
use of the spectral identity (\ref{eid}). Shifting $b$, and using the
symmetry of the model about $b=2$, the identity can be written as
\eq
\{  E^{\rm strip}_n(M,R)   \}_{\Phi(h(-b)),\Phi(h(b))} = 
\{  E^{\rm strip}_n(M,R)   \}_{\One,\One}
\cup
\{  E^{\rm strip}_n(M,R)   \}_{\One,\Phi(h(2{-}b))} \, .
\label{eidd}
\en
Consider now the two ground state energies 
which appear on
the RHS of this
equation. Depending on their relative values, one or other will
correspond to
ground state of the model on the LHS.
Two limits are easily analysed. As $R\to 0$, the fact that the
$(\One,\Phi)$ conformal Hilbert space contains the field $\psi$ while the
$(\One,\One)$ does not shows that
\bea
R\to 0~:
\qquad
\qquad
\qquad
E^{\rm strip}_0|_{(\One,\One)}&\sim & +\infty 
\qquad \qquad \qquad \qquad
\qquad \qquad \qquad
\nn\\
E^{\rm strip}_0|_{(\One,\Phi(h(2{-}b)))}&\sim &-\infty
\label{levdegns}
\eea
In the opposite,
large-$R$, limit, the behaviours follow from (\ref{largeR}):
\bea
R\to \infty~:
\qquad
\qquad
\qquad
E^{\rm strip}_0|_{(\One,\One)}
&\sim & 
\Eblk R+ (\sqrt{3}-1)M~,\nn\\
E^{\rm strip}_0|_{(\One,\Phi(h(2{-}b)))}
&\sim & 
\Eblk R+ (\sqrt{3}-1 +\sin\fract{(2{-}b)\pi}{6})M~.
\qquad
\label{levdegn}
\eea
In the latter formula, without loss of generality, we
took $b$ to be positive. 
Comparing (\ref{levdegns}) with (\ref{levdegn}) shows that the relative
values of the two ground states on the RHS of (\ref{eidd}) swap over 
when going from small to large values of $R$ whenever $b$ is less than
$2$.
Since these states can be identified with the two lowest-lying levels of
the $(\Phi,\Phi)$ model on the LHS of
(\ref{eidd}), we see that a level-crossing in this model is inevitable
in all such cases.
Thus the point $(0,0)$ in the $(b_l,b_r)$ plane, shown in
figure \ref{fig:Graph4b},
belongs to a whole line of points $(b,-b)$, $|b|<2$
which also exhibit a level crossing.
The fact that the relevant two states are taken from the
spectra of distinct models on the RHS of (\ref{eidd}) prohibits their
mixing and ensures that the crossing will be exact. 
Once the line $b_l+b_r=0$
is left, the identity (\ref{eidd}) can no longer be invoked and
the exact level crossing is lost, as can be seen in
figures \ref{fig:Graph4c} and \ref{fig:Graph4}.
Observe in the first of these the lowest two levels remain real, 
while in the second there is an intermediate range of $R$
for which their energies become complex. Physically this can be
explained as follows: 
in the first situation, $b_l+b_r<0$ and the boundary fields are
less strong than in the `marginal' case of $b_l+b_r=0$, and hence have less
chance to destabilise the model; and in the second the story is reversed,
the boundary fields are stronger, and there is therefore a possibility of
a vacuum instability for some finite values of $R$.
Once this has been understood it is reasonable to conjecture that for all
points $(b_l,b_r)$ in the fundamental domain $-3\le b_l,b_r\le 2$ with
$b_l+b_r>0$ the model with
$(\Phi(h(b_l)),\Phi(h(b_r)))$ boundary conditions
exhibits a vacuum
instability for some range of system sizes, 
while 
for the points below this line,
the spectrum remains entirely real at all values of $R$. 

This picture is confirmed by TCSA
plots analogous to figures~\ref{fig:Graph4c}--\ref{fig:Graph4}
taken at various other values of $b_l$ and $b_r$, and
leads to the
phase diagram for the model shown in figures~\ref{fig:phbb}
and~\ref{fig:phhh}.
The change in coordinates from $(b_l,b_r)$ to $(h_l,h_r)$ 
in passing between the two figures transforms the line segment
$(b,-b)$, $|b|<2$ into the  portion of the ellipse 
$  (h_l{+}h_r)^2/\sin^2\frac{\pi}{10}
 + (h_l{-}h_r)^2/\cos^2\frac{\pi}{10}
 = h_{\rm crit}^2$ 
on figure \ref{fig:phhh} which touches the (shaded) region of instability.
(Along the rest of the ellipse, the spectral identity (\ref{eidd})
also holds, but does not imply a level crossing.)
Note also that whenever either $h_l<\hc$ or $h_r<\hc$, the vacuum is
already unstable in infinite volume and so this region can  immediately
be shaded in, without the need to appeal to more subtle arguments. 
No markedly new features emerge in the spectra for $b_l\neq b_r$, so
we will not  show any further plots.
However, we note that the value of $R$
at which the level-crossing occurs for models on the line $b_l+b_r=0$ 
diverges as $b_l$ (or $b_r$) tends to $2$. This is as one would expect,
since in this limit the value of one of the
boundary fields is approaching $-|\hc|$, and the
corresponding boundary bound
state
is becoming degenerate with the vacuum in infinite
volume.

\[
\begin{array}{cc}
\refstepcounter{figure}
\label{fig:phbb}
\epsfxsize=.4\linewidth
\epsfbox[0 0 389 387]{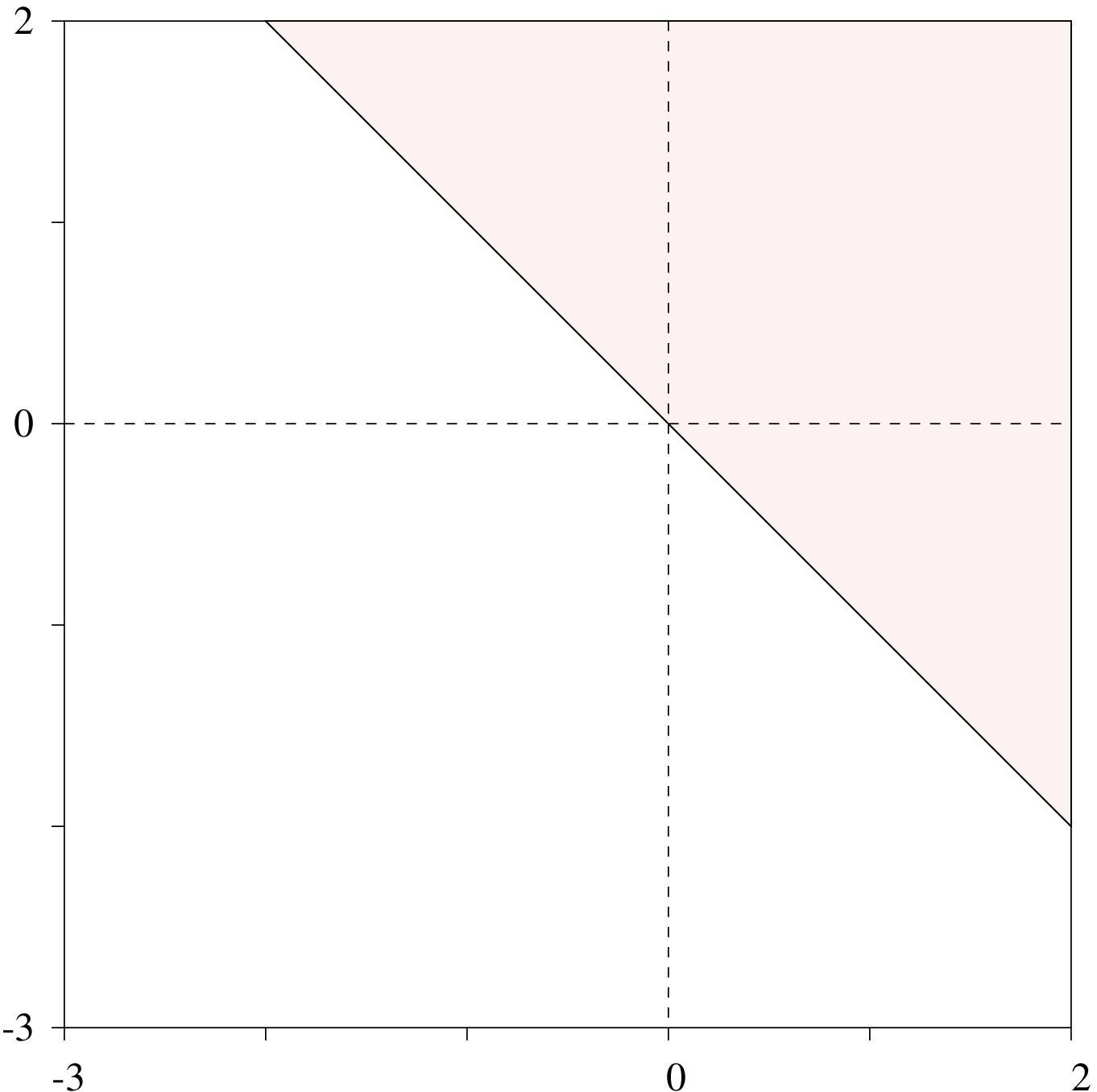}{~~~~~~~} 
  &
\refstepcounter{figure}
\label{fig:phhh}
\epsfxsize=.4\linewidth
{}\epsfbox[0 -12 378 376]{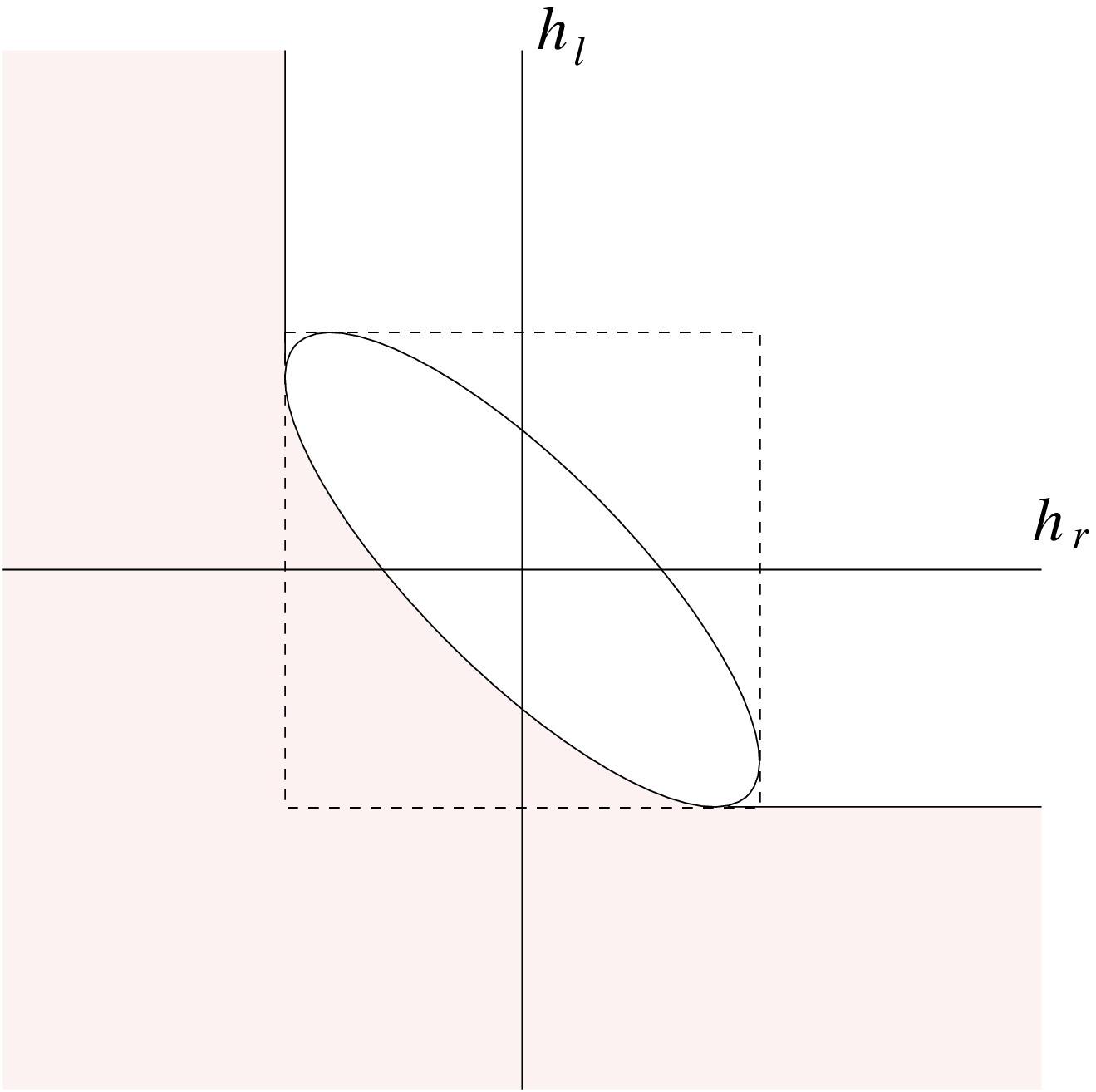}
\\
\parbox[t]{.47\linewidth}{\small\raggedright%
Figure \ref{fig:phbb}:\\ The fundamental region in the $(b_l,b_r)$
plane, showing (shaded) the region within which, for
at least one value of the
strip width, the model exhibits a boundary-induced instability.
}~
  &
\parbox[t]{.47\linewidth}{\small\raggedright%
Figure \ref{fig:phhh}:\\
As figure~\ref{fig:phbb}, but plotted on the $(h_l,h_r)$ plane.
The inner dashed line, a symmetrically-placed square of size $2\hc$,
indicates the extent of the region covered for real
values of $b_l$ and $b_r$.
}
\end{array}
\]

\resection{Conclusions}
\label{concl}

In this paper we have given a rather detailed analysis of the one-point
functions of both bulk and boundary fields in a simple but nonetheless
non-trivial integrable quantum field theory, the scaling Lee-Yang model.
A number of different techniques have been explored, with results which
have been shown to be in good accord. Previous work on this topic has
tended to use just one method (most usually the form-factor expansion) 
and thus the consistency that we have found with between the different
approaches is an important confirmation that the earlier studies have 
been well-founded. For this purpose the use of the scaling Lee-Yang model
as a testing-ground is very natural, but we feel that it would be very
worthwhile to extend this work to further theories, and work on such
matters is currently in progress. 

There are a number of open questions that arise.
The modification to Ghoshal and Zamolodchikov's boundary state 
that we were forced to make in order
to reconcile FF and TCSA results remains a
numerical observation, for which we have no terribly compelling physical
argument. Furthermore, in section 4.2 we noted that the 
the boundary states could sometimes be analytically continued, yielding
`excited' boundary states associated with the boundary bound states of
the semi-infinite line. A systematic understanding of off-critical
boundary states, both  in infinite and finite geometries, is
still lacking and would presumably shed some light on the questions
raised by our observations. It also might help to understand at a more
fundamental level how the
modifications to the boundary TBA equations of \cite{LMSS}, found in
\cite{Us1} by an indirect method of analytic continuation, arise.

It would also be worth investigating boundary-to-boundary correlation
functions such as
\eq
{1 \over L^2} \vev{ (\int\!\! \phi_l) (\int\!\! \phi_r)} -
\vev{\phi_l} \vev{\phi_r}
\;.
\en
These are accessible by a simple generalisation of the techniques
explained in section 6.1 above, and might be amenable to
comparison with, for example, the results of lattice simulations.
We also remark that the one-point functions at the end of an infinite
cylinder calculated in section 6.2 are essentially the analogues, for 
the boundary fields of a semi-infinite system, of the finite-temperature
expectation values discussed in \cite{LMa,Luka}. In the model we
discussed these were relatively easy to obtain as the field under
consideration was also the boundary perturbing operator.  For more
general fields in more complicated models it would presumably be
necessary to develop the discussion more along the lines of the work in
\cite{Luka}. This seems to us an important open problem which should
certainly be investigated further.

Finally, we remark that the partition function identity discovered in
section 6.2 merits further study. In some senses it can be considered as
a first off-critical extension of the identities between (sums of)
conformal partition functions that can be observed on examining the lists
of such objects provided in, for example, \cite{BPZb}.
A physical understanding of why such 
identities  should exist is lacking, even in the conformal cases, and
perhaps the broader perspective provided by the off-critical results will
help towards this end. This alone should motivate the extension our
work on off-critical boundary integrable models to further examples.


\bigskip

\noindent{\bf Acknowledgements --- }
{
\parindent 0pt
\parskip 3pt
\raggedright

The work was supported in part by a TMR grant of the European
Commission, contract reference ERBFMRXCT960012, in part by a NATO
grant, number CRG950751, and in part by EPSRC grants GR/K30667 and 
GR/L26216.

PED and GMTW thank the EPSRC for Advanced Fellowships, 
MP thanks the EPSRC for support and 
RT thanks the Universiteit van Amsterdam for a post-doctoral
fellowship.  

We would like to thank 
Ph.~Di~Francesco,
A.~Leclair,
G.~Mussardo, I.~Runkel, G.~Tak\'acs,
Al.~Zamolodchikov and J.-B.~Zuber for helpful discussions.
GMTW would also like to thank 
R.~Guida and N.~Magnoli for many discussions of TCSA methods,
Brian Davies for very helpful discussions on the numerical
analysis of the results,
and SPhT for hospitality during the final stages of this work.

}

\vskip 1.5cm

\newpage
\small
\renewcommand\baselinestretch{0.95}


\begin{thebibliography}{99}
\raggedright
\parskip 1pt

\bibitem{BLZ1}
V.V.\ Bazhanov, S.L.\ Lukyanov and A.B.\ Zamolodchikov,
{\em Integrable Structure of Conformal Field Theory, Quantum KdV Theory and 
Thermodynamic Bethe Ansatz},
Commun. Math. Phys. 177 (1996) 381-398\xtra{9412229}

\bibitem{BLZ2}
V.V.\ Bazhanov, S.L.\ Lukyanov and A.B.\ Zamolodchikov,
{\em Integrable quantum field theories in finite volume: excited state
energies},
\NP{B489} (1997) 487--531\xtra{9607099}

\bibitem{BPZb}
R.E.~Behrend, P.A.~Pearce and J.-B.~Zuber,
{\em Integrable boundaries, conformal boundary conditions and A-D-E
fusion rules},
\JP{A31} (1998) L763--L770\xtra{9807142}

\bibitem{CM}
J.L. Cardy and G.Mussardo,
\newblock
{\it S matrix of the Yang-Lee edge singularity in two dimensions},
\newblock  Phys.~Lett.~{\bf B225} (1989) 275--278.

\bibitem{Us1}
P.~Dorey, A.~Pocklington, R.~Tateo and G.~Watts,
{\it
TBA and TCSA with boundaries and excited states},
\NP{B525} (1998) 641--663\xtra{9712197}

\bibitem{Us3}
P.~Dorey, I.~Runkel, R.~Tateo and G.~Watts,
{\em
$g$--function flow in perturbed boundary conformal field theories},
\newblock \NP{B578} (2000) 85--122\xtra{9909216}

\bibitem{DTa}
P.\ Dorey and R.\ Tateo,
{\em Excited states by analytic continuation of TBA equations},
\NP{B482} (1996) 639--659\xxtra{9607167};\\
{}~~~---~~
{\em Excited states in some simple perturbed conformal field theories},
\NP{B489} (1998) 575-623\xtra{9706140}


\bibitem{Us2}
P.~Dorey, R.~Tateo and G.~Watts,
{\em
Generalisations of the Coleman-Thun mechanism and boundary reflection
factors},
\newblock \PL{B448} (1999) 249-256\xtra{9810098}

\bibitem{Usnext}
P.~Dorey, R.~Tateo and G.~Watts,
{\em in preparation.}


\bibitem{FLZZ} 
V.A.~Fateev, S.~Lukyanov, A.B.~Zamolodchikov and Al.B.~Zamolodchikov,
{\em
Expectation values of local fields in Bullough-Dodd model and integrable
perturbed conformal field theories},
\NP{B516} (1998) 652-674\xtra{9709034}

\bibitem{GZ} 
S.~Ghoshal, A.B.~Zamolodchikov,
{\em Boundary S matrix and boundary state in two-dimensional integrable
quantum field theory},
\IJMP{A9} (1994) 3841-3886, 
erratum \IJMP{A9} (1994) 4353\xtra{9306002}

\bibitem{GMagn1}
R.~Guida and N~Magnoli,
{\em
Vacuum expectation values from a variational approach},
\newline
\PL{B411} (1997) 127-133\xtra{9706017}

\bibitem{KLMu1}
R.~Konik, A.~Leclair and G.~Mussardo,
{\em 
On Ising correlation functions with boundary magnetic field},
\newblock
\IJMP{A11} (1996) 2765\xtra{9508099}

\bibitem{FFextrab} 
A.~Leclair, F.~Lesage and H.~Saleur
{\em Exact Friedel oscillations in the g=1/2 Luttinger liquid},
\PR{B54} (1996) 13597--13603\xtrac{9606124}

\bibitem{LMa}
A.~Leclair and G.~Mussardo,
{\em Finite temperature correlation functions in integrable QFT},
\NP{B552} (1999) 624--642\xtra{9902075}

\bibitem{LMSS}
A.~Leclair, G.~Mussardo, H.~Saleur and S.~Skorik,
{\em Boundary energy and boundary states in integrable quantum field
theories},
\NP{B453} (1995) 581-618\xtra{9503227}


\bibitem{FFextrac} 
F.~Lesage and H.~Saleur,
{\em Form-factors computation of Friedel oscillations in Luttinger liquids},
\JP{A30} (1997) L457--L463\xtrac{9608112}

\bibitem{Luka}
S.~Lukyanov,
{\em Finite temperature expectation values of local fields in the
sinh-Gordon model}\xtra{0005027}


\bibitem{FFextraa} G.~Mussardo,
{\em Sprectral  representation of correlation functions in
two-dimensional quantum field theories},
Talk given at the international colloquium on modern QFT II, Tata
Institute, Bombay, January 1994\xtra{9405128}

\bibitem{MP} M.~Pillin,
{\em The form factors in the Sinh-Gordon model}, 
\newline
\IJMP{A13} (1998) 4469-4486\xtra{9712033}

\bibitem{SM} 
F.A.~Smirnov,
{\em 
The perturbated $C < 1$ conformal field theories as reductions of
sine-Gordon model},
\IJMP{A4} (1989) 4213-4220;\\
{}~~~---~~
{\em Reductions of the sine-Gordon model as a perturbation of minimal models
of conformal field theory},
\NP{B337} (1990) 156-180.

\bibitem{SM2} F.A. Smirnov, {\it Form factors in completely 
integrable models of quantum field theory}, 
\newline
Adv.~Series in Math.~Phys.~14 (World Scientific, Singapore, 1992).


\bibitem{YZ}
V.P.Yurov and \AlBZ,
{\em Truncated conformal space approach to the scaling Lee-Yang model,}
\IJMP{A5} (1990) 3221.


\bibitem{Zb}
\AlBZ, 
{\it
Thermodynamic Bethe Ansatz in Relativistic Models. Scaling
 3-state Potts and Lee-Yang Models}, 
\newblock
\NP{B342} (1990) 695-720.

\bibitem{ZAM} 
Al.B.~Zamolodchikov,
{\em Two point correlation function in scaling Lee-Yang model},
\newline
\NP{B348} (1991) 619.

\bibitem{Zg}
\AlBZ, 
{\em
Mass scale in sine-Gordon model and its reductions},
\newline
\IJMP{A10} (1995) 1125-1150.

\bibitem{AlZY}
\AlBZ,
{\em On the thermodynamic Bethe ansatz equations for the reflectionless
ADE scattering theories}
\PL{B253} (1991) 391-394.


\end{thebibliography}
\end{document}